\documentclass[10pt,conference]{IEEEtran}
\usepackage{makeidx,epsfig}
\usepackage{graphicx}
\usepackage{setspace,graphicx}

\usepackage{booktabs}
\usepackage{multirow}
\usepackage{hyperref}

\usepackage{comment} 

\usepackage{amsmath}
\usepackage{algorithm}
\usepackage[noend]{algpseudocode}

\usepackage[normalem]{ulem} 

\usepackage{threeparttable}
\usepackage{pifont}

\newcommand\eat[1]{}

\makeatletter
\renewcommand{\@makefnmark}{\hbox{\textsuperscript{\tiny{\@thefnmark}}}}
\makeatother

\usepackage{footnote}
\makesavenoteenv{tabular}
\makesavenoteenv{table}

\usepackage{amsmath}
\usepackage{cite}
\usepackage{url}
\usepackage[table,xcdraw]{xcolor}

\usepackage{subfigure}

\newcommand{\subparagraph}\textit{{}}

\usepackage{lipsum}

\begin{document}
\title{\huge A Survey on Industrial Internet of Things (IIoT) Testbeds for Connectivity Research}

\author{
\IEEEauthorblockN{
Tianyu Zhang, 
Chuanyu Xue, 
Jiachen Wang, 
Zelin Yun,
Natong Lin,
Song Han
}

\IEEEauthorblockA{School of Computing, University of Connecticut}  
Email: \{tianyu.zhang, chuanyu.xue, jc.wang, zelin.yun, natong.lin, song.han\}@uconn.edu
}

\maketitle

\thispagestyle{plain}
\pagestyle{plain}

\begin{abstract}
Industrial Internet of Things (IIoT) technologies have revolutionized industrial processes, enabling smart automation, real-time data analytics, and improved operational efficiency across diverse industry sectors. IIoT testbeds play a critical role in advancing IIoT research and development (R\&D) to provide controlled environments for  technology evaluation before their real-world deployment.
In this article, we conduct a comprehensive literature review on existing IIoT testbeds, aiming to identify benchmark performance, research gaps and explore emerging trends in IIoT systems. 
We first review the state-of-the-art resource management solutions proposed for IIoT applications. 
We then categorize the reviewed testbeds according to their deployed communication protocols (including TSN, IEEE 802.15.4, IEEE 802.11 and 5G) and discuss the design and usage of each testbed. 
Driven by the knowledge gained during this study, we present suggestions and good practices for researchers and practitioners who are planning to design and develop IIoT testbeds for connectivity research. 
\end{abstract}

\vspace{0.05in}

\IEEEpeerreviewmaketitle

 \section{Introduction}
\label{Sec:intro}

Internet of Things (IoT), the technologies to interconnect physical devices with computing and networking capabilities, has been referred to as ``the infrastructure of the information society"~\cite{martynov2019information}. Not surprisingly, there have been various extensions of IoT technologies to different industry sectors such as healthcare~\cite{dimitrov2016medical}, military~\cite{gotarane2019iot}) and industrial automation~\cite{sisinni2018industrial}. 

Industrial IoT (IIoT), the focus of this article, is the application of IoT technologies in industrial automation environments, including advanced process automation and factory automation~\cite{da2014internet,breivold2015internet,sisinni2018industrial}. Compared with IoT systems designed for consumer applications, IIoT systems are usually deployed in harsh and complex environments, and have stringent dependability, timing performance (e.g., latency and jitter), energy-efficiency and security requirements to optimize manufacturing quality and productivity, and to avoid potentially catastrophic consequences. A more recent trend in advanced industrial automation is to connect interdependent factories together through IIoT technologies to provide decentralized, collaborative and sometimes immutable and verifiable services (e.g., distributed supply chain management). These unique requirements on IIoT systems pose many challenges in their communication fabric design, distributed data management, analysis and decision making, and security protection for both the communication and data infrastructure.

To meet the stringent performance requirements of IIoT solutions, it is fundamental to develop novel resource management approaches, analysis and decision making strategies, and security protection methods. However, to validate the correctness and evaluate the performance of these methodologies, it requires a comprehensive testing infrastructure.  Researchers generally rely on a controlled experimental environment, called \textit{testbed}, designed to conduct well-defined experiments, test new technologies, evaluate systems' performance, and validate solutions before their real-world deployment. 

In the context of IIoT, testbeds may include physical setups comprising a network of interconnected sensors, actuators, and devices to represent an industrial ecosystem. Physical testbeds provide a high level of realism, closely mimicking real-world IIoT scenarios and challenges. But setting up and maintaining physical testbeds can be costly and complex, especially for large-scale deployments. 
Alternatively, testbeds can be virtual or cloud-based environments where IIoT solutions are tested using simulations to reduce cost and increase flexibility and scalability. 
Hybrid testbeds may also exist, combining elements of both physical and virtual testbeds, offering a flexible and adaptable environment that blends real hardware with virtual components.

Testbeds offer many benefits to the design of IIoT systems, enabling researchers, engineers, and developers to explore, validate, and optimize their IIoT solutions. 
However, the development of a new testbed is not straightforward, instead it is challenging from different points of view, ranging from implementation costs, sharing capability, and fidelity. 
For example, the testbed must accurately represents real-world industrial scenarios and mimic the complexities of industrial settings, including diverse devices, heterogeneous networks, and varying environmental conditions. 
Testbeds should adhere to standardized methodologies and configurations to ensure that experiments can be replicated and compared across different studies. Managing and analyzing the vast amount of data generated by IIoT testbeds can also be challenging where efficient data storage, processing, and analytics mechanisms is desired to derive valuable insights from testbed experiments. Last but not least, it is always desired to reduce the cost associated with acquiring hardware, software, and infrastructure components to build and maintain an IIoT testbed. 

However, \textbf{to the best of our knowledge, there lacks a comprehensive literature review of the existing IIoT testbeds}, which motivates us to fill this gap. To this end, this survey paper aims to explore emerging trends in IIoT technology and deployment by understanding what has been done and what areas need further exploration to build a valuable IIoT testbed. A comprehensive study enables researchers to benchmark the performance of different IIoT testbeds and compare their strengths and weaknesses. This comparative analysis aids in understanding the best practices and challenges in designing effective IIoT testbeds. 
A thorough literature study can also serve as a valuable resource for educational purposes to teach students and professionals about IIoT testbeds, their significance, and their role in IIoT research and development. 
Note that, the definition of IIoT itself encompasses a wide range of applications and scenarios. However, this article focuses on the connectivity aspect of IIoT testbeds, encompassing all components involved in data transmission and connectivity within the implemented systems. This includes transmission devices, communication protocols, network architectures, as well as resource management, among others. 

The remainder of this paper is organized as follows. Section~\ref{sec:related} gives an overview of the existing survey papers in this field, and highlights how our study differs from them. Section~\ref{sec:overview} introduces the IIoT reference architecture and background of IIoT communication protocols. 
Section~\ref{sec:tsn}, Section~\ref{sec:15.4}, Section~\ref{sec:11} and Section~\ref{sec:5g} present the details of existing IIoT testbeds deploying TSN, IEEE 802.15.4, IEEE 802.11 and 5G as the main communication protocols, respectively. Based on the extensive analysis of all the reviewed IIoT testbeds, in Sec.~\ref{sec:lesson} we present some advice and good practices for researchers that want to design and develop IIoT testbeds for connectivity research. 
Finally, we conclude the paper in Section~\ref{sec:concl}. 
 \section{Related IIoT Testbed Surveys}\label{sec:related}

The connectivity of IIoT, cyber-physical systems (CPSs) and wireless sensor-actuator networks (WSANs) environments is a very broad field of research, and one may find a myriad of related surveys (e.g., \cite{sisinni2018industrial,sanchez2016state,da2014internet,xu2018survey,mirani2022key,shi2011survey,humayed2017cyber,gunes2014survey,giraldo2017security,chen2017applications,yick2008wireless,akyildiz2002wireless,ramson2017applications,ojha2015wireless}). Most of these surveys in the literature mainly focus on summarizing the research activities on the definition/concept of IIoT architecture, protocol stack, standardization, as well as identifying research trends and challenges in this area. The contribution of these works lies on providing a systematic overview of the state-of-the-art research efforts and potential research directions to solve Industrial IoT challenges. 
However, to the best of our knowledge, there is no comprehensive survey gathering and summarizing the landscape of the developed testbed facilities in industrial scenarios in a systematic manner. 

In this section, we perform the literature study on the existing in-depth surveys on testbeds constructed for industrial-related use cases. Due to the existence of many closely related concepts bearing some similarities and relevance to IIoT, e.g., CPS, WSAN, and industrial control systems (ICSs), we encompass all relevant surveys in our literature study. 
To provide the readers a better understanding of the proposed review, we first discuss these related concepts and then provide the existing surveys reviewing the industrial testbeds in the scenarios corrsponding to each concept. 
A summary of the existing testbed surveys can be found in Table~\ref{tab:surveys}. 

\begin{figure}[tb]
\centering\includegraphics[width=0.45\linewidth]{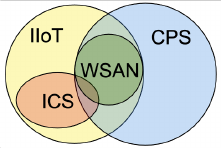}
  \caption{The relationship between IIoT, CPS, ICS, and WSAN.}
  \label{Fig:terms}
  \vspace{-0.1in}
\end{figure}

\subsection{IIoT, CPS, ICS and WSAN}

IIoT, CPS, ICS, and WSAN are closely related concepts, but cannot be used interchangeably. Although there does not exist unanimously accepted and authoritative definition of each concept, we try to  provide a rough classification of these terms. 

The Industrial Internet of Things (IIoT) refers to the network of interconnected devices and sensors used in industrial settings to collect, exchange, and analyze data~\cite{sisinni2018industrial}. It is an extension of the Internet of Things (IoT) specifically tailored for industrial applications. IIoT involves a wide range of devices, such as sensors, actuators, industrial machines, and other smart equipment, that are connected to the internet or an internal network. These devices can communicate with each other and with centralized systems to facilitate automation, data analytics, and intelligent decision-making. 

Cyber-Physical Systems (CPS) refer to integrated systems where physical components, such as sensors, actuators, and machines, are tightly coupled with computational and communication elements~\cite{shi2011survey}. The goal of CPS is to create intelligent and autonomous systems that interact with the physical world in real time, often with minimal human intervention. CPS bridges the gap between digital and physical realms, enabling seamless interaction and feedback loops between the two.

Industrial Control Systems (ICS) are specialized systems used to control and monitor industrial processes and equipment in sectors like manufacturing, energy, transportation, and more~\cite{stouffer2011guide}. ICS encompasses various control systems, including Supervisory Control and Data Acquisition (SCADA), Distributed Control Systems (DCS), and Programmable Logic Controllers (PLCs). These systems help manage complex industrial operations and ensure that processes run smoothly and efficiently.

A Wireless Sensor-Actuator Networks (WSAN) typically consists of a group of interconnected sensor and actuator nodes with limited computing resources, capable of wirelessly communicating with each other to gather and transmit data from the surrounding environment~\cite{yick2008wireless}. Early success of industrial WSANs mainly focused on monitoring applications, while significant value has been explored in WSANs for process control applications to take full advantage of wireless technology in industrial plants\cite{lu2015real}.
Thus, WSANs nowadays have a wide range of applications, including environmental monitoring, industrial automation, healthcare, agriculture, and many other fields where real-time data collection and communication are essential.

The relationship between IIoT, CPS, ICS and WSAN is interconnected and illustrated in Fig.~\ref{Fig:terms}. IIoT often serves as the overarching framework that connects various devices, sensors, and systems in an industrial environment. CPS provides the underlying architecture that enables the integration of physical processes with computing and communication systems. WSAN facilitates the wireless communication between sensors and actuators, forming a crucial part of the connectivity within IIoT and CPS. ICS is the traditional control infrastructure in industrial settings, but it can be enhanced and integrated with IIoT, CPS, and WSAN technologies for more intelligent and responsive control of industrial processes. In summary, these terms together contribute to the development of smart and interconnected industrial systems, where model-driven and/or data-driven decision-making, automation, and real-time control play significant roles. 

\begin{table*}[htb]
\caption{Comparison of existing CPS/ICS/WSAN/IIoT testbed surveys. A survey is comprehensive if it includes detailed discussions on each testbed.}
\vspace{-0.2in}
\label{tab:surveys}
\begin{center}
\resizebox{\linewidth}{!}{
\begin{tabular}{|c|c|c|c|c|c|c|}
\hline
                     & Authors           & Year & \begin{tabular}[c]{@{}c@{}}\# of \\ Testbeds\end{tabular} & \begin{tabular}[c]{@{}c@{}}Communication\\ Protocols\end{tabular} & Research Focus                      & Comprehensive \\ \hline
\multirow{4}{*}{CPS} & Cintuglu et al.   & 2016 & 28                                                        & Modbus, DNP3, IEC 61850, OPC UA, C37.118, Zigbee                  & Smart grid                          & Y             \\ \cline{2-7} 
                     & Salunkhe et al.   & 2018 & 97                                                        & OPC UA, IEC 61499                                                 & Smart grid, security, communication & N             \\ \cline{2-7} 
                     & Zhou et al.       & 2018 & 39                                                        & Modbus, DNP3, IEC 61850, OPC UA, C37.118, Openflow, Zigbee        & Security, communication             & N             \\ \cline{2-7} 
                     & Smadi et al.      & 2021 & 19                                                        & Modbus, DNP3, IEC 61850, C37.118, GOOSE, TCP/IP                   & Smart grid                          & N             \\ \hline
\multirow{5}{*}{ICS} & Holm et al.       & 2015 & 30                                                        & Modbus, DNP3, IEC 61850, IEC 60870, Profibus                      & Security                            & N             \\ \cline{2-7} 
                     & McLaughlin et al. & 2016 & 5+                                                        & NA                                                                & Security                            & N             \\ \cline{2-7} 
                     & Geng et al.       & 2019 & 20+                                                       & Modbus, DNP3, OPC UA                                              & General design                      & N             \\ \cline{2-7} 
                     & Conti et al.      & 2019 & 86                                                        & Modbus, S7Comm, Ethernet/IP, DNP3, Logs, Phy., et al.             & Security                            & Y             \\ \cline{2-7} 
                     & Ani et al.        & 2021 & 57                                                        & TCP, UDP, OPC, Modbus/TCP, DNP3, Ethernet/IP                      & Security                            & N             \\ \hline
\multirow{3}{*}{WSAN} & Horneber et al.   & 2014 & 46                                                        & IEEE 802.15.4 (no details)                                        & Energy, security, scalability       & N             \\ \cline{2-7} 
                     & Park et al.       & 2017 & 3                                                         & IEEE 802.15.4, WirelessHART, ISA-100, 6TiSCH                      & Communication                       & Y             \\ \cline{2-7} 
                     & Judvaitis et al.  & 2023 & 32                                                        & NA                                                                & General design                      & N             \\ \hline
IIoT                 & Ours              & 2023 & 83                                                        & TSN, IEEE 802.11, IEEE 802.15.4, 5G                               & Communication                       & Y             \\ \hline
\end{tabular}
}
\end{center}
\end{table*}

\subsection{CPS Testbed Surveys}
In 2016, Cintuglu et al.~\cite{cintuglu2016survey} presented a comprehensive survey on cyber-physical smart grid testbeds aiming to provide a taxonomy and insightful guidelines for the development as well as to identify the key features and design decisions while developing smart grid testbeds. The authors provided detailed discussions on 28 existing smart grid cyber-physical testbeds with a focus on their domains, research goals, test platforms, and communication infrastructure. Furthermore, the authors evaluated the testbeds on research support capacity, communication capability, security and privacy awareness, protocol support and remote access capability. 
However, since the paper is not recent, most of the present testbeds are old and constructed using mature but outdated wired communication protocols, e.g., Modbus and OPC UA. 

Salunkhe et al.~\cite{salunkhe2018cyber} presented a literature review, from 2007 to 2017, on the use of CPS testbeds in research. CPS testbeds from the findings are divided into three categories according to their application areas, including smart power grid, cyber security, and communication. 
In fact, this paper only briefly presents a small list of some testbeds which can be used in UAV and manufacturing sites, and all the information is provided in an aggregate manner without detailed analysis.  

In 2018, Zhou et al.~\cite{zhou2018review} investigated the advances in CPS testing methods from ten aspects, including different testing paradigms, technologies, and some non-functional testing methods (including security testing, robust testing, and fragility testing). 
The authors further elaborated on the infrastructures of CPS testbeds from the perspectives of their architecture and the corresponding function analyses. 
However, the majority of described CPS testbeds are simulation-based or semi-simulated. The number of fully hardware-based testbeds is very small, and they are usually applied to small-scale CPSs, e.g., robot and unmanned aerial vehicle systems. 
Furthermore, the communication protocols employed in these testbeds are either outdated (e.g., Modbus and OPC UA) or very specific standards (e.g., DNP3 and IEC 61850 used in utilities). 

Smadi et al.~\cite{smadi2021comprehensive} performed a comprehensive review  of the advancement of Cyber-Physical-Smart Grids (CPSG) testbeds including diverse testing paradigms. Particularly, the authors broadly discussed CPSG testbed architectures along with the associated functions, main vulnerabilities, testbed requirements, constraints, and applications. 
However, this work is limited to specific power grid systems where specific communication protocols, e.g., Modbus and DNP3, are adopted in the testbeds. Notably, this paper occasionally uses ICS to refer to CPS systems, highlighting the evident similarities and overlaps between these two closely related terms.

\subsection{ICS Testbed Surveys}
In 2015, Holm et al.~\cite{holm2015survey} surveyed 30 ICS testbeds that have been proposed for scientific research, most of which facilitate vulnerability analysis, education and tests of defense mechanisms. 
The authors described the implementation of each ICS testbed from four general areas, including the control center, the communication architecture, the field devices and the physical process itself. The communication architecture of these ICS testbeds are mostly based on Ethernet, which is often virtualized through, e.g., VirtualBox. This is a very short survey paper and no detailed information regarding each testbed is offered.  

In 2016, McLaughlin et al.~\cite{mclaughlin2016cybersecurity} explored the ICS cybersecurity landscape and a survey of ICS testbeds that capture the interactions between the various layers of an ICS is given in the paper. The authors only focused on lab-based ICS testbeds and briefly discussed several smart grid testbeds have been developed in the United States. 

Geng et al.~\cite{geng2019survey}, in 2019, presented a survey of ICS testbeds and classified ICS testbeds into four categories according to the different implementation and configuration methods of the testbed: physical simulation testbed, software simulation testbed, semi-physical simulation testbed and virtualized testbed. However, this paper only briefly discuss several example ICS testbeds to describe the testbed implementation architecture of each category. 

In 2021, Conti et al.~\cite{conti2021survey} provided a deep and comprehensive overview of ICSs with a focus of security research and collected, compared, and described ICS testbeds and datasets in the literature. Specifically, the authors proposed a detailed description of a set of testbeds they are interested in, dividing them into the three categories: physical testbeds, simulated testbeds, and hybrid testbeds. 
The authors also reported the best performing Intrusion Detection Systems (IDS) algorithms tested on every dataset to create a baseline in state of the art for this field. 

In the same year, Ani et al.~\cite{ani2021design} presented a mapping framework for design factors and an implementation process for building credible ICS security testbeds. Specifically, the main focus of this paper is to identify relevant design factors that can provide guidance on security testbed development and use. Except for a table that indicates each testbed’s basic information (e.g., institution, country, and objectives), this paper does not provide detailed discussion on each testbed but only delivers some qualitative study results. 

\subsection{WSAN Testbed Surveys}

In 2014, Horneber et al.~\cite{horneber2014survey} presented an investigations of WSN testbeds, ranging from discontinued testbeds over existing work to current trends in WSN experimentation and testbed research. The authors discussed the underlying research foci and design considerations, and identified common aspects and requirements for experimentation and analyzed current solutions and testbed architectures. This paper put a lot of effort on discussing the energy efficiency of WSN testbeds, as well as security and scalability researches, while it lacks in-depth study on the connectivity and resource management solutions. 


Park et al.~\cite{park2017wireless}, in 2017, presented an exhaustive review of the literature on wireless network design and optimization for wireless networked control systems (WNCSs), a specific implementation of WSAN with a focus on industrial control performance. For the reviewed experimental testbeds, this paper only introduces three representative large-scale and hardware-based WNCS testbeds. 

Recently, Judvaitis et al.~\cite{judvaitis2023testbed} provided a systematic review of the availability and usage of 32 IoT and WSN testbed facilities. However, this survey paper mainly focuses on the hardware facilities and highlights some challenges in the testbed development. The authors did not describe the communication techniques used in the considered testbeds. 

Different from previous works, our survey aims to collect all the IIoT testbeds useful for connectivity research. We base the existing literature to provide a detailed review of the current IIoT testbed systems conducted based on individual communication protocols and highlight the resource management (if any) deployed on the testbeds. 
 \section{Industrial IoT Overview}\label{sec:overview}

In this section, we provide a background on IIoT systems for readers starting approaching this field and understanding the rest of this paper. Given that IIoT interconnects a large number of components in terms of sensing, communication, resource management technologies, and security, it is impossible have a comprehensive description on all the recent advancements in such a diverse field. However, since this survey paper focuses on the connectivity research of IIoT testbeds, we only highlight some foundational aspects, including the architecture, the connectivity, and the communication protocols.

\subsection{IIoT Architecture}\label{ssec:arch}
A basic understanding of IIoT architecture is required in order to have a higher level of abstraction and help identify issues and challenges for different IIoT layers. In the literature, there exist several reference architecture frameworks originated in the past in different application contexts for IIoT~\cite{mirani2022key,qiu2020edge,raposo2018industrial,mahmood2021industrial}.
In this section, we briefly discuss the general and typical IIoT reference architecture 
which is divided into three layers: sensing layer, network layer, and application layer~\cite{jia2012rfid,domingo2012overview,atzori2010internet}, each serving distinct functionalities in the data collection, communication, and decision-making process. 

\subsubsection{Sensing Layer}
The sensing layer or perception layer is the foundation of the IIoT architecture, comprising sensor nodes, actuators, controllers and other industrial machines deployed throughout the industrial environment. These devices are responsible for collecting raw data from various physical parameters, e.g., temperature, pressure, humidity, and vibration, etc. Leveraging cutting-edge sensor technologies, the sensing layer ensures continuous data acquisition, forming the basis for subsequent data processing and analysis. 

\subsubsection{Network Layer}
The network layer facilitates the seamless communication and data transfer between the sensing layer and the higher layers of the IIoT architecture. This layer encompasses diverse communication protocols both wired or wireless, such as Ethernet~\cite{marshall2004industrial}, TSN~\cite{wang2023time}, Wi-Fi~\cite{rincon2018impact}, Zigbee~\cite{somani2012zigbee}, and 5G~\cite{aijaz2020private}, tailored to suit the specific requirements of industrial environments. Moreover, edge computing devices within the network layer perform resource management including data pre-processing, filtering, and scheduling, to reduce latency and bandwidth usage while optimizing data flow. 

\subsubsection{Application Layer}
The application layer represents the highest tier of the IIoT architecture, encompassing cloud platforms, centralized servers, and data centers. Here, data collected from the Perception Layer is transmitted through the Network Layer and undergoes advanced processing, analytics, and artificial intelligence algorithms. This layer transforms raw data into valuable insights, empowering end-users with real-time monitoring, predictive maintenance, and intelligent decision-making through user-friendly applications, dashboards, and visualization tools.


\subsection{IIoT Communication Protocols}\label{ssec:protocols}
Communication protocols play a pivotal role in the success and efficiency of IIoT deployments. Effective communication protocols optimize data exchange between devices, sensors, and central systems, which is crucial in IIoT applications where large amounts of data are generated and transmitted in real-time. 
Various communication protocols are used to facilitate seamless and efficient data exchange between IIoT devices and each communication protocol is designed to suit specific requirements, such as range, data rate, power consumption, and scalability. IIoT communication protocols can be classified into different categories based on various criteria, e.g., messaging protocols (including MQTT (Message Queuing Telemetry Transport)~\cite{quincozes2019mqtt}, CoAP (Constrained Application Protocol)~\cite{shelby2014constrained}, and DDS (Data Distribution Service)~\cite{kaushik2017comparative} etc.), transport layer protocols (including TCP/UDP), fieldbus protocols (including Modbus~\cite{swales1999open}, PROFIBUS~\cite{bender1993profibus}), industrial Ethernet protocols (including Ethernet/IP~\cite{brooks2001ethernet}, PROFINET~\cite{feld2004profinet}). 

Besides, to meet many stringent requirements in terms of timing, scalability and reliability raised by emerging industrial applications, many advancing communication protocols are being deployed in the IIoT systems, e.g., wireless protocols (including IEEE 802.15.4~\cite{petrova2006performance}, IEEE 802.11~\cite{hiertz2010ieee}), cellular networks (4G-LTE, 5G~\cite{rao2018impact}), and TSN networks~\cite{wang2023time}. 
According to the discussion in Sec.~\ref{sec:related}, many comprehensive literature studies have been explored on the systematical reviews of industrial testbeds based on the conventional IIoT communication protocols. This paper, thus, mainly focuses on researching the aforementioned IIoT communication protocols and reviewing their deployment and implementation in various industrial applications. 

\subsubsection{Time-Sensitive Networking}
TSN is a set of IEEE standards (IEEE 802.1Q) that provides deterministic, low-latency, and time-sensitive communication in Ethernet networks. TSN is specifically designed to address the challenges of real-time communication and time-critical applications in various industrial domains. TSN enables the convergence of IT (Information Technology) and OT (Operational Technology) networks, making it an ideal solution for Industry 4.0 and IIoT implementations.

TSN introduces time synchronization mechanisms and precise scheduling of data traffic to achieve deterministic communication. By using time-awareness features, TSN ensures that critical data packets are delivered within specified time bounds, reducing packet delay variation and eliminating jitter. This capability is crucial in real-time control systems, where consistent and predictable communication is essential to maintain process stability and accuracy.

TSN operates on standard Ethernet infrastructure, making it compatible with existing Ethernet networks commonly found in industrial settings. The adherence to IEEE 802.1Q standards ensures interoperability among TSN-capable devices and switches from different vendors. This allows for gradual integration of TSN into existing networks, making it a cost-effective solution for industrial applications.

\subsubsection{IEEE 802.15.4-based Protocols}
IEEE 802.15.4~\cite{koubaa2005ieee} is a set of communication standards designed for low-power, low-data-rate wireless communications. These standards are widely used in various industrial applications where devices need to communicate efficiently in a constrained and energy-efficient manner. IEEE 802.15.4-based protocols offer advantages such as low power consumption, long battery life, and support for mesh networking, making them well-suited for IIoT and industrial automation deployments.
Several representative 802.15.4-based protocols used in industrial applications include:
\begin{itemize}
    \item \textit{ZigBee}~\cite{zheng2006zigbee} supports mesh networking, enabling devices to communicate with each other in a self-organizing, self-healing network. In 2004 and 2005, ZigBee was the buzz of the industry. However, The industry demanded secure and reliable communication, but static and multi-path fading sometimes blocked ZigBee due to its use of one static channel~\cite{lennvall2008comparison}. 
    \item \textit{WirelessHART}~\cite{song2008wirelesshart} extends the capabilities of the HART (Highway Addressable Remote Transducer) protocol and is an industrial wireless communication protocol specifically designed for process automation applications in industries such as oil and gas, chemical, pharmaceutical, and manufacturing. WirelessHART is a Time Division Multiple Access (TDMA) based network where all devices are time synchronized and communicates in pre-scheduled fixed length time-slots. TDMA minimizes collisions and reduces the power consumption of the devices. 
    \item \textit{ISA100.11a}~\cite{petersen2011wirelesshart} also known as ISA100 Wireless, is an open, standards-based protocol built on IEEE 802.15.4. It is designed for industrial automation and control systems, offering robust and secure wireless communication. ISA100.11a supports time-synchronized mesh networking and can coexist with other wireless technologies, making it a suitable choice for complex and interconnected industrial applications.
    \item \textit{6TiSCH}~\cite{dujovne20146tisch} stands for IPv6 over the Time-Slotted Channel Hopping (TSCH) mode of IEEE 802.15.4e. It is a protocol stack that enables IPv6 communication over low-power and lossy industrial wireless networks. 6TiSCH is designed to provide determinism, reliability, and energy efficiency, making it suitable for various industrial applications where reliable communication with low power consumption is essential.
\end{itemize}

\subsubsection{IEEE 802.11-based Protocols}
IEEE 802.11-based protocols~\cite{banerji2013ieee}, commonly known as Wi-Fi, is another widely used wireless communication in various industrial domains. While Wi-Fi is primarily associated with consumer and enterprise networks, many Wi-Fi-based networks are designed for real-time application and be able to provide deterministic feature (e.g., Det-WiFi~\cite{cheng2017det} and RT-WiFi~\cite{wei2013rt}). These protocols combine the advantages of high-speed IEEE 802.11 physical layer and a software TDMA based MAC layer. 

\subsubsection{5G}
5G networks, the fifth generation of cellular technology, have the potential to revolutionize industrial applications with their high data rates, low latency, massive connectivity, and reliability. 5G's capabilities make it well-suited for various industrial use cases, driving the development of new applications and enhancing existing ones. 

5G's URLLC (Ultra-Reliable Low-Latency Communication) feature ensures ultra-reliable and low-latency communication, making it ideal for time-critical industrial applications. 5G's mMTC capability allows for connecting a vast number of devices simultaneously. This is especially beneficial for industrial IoT (IIoT) deployments where numerous sensors, actuators, and machines need to communicate and share data efficiently. 5G's eMBB feature provides significantly higher data rates compared to previous generations of cellular technology. In industrial settings, this can facilitate high-quality video streaming for surveillance and monitoring purposes.

In some industrial settings, private 5G networks are deployed to meet specific requirements related to security, performance, and control. Private 5G networks enable industries to have full control over their network infrastructure and tailor it to their specific needs, making them an attractive option for sensitive and mission-critical applications~\cite{aijaz2020private}.

\subsection{Resource Management in IIoT Systems}
Resource management in IIoT systems involves efficiently utilizing communication resources, such as bandwidth, channels, and time resources, to optimize data exchange and ensure reliable and timely communication. Different communication protocols used in IIoT have specific resource management mechanisms to address the unique challenges posed by industrial environments and industrial application requirements. 

\subsubsection{QoS Prioritization} It is crucial to understand the need for Quality of Service (QoS) suite/metrics in industrial applications for deploying the communication framework of Industry 4.0. The QoS sensitive communication in IoT applications such as smart grids is reliant on various performance measures/metrics, e.g., timing~\cite{shen2022qos}, stability~\cite{khalid2021qos}, network lifetime~\cite{kharche2020optimizing}, and complexity~\cite{rani2021optimized}. 
Some main QoS optimization objectives (e.g., reliability, average latency, and bandwidth efficiency) focus on ensuring efficient, reliable, and timely data transmission while effectively managing network resources. 

\subsubsection{Real-Time Communication}
Many IIoT applications have stringent timing and reliability requirements on timely collection of environmental data and proper delivery of control decisions\cite{sisinni2018industrial,zhang2021dynamic}. The QoS offered by IIoT is thus often measured by how well it satisfies the end-to-end (e2e) deadlines of the real-time sensing and control tasks executed in the system~\cite{zhang2017distributed,zhang2018distributed}. To meet the real-time communication, many IIoT protocols (e.g., TSN~\cite{lin2021queue} and 6TiSCH~\cite{wang2023resource}) incorporate time synchronization mechanisms to ensure that data packets are sent and received at precise time intervals. 

\subsubsection{Dynamic Reconfiguration}
In harsh industrial environments, dynamic network reconfiguration is needed due to factors such as device mobility, node additions or removals, changes in network topology, or link quality variations. Thus, resource management strategies aim to handle these dynamic scenarios effectively. Some potential dynamic reconfiguration techniques include dynamic routing and topology adaptation, dynamic scheduling, link quality monitoring and adaptation, network virtualization and resource partitioning, and fault tolerance, etc.

\subsubsection{Power Management}
Many IIoT devices are battery-powered or operate on limited power sources. Efficient power management techniques, including duty cycling, sleep modes, and energy harvesting, are desired to extend device lifespans and reduce maintenance requirements~\cite{salvadori2009monitoring}. In order to further increase node lifespan, energy harvesting has been used as an alternative to supplement batteries where the power manager plays the crucial role of balancing the energy harvested from the environment with the energy consumed by the node~\cite{castagnetti2014joint}. 

\subsubsection{Security}
Given that many IIoT systems control physical and sometimes dangerous processes, security issues have received significant attention from both industrial manufactures and users~\cite{jiang2020experimental}. 
Resource management includes implementing robust security mechanisms to safeguard IIoT devices, data, and communication. This involves secure device authentication, data encryption, and access control to protect against unauthorized access, data breaches, and cyber threats. 

\subsection{IIoT Testbed Classification}\label{ssec:tb-classification}
In this section, we introduce the classification method we employ in this work for the comprehensive IIoT testbed review. 
There are multiple possible classifications of an IIoT testbed, e.g., according to its institution (industry or academia), involved functional elements (physical or simulated), and orientation (security or connectivity).
In this survey, we describe the considered testbeds by classifying them according to their deployed communication protocols in Sec.~\ref{ssec:protocols}, covering both wired and wireless emerging technologies. Note that, some of the testbeds may integrate multiple communication protocols to combine the advantages of individual technology (e.g., 6TiSCH+RT-WiFi, 5G+TSN). In this case, we classify the testbed according to the testing techniques that are of primary concern for this testbed. 

For each testbeds, we reported the following information: 
\begin{itemize}
    \item Testbed name or of the authors if a name is not provided;
    \item Institution in which the testbed has been developed;
    \item Region where the testbed has been deployed;
    \item Use case (or scenario) indicates the key industrial applications considered in the testbed. 
\end{itemize}

For each particular testbed category, there are some additional information summarized, e.g., number of switches in a TSN testbed, specific protocol used in an IEEE 802.15.4-based testbed, and frequency band used in a 5G testbed, etc. 
In the following sections, we provide the detailed descriptions of the IIoT testbeds in each category based on our classification method. 
 
\section{TSN-based IIoT Testbeds}\label{sec:tsn}

In this section, we describe the recent progress on the development of TSN and Ethernet-based industry testbeds. These testbeds are divided into three categories: general TSN testbeds, OPC-UA-TSN testbeds and wireless testbeds. A summary of all the reviewed TSN testbeds using real TSN hardware devices is given in Table.~\ref{tab:tsn}.

\subsection{General TSN Testbeds}

General TSN testbeds support the fundamental TSN functionalities defined by the TSN standard, e.g., scheduled traffic, credit based shaper, and time synchronization, to achieve real-time communication and deterministic behavior for industrial applications.

\begin{table*}[ht]
\caption{Summary of Hardware-based TSN Testbeds.}
\label{tab:tsn}
\resizebox{\linewidth}{!}{
\begin{tabular}{|l|l|l|l|l|l|l|l|l|}
\hline
\rowcolor[HTML]{EFEFEF} 
\textbf{Name (Authors.)}      & \textbf{Inst.}                                                     & \textbf{Region} & \textbf{Scenario}                                             & \textbf{Func.}                                                              & \textbf{SW Impl.}                                            & \textbf{ES Impl}   & \begin{tabular}[c]{@{}l@{}}\textbf{Num.}\\\textbf{ES}\end{tabular} & \begin{tabular}[c]{@{}l@{}}\textbf{Num.}\\\textbf{Bridge}\end{tabular} \\ \hline
IIC  (Didier et al.)           & IIC                                                                & Europe          & General                                                       & \begin{tabular}[c]{@{}l@{}}IEEE 802.1AS, \\ Qbv, Qav, Qbu, CB*\end{tabular} & COTS (TTTech)                                                & Linux Stack        & 200              & 1                    \\ \hline
B\&R  (Bruckner  et al.)       & IIC                                                                & Europe          & OPC-UA                                                        & \begin{tabular}[c]{@{}l@{}}IEEE 802.1AS, \\ Qbv, Qav, Qbu, CB*\end{tabular} & COTS (TTTech)                                                & Linux Stack        & 200              & 1                    \\ \hline
OpenTSN     (Quan et al.)      & NUDT                                                               & China           & General                                                       & \begin{tabular}[c]{@{}l@{}}IEEE 802.1AS, \\ Qci, Qav, Qbv, Qch\end{tabular} & \begin{tabular}[c]{@{}l@{}}FPGA\\ (Intel / AMD)\end{tabular} & FPGA NIC           & 1                & 5                    \\ \hline
TSN-Flex    (Ulbricht et al.)  & TUD                                                                & Europe          & General                                                       & \begin{tabular}[c]{@{}l@{}}IEEE 802.1AS,\\ Qbv\end{tabular}                 & \begin{tabular}[c]{@{}l@{}}COTS\\ (FibroLAN)\end{tabular}    & Moongen            & 5                & 1                    \\ \hline
(Schriegel et al.)             & IOSB-INA                                                           & Europe          & OPC-UA                                                        & \begin{tabular}[c]{@{}l@{}}IEEE 802.1AS, \\ Qbv, Qav\end{tabular}           & \begin{tabular}[c]{@{}l@{}}COTS\\ (NXP)\end{tabular}         & PROFINET           & 5                & 2                    \\ \hline
ZIGGO       (He et al.)        & THU                                                                & China           & \begin{tabular}[c]{@{}l@{}}Event triggered\\ Deeplearning\end{tabular}                                                  & \begin{tabular}[c]{@{}l@{}}IEEE 802.1AS, \\ Qbv\end{tabular}                & \begin{tabular}[c]{@{}l@{}}FPGA\\ (AMD)\end{tabular}         & PetaLinux          & 4                & 2                    \\ \hline
NEON        (Thi et al)        & LIST                                                               & Europe          & Synchronization                                               & IEEE 802.1AS                                                                & \begin{tabular}[c]{@{}l@{}}COTS\\ (NXP)\end{tabular}         & Linux stack        & 2                & 2                    \\ \hline
(Miranda et al.)               & \begin{tabular}[c]{@{}l@{}}University \\of Antwerp \end{tabular}   & Europe          & General                                                       & \begin{tabular}[c]{@{}l@{}}IEEE 802.1AS, \\ Qbv\end{tabular}                & Linux                                                        & Linux stack        & 4                & 3                    \\ \hline
(Jiang et al.)                 & \begin{tabular}[c]{@{}l@{}}Hanyang\\ University\end{tabular}       & Korea           & General                                                       & \begin{tabular}[c]{@{}l@{}}IEEE 8021AS, \\ Qbv\end{tabular}                 & \begin{tabular}[c]{@{}l@{}}COTS\\ (Cisco)\end{tabular}       & Linux stack        & 8                & 2                    \\ \hline
(Xue et al.)                   & UConn                                                              & NA              & General                                                       & \begin{tabular}[c]{@{}l@{}}IEEE 802.1AS, \\ Qbv, Qav, Qbu, CB\end{tabular}  & \begin{tabular}[c]{@{}l@{}}COTS\\ (TTTech)\end{tabular}      & Linux stack        & 8                & 8                    \\ \hline \hline
w-iLab.t (Sudhakaran et al.)            & Intel                                                              & NA               & Wireless                                                     & \begin{tabular}[c]{@{}l@{}}IEEE 802.1AS, \\ Qbv\end{tabular}                &                                                              &                    & 3                & 1                    \\ \hline
(Kehl et al.)                  & FIPT                                                               & Europe (Germany) & Wireless                                                     & \begin{tabular}[c]{@{}l@{}}IEEE 802.1AS, \\ Qbv\end{tabular}                & COTS                                                         &                    & 2                & 2                    \\ \hline
(Miranda et al.)               & \begin{tabular}[c]{@{}l@{}}University \\of Antwerp \end{tabular}   & Europe (Belgium) & Wireless                                                     & \begin{tabular}[c]{@{}l@{}}IEEE 802.1AS, \\ Qbv\end{tabular}                & FPGA                                                         &                    & 2                & 1                    \\ \hline
\end{tabular}
}
\end{table*}

\subsubsection{IIC}
\begin{figure}[tb]
  \centering\includegraphics[width=.9\linewidth]{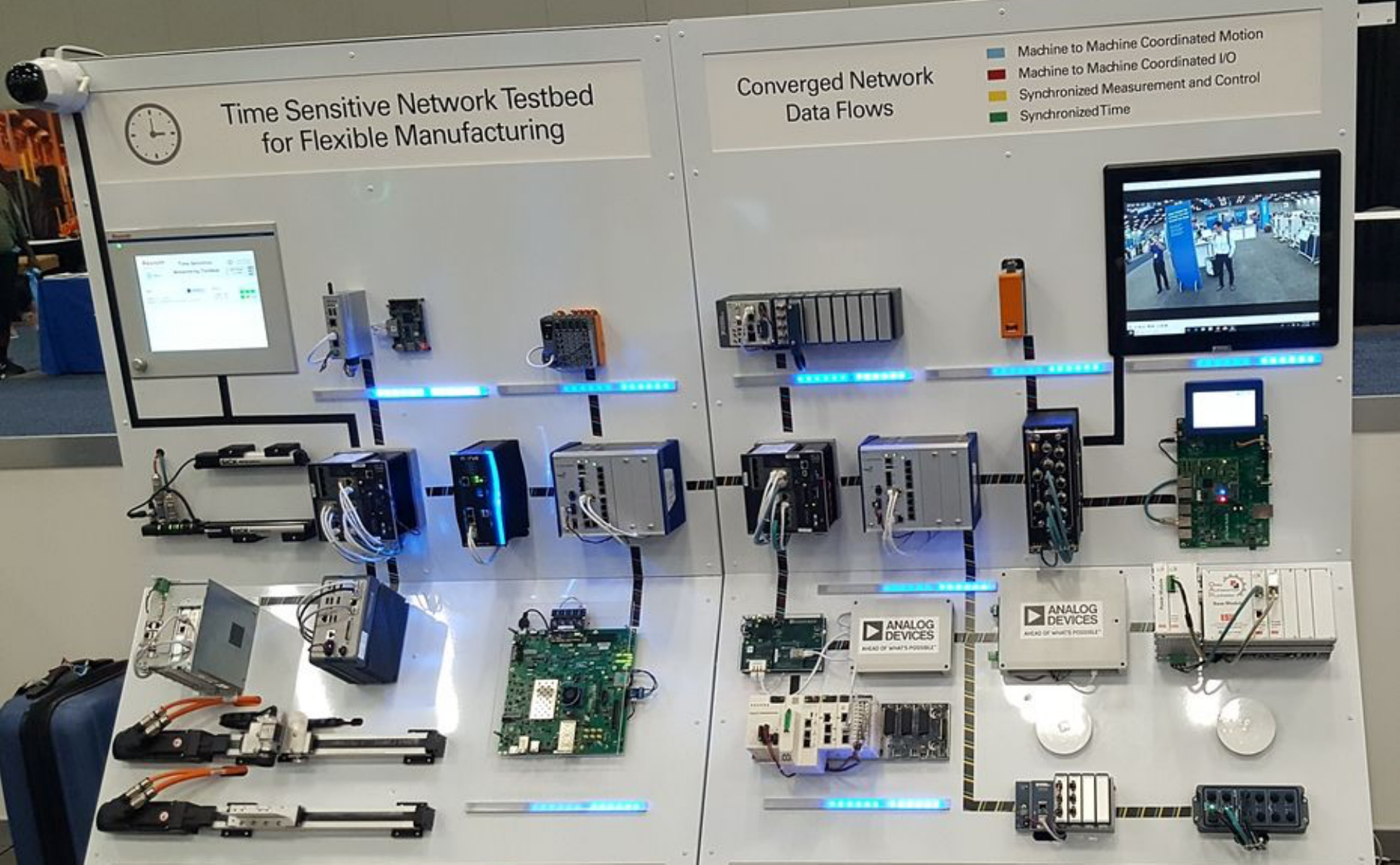}
  \caption{IIC TSN testbed located at Stuttgart University.}
  \label{Fig:TSN:IICEU}
  \vspace{-0.2in}
\end{figure}

The Industrial Internet Consortium (IIC) establishes two physical testbeds to demonstrate TSN capabilities and applicability in industrial control systems~\cite{didier2017results}. The first testbed sits at the National Instruments headquarters in North America, while the second one locates at the University of Stuttgart in Germany. 
The testbed at the National Instruments features a ring topology with six bridges and eight end stations directly network-connected. The testbed at Stuttgart University forms a line topology with 6 bridges and 12 end stations, as shown in Fig.~\ref{Fig:TSN:IICEU}. The testbeds present several use cases from industrial automation and control applications. For example, one use case involving a leading robot supplier and an automation control provider aims to leverage TSN technology for intercommunication between their devices and systems. This includes coordination and control of PLCs, along with sensor data monitoring between robots and PLCs. This scenario integrates receiving time-sensitive, synchronized data from sensors, and transmitting control or actuation commands back to other devices, e.g., robots.

\subsubsection{TSN-Flex}
Constructed in 2022, TSN-FlexTest~\cite{ulbricht2022tsn} serves as a testbed designed to assess the real-time performance of TSN. It primarily studies TSN attributes including synchronization quality, latency, and packet delay variation reduction via flow control, ultra-reliability, and resource management. The testbed employs a star topology with five end stations and a bridge, incorporating a Commercial Off-The-Shelf (COTS) TSN bridge from FibroLAN that supports cut-through capability and IEEE 802.1 standards based on a Hybrid ASIC-FPGA architecture. The end stations utilize x86 CPUs and multiple Network Interface Cards (NICs), with a Linux stack and MoonGen for precise packet injection.

\eat{\begin{figure}[tb]
    \centering
    \includegraphics[width=\linewidth]{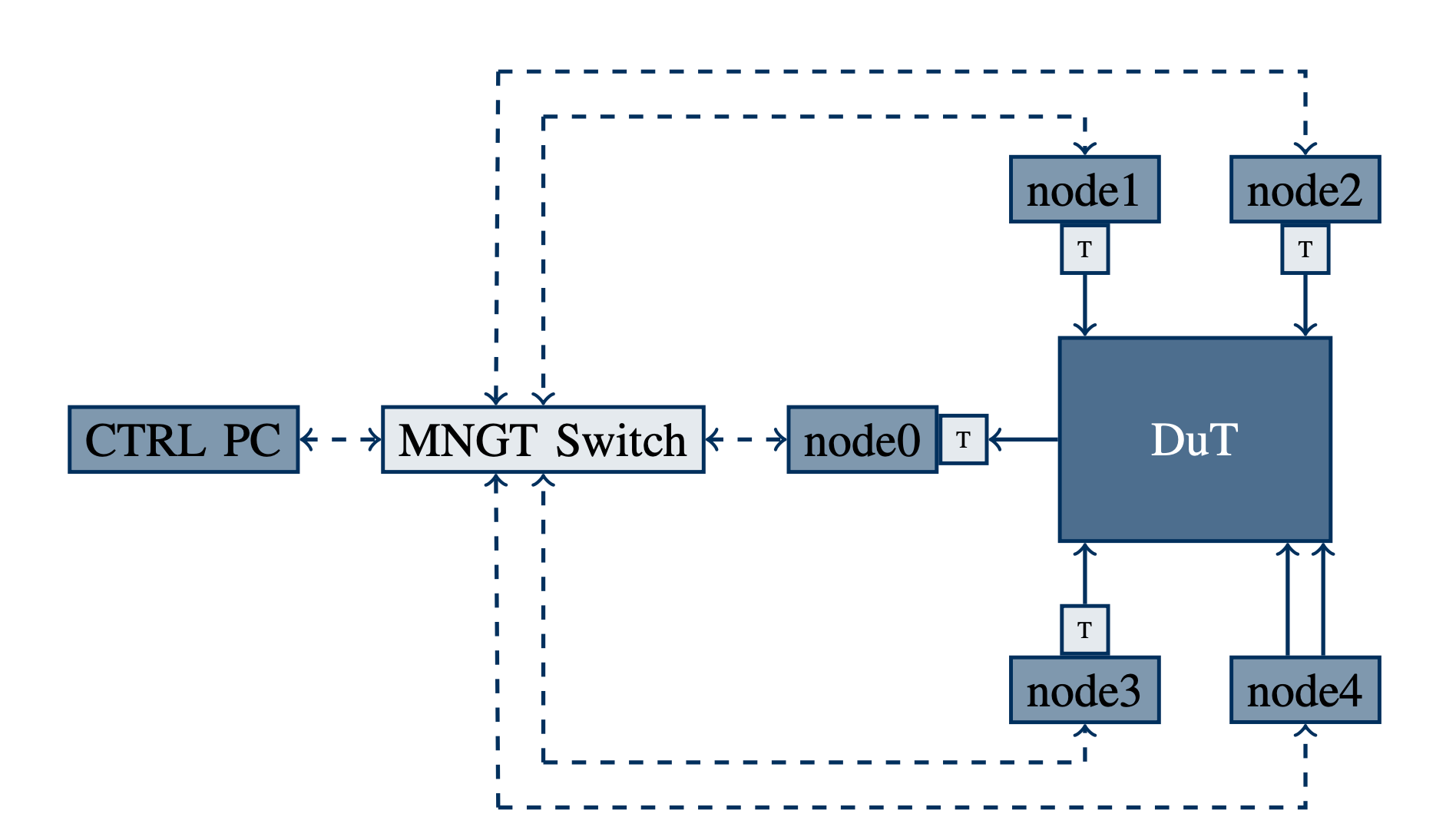}
    \caption{TSN-Flex testbed architecture.}
    \label{Fig:TSN:Flex}
  \end{figure}}

Experimental evaluation focusing on three metrics is conducted utilizing the testbed. 1) Achievable precision via the Precision Time Protocol: it demonstrate that the synchronization precision is approximately 11$ns$ with 64 synchronization messages per second. 2) The accuracy of cyclic data transmissions: it shows that MoonGen sending is more precise than Linux API, achieving jitter within 7$us$. 3) The impact of the cut-through feature on transmission delay modeling. 
The study finds that modifying the Gate Control List (GCL) and inserting guard bands can mitigate Packet Delay Variation (PDV) to certain extent, with the talker's erratic behavior significantly contributing to high PDV.

\subsubsection{OpenTSN}
OpenTSN \cite{quan2020opentsn}, established in 2020, aims to bridge the gap between TSN standards and specific TSN system applications by providing an accessible platform for rapid TSN system prototyping and evaluation. It features a Software-Defined Networking (SDN)-based TSN network control mechanism, a time-sensitive management protocol, and a time-sensitive switching model. The paper presents two FPGA-based prototyping examples in star topology and ring topology. 
While, some specific testbed specification information, e.g., the number of end stations and bridges, is not described. 
The star network components are implemented based on the Xilinx Zynq 7020 FPGA SoC. The bridges in the ring network use the Altera Arria 10 based FPGA platform. The prototypes include IEEE 802.1AS, Qci, Qav, Qbv, Qch in the bridge, with further developments beyond TSN standards like AS6802, EMF, and EOS-PIFO underway. The development stack is openly accessible on their GitHub repository. Testbed evaluation reveals synchronization precision below 32 $ns$, with transmission performance aligning with theoretical analysis of the Cyclic Queue and Forwarding based TSN networks.

\eat{
  \begin{figure}[tb]
    \centering\includegraphics[width=\linewidth]{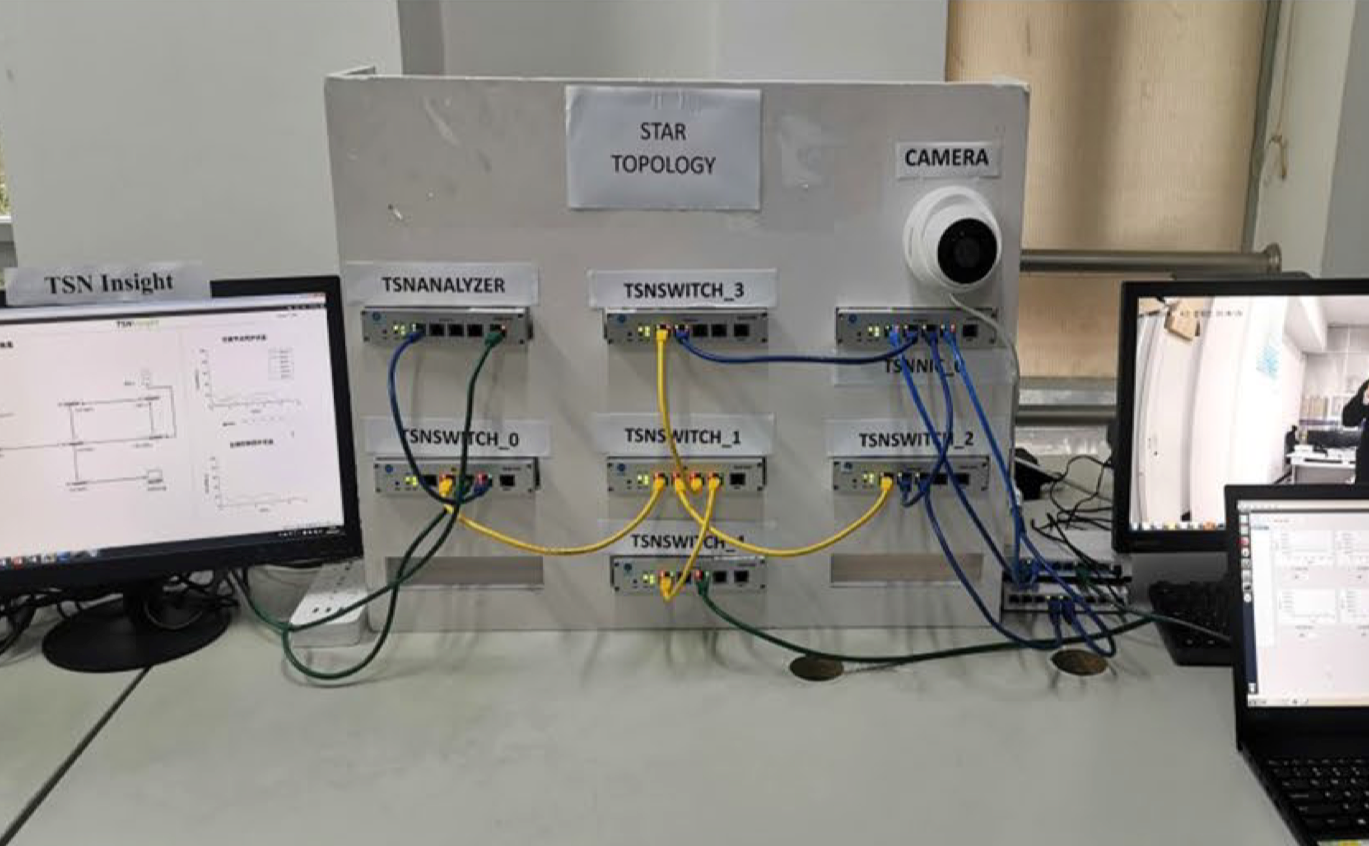}
    \caption{OpenTSN testbed in star topology.}
    \label{Fig:TSN:OpenStar}
  \end{figure}

  \begin{figure}[tb]
    \centering\includegraphics[width=\linewidth]{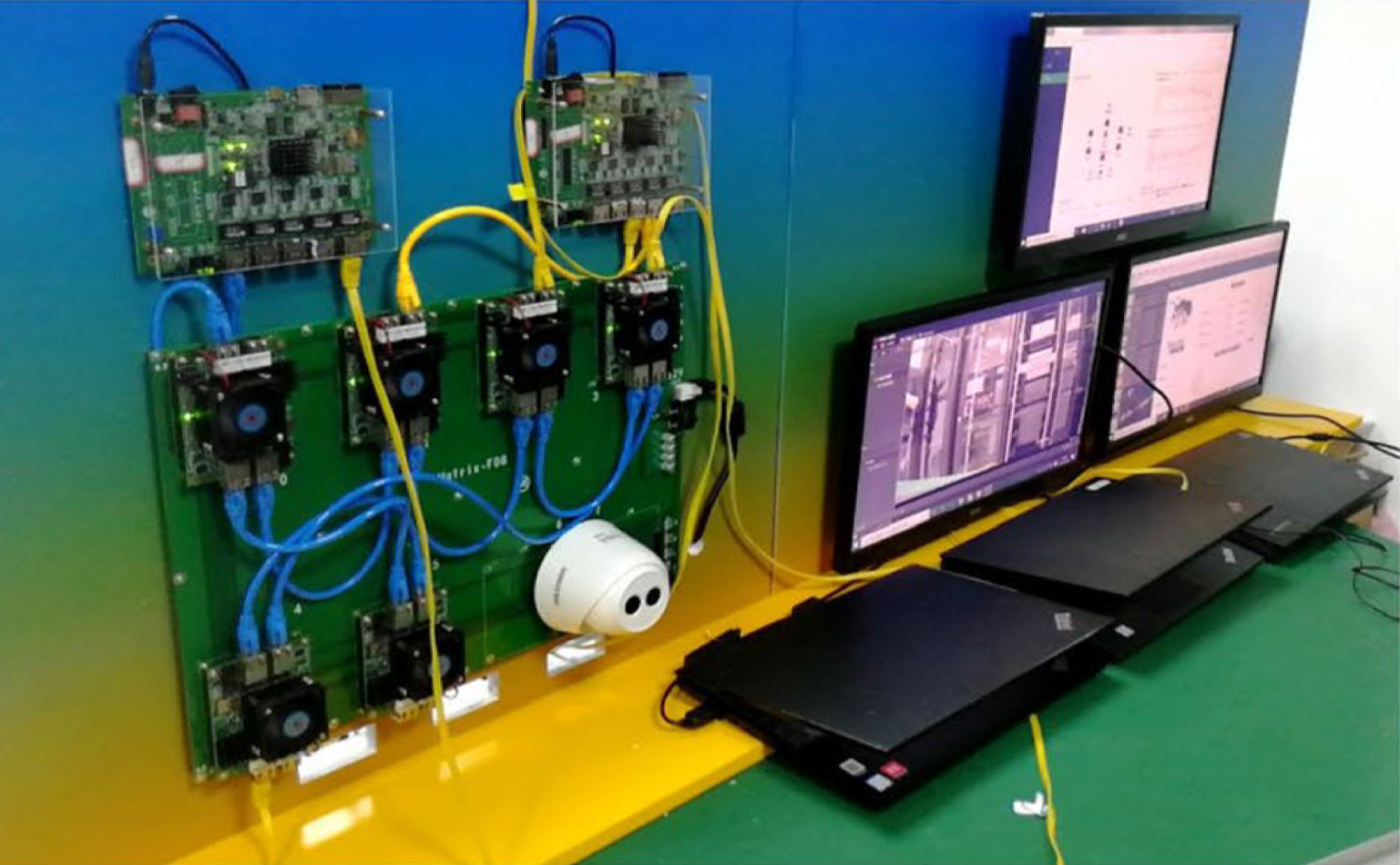}
    \caption{OpenTSN testbed in ring topology.}
    \label{Fig:TSN:OpenRing}
  \end{figure}
}

Research utilizing OpenTSN testbeds leads to several advancements. \cite{bu2019tsn} extends the PTP packet formats for improved testbed status monitoring. This enables accurate real-time monitoring of the time synchronization accuracy, the bridge status, and the link status. Another study \cite{9218753} proposes a template-based development model for swiftly customizing resource-efficient TSN bridges to meet application-dependent requirements. This approach, tested on the OpenTSN platform, shows a significant reduction in on-chip memory usage, up to 80.53\%, maintaining the same Quality-of-Service compared to the resource configuration on the COTS bridge. \cite{yang2021cames} addresses challenges related to self-driving vehicles, including high-bandwidth deterministic interconnection, high-performance computing for AI-powered inference, and over-the-air (OTA) updates, and demonstrates their solutions using OpenTSN.

\subsubsection{Ziggo}
Ziggo \cite{ziggo}, a flexible TSN testbed suitable for industrial control, automotive electronics, and other time-sensitive applications, offers precise time synchronization and ultra-low network delay. Compliant with IEEE 802.1AS, Qav, Qbv, and Qcc standards, it includes a TSN bridge implemented on ZYNQ-7000 SoC, exploiting both its hardware and software programmability. An evaluation toolkit, featuring a similar architecture, is used to assess the network performance. A testbed incorporating 2 bridges and 4 end stations is set up based on the Ziggo platform. 

\eat{
  \begin{figure}[tb]
    \centering  \includegraphics[width=\linewidth]{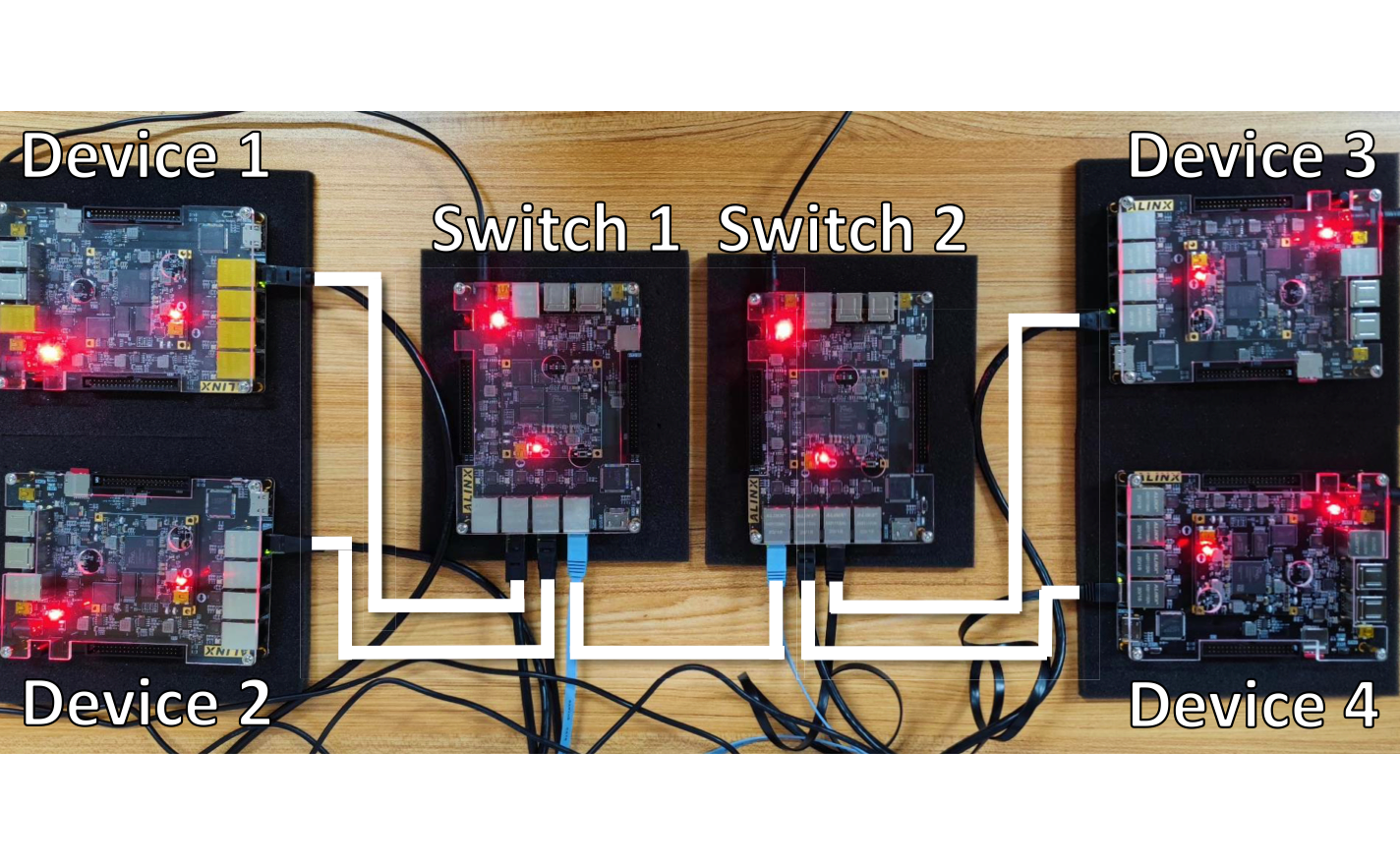}
    \caption{Ziggo testbed with 2 bridges and 4 end stations.}
    \label{Fig:TSN:Ziggo}
  \end{figure}
}

Several studies perform experimental research on Ziggo platform. For instance, \cite{zhao2022tsn} pursues deterministic transmissions of event-triggered critical traffic alongside time-triggered traffic. Through probabilistic stream, prioritized slot sharing, and prudent reservation techniques, experiments on Ziggo demonstrate that a new paradigm for TSN scheduling significantly reduces the event triggered critical traffic's latency and jitter compared to the conventional methods. \cite{he2023deepscheduler} employs deep learning methods to discern effective scheduling policies amidst complex data flow dependencies, providing a fast and scalable solution that greatly enhances schedulability. \cite{yang2023caas} proposes a novel Control-as-a-Service (CaaS) architecture for industrial control systems, integrating control tasks within TSN bridges rather than dedicated controllers. Implemented on Ziggo, CaaS virtualizes the industrial TSN network into a single Programmable Logic Controller (PLC).

\subsubsection{Jiang et al.}
\begin{figure}[tb]
  \centering
  \includegraphics[width=\linewidth]{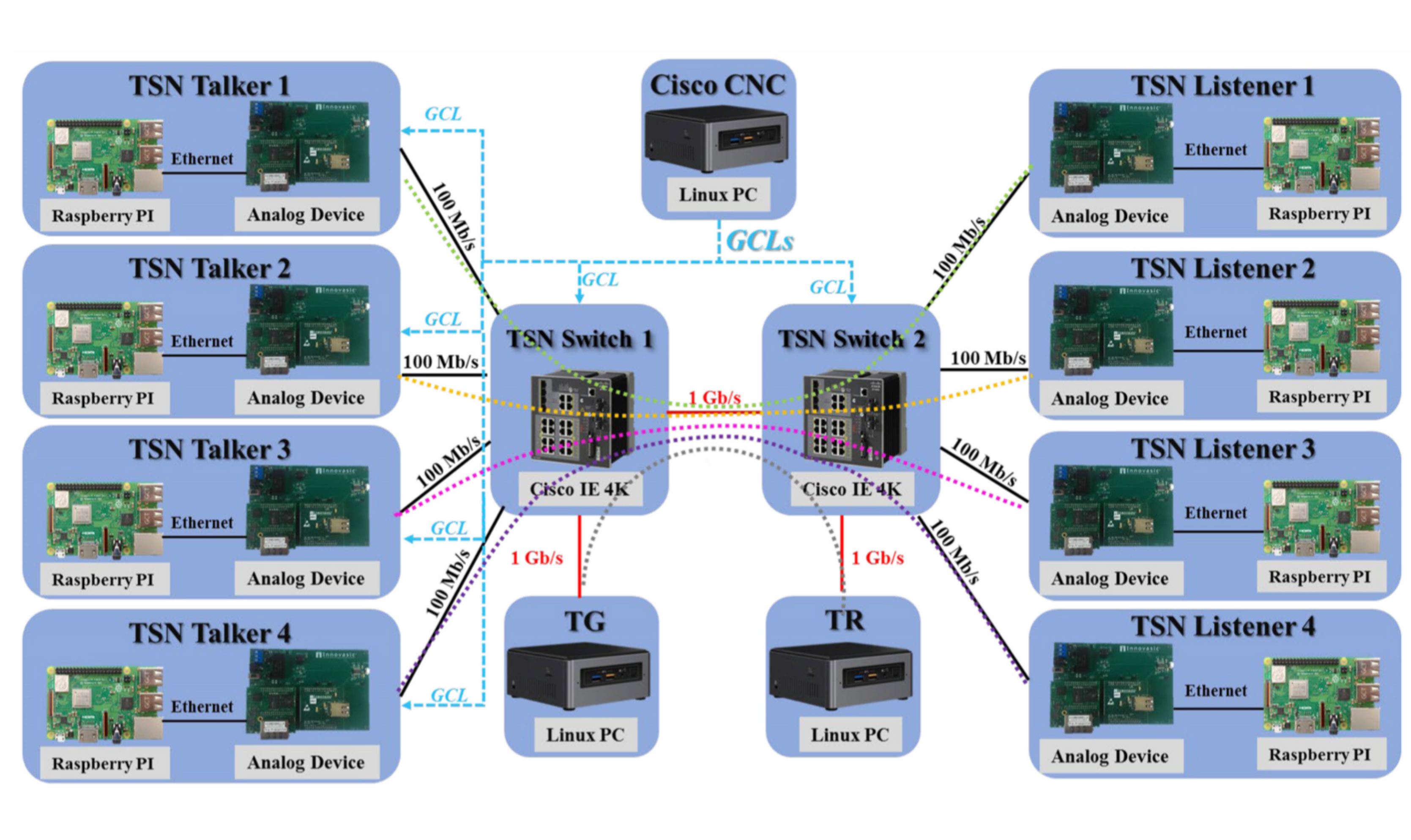}
  \caption{TSN testbed with Cisco IE 4000 COTS bridge used to validate the simulation model.}
  \label{Fig:TSN:Dual}
  \vspace{-0.1in}
\end{figure}

\cite{jiang2019simulation} establishes a TSN testbed to validate the experimental observations in the simulation model. The goal is to offer a simulation tool verified by real testbed to assist in the design and evaluation of TSN networks in industries like automotive, industrial automation, and telecommunications. The testbed (shown in Fig.~\ref{Fig:TSN:Dual}) incorporates two COTS TSN bridges (Cisco IE 4000) supporting IEEE 802.1AS and Qbv. Eight Raspberry Pis, each equipped with an analog device, function as end stations, utilizing the Analog Device (AD) module as the TSN agent for time synchronization and traffic scheduling. 

\subsubsection{NEON}
NEON, an SDN-enabled testbed, is constructed to verify and analyze the network synchronization management solutions proposed in~\cite{thi2020sdn}. The goal of the solutions is to leverage SDN techniques for flexible self-configuration of IEEE 802.1AS, accommodating traffic changes and link breakdown. NEON's primary components include a southbound API, a controller, and services. The southbound API allows the management of network devices such as access points, switches/bridges, and gateways. The testbed incorporates two NXP SJA1105 Ethernet bridges, which comply with the IEEE 802.1AS standard. End stations are implemented using the linuxptp stack. Evaluation results demonstrate that the nodes attain highly precise time synchronization and swift synchronization recovery.

\eat{
  \begin{figure}[tb]
    \centering
    \includegraphics[width=\linewidth]{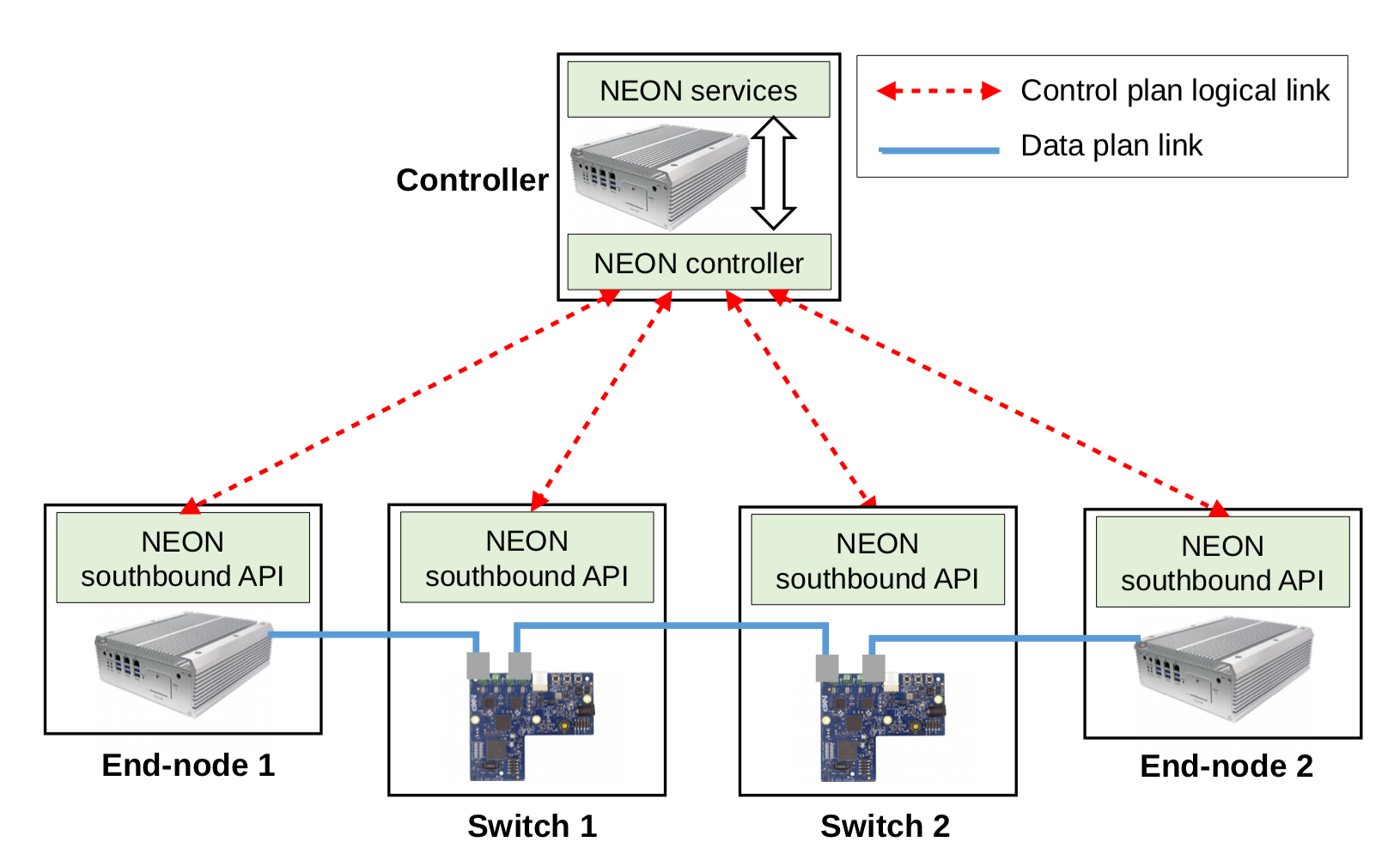}
    \caption{The architecture of SDN-enabled NEON TSN testbed.}
    \label{Fig:TSN:neon}
  \end{figure}
}

\subsubsection{Miranda et al.}
A TSN testbed is presented in~\cite{9798368} to address the lack of flexibility in current TSN feature experiments which require specific hardware support and testbed infrastructure. The study evaluates two different cloud testbeds for TSN experimentation, investigating their hardware features, testbed management infrastructure's influence, and data plane performance. The authors set up online cloud testbeds based on CloudLab and Virtual Wall, maintaining identical topologies across both for accurate comparative analysis. IEEE 802.1AS synchronization is implemented using ethtool and linuxptp, without hardware timestamping support, while IEEE 802.1Qbv is achieved via the Linux TAPRIO qdisc.


\subsubsection{Xue et al.}
\begin{figure}[tb]
  \centering
  \includegraphics[width=.9\linewidth]{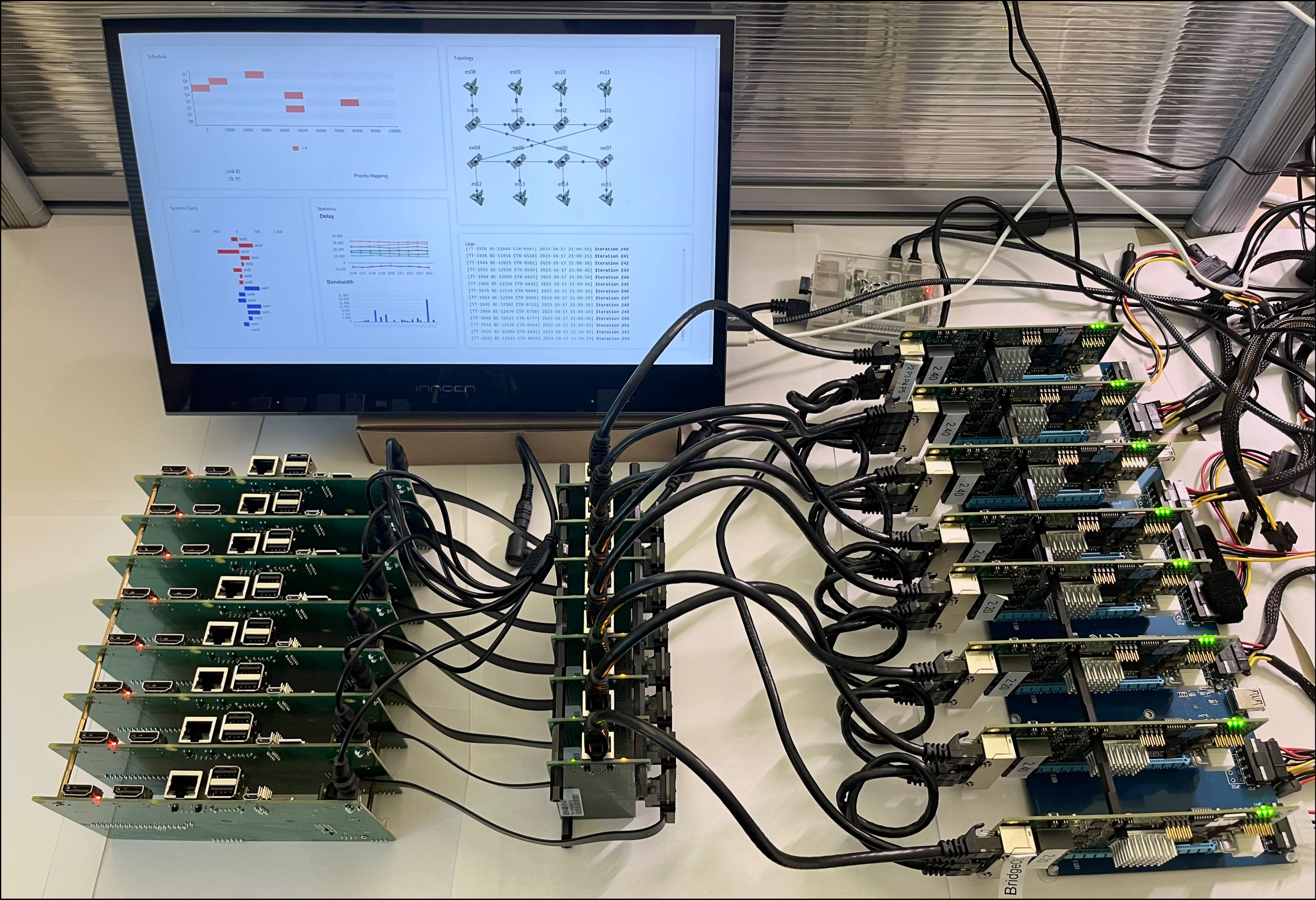}
  \caption{The TSN testbed implemented with COTS TTTech TSN board used to evaluate TAS-based scheduling methods.}
  \label{Fig:TSN:Sche}
  \vspace{-0.1in}
\end{figure}

\cite{xue2023real} presents a TSN testbed designed to validate scheduling model assumptions via real-world experimental measurements. The testbed, depicted in Fig. \ref{Fig:TSN:Sche}, comprises eight bridges and eight end stations in a ring topology. Bridges employ the COTS TTTech TSN Evaluation board for implementation. End stations are implemented using two approaches: one based on the Linux Ethernet stack, and the other based on the Data Plane Development Kit (DPDK) with NIC equipped with specific hardware functions~\cite{xue2024es}. Measurements reveal the propagation delay, processing delay, and synchronization error to be bounded within 6$ns$, 1.9$\mu s$, and 10$ns$ respectively. The testbed is also used to validate the correctness of the scheduling methods by comparing the measured end-to-end delay with the analytically derived delay bounds.

\subsection{OPC-UA-TSN Testbeds} 

OPC-UA-TSN testbeds support OPC-UA protocol which is platform-independent and ensures the seamless flow of information among devices from multiple vendors. In the OPC-UA-TSN testbeds, OPC UA usually serves as an IoT enabler for higher-level applications while TSN can provide the low-level data transport. The combination of OPC-UA and TSN aims to enhance the convergence of Information Technology (IT) and Operational Technology (OT), and has gained significant attention in the field of industrial automation recently.

\subsubsection{R\&B}
The R\&B testbed, built for the SPS IPC Drives 2017 in Nuremberg, proposes a new solution for industrial communication by integrating OPC Unified Architecture (UA) and TSN~\cite{bruckner2019introduction}. The goal of the solution is to overcome the limitations of traditional industrial communication systems on interoperability, scalability, and real-time constraints. The authors argue that combining OPC UA and TSN can deliver a unified, scalable, and real-time communication solution suitable for Industry 4.0.
The testbed involves both hardware and software components, including a TSN bridge, 50 I/O nodes, an OPC UA server, OPC UA clients, and a TSN configuration tool. Measurement results indicate that the forwarding latency of OPC UA on a 1Gbps TSN bridge is 780 $ns$, significantly less than the forwarding latency in existing networks such as traditional Ethernet (2 $\mu s$), POWERLINK (0.76 $\mu s$), EtherCAT (0.85 $\mu s$), and SERCOS III (0.63 $\mu s$).

\eat{
  \begin{figure}[tb]
    \centering
    \includegraphics[width=\linewidth]{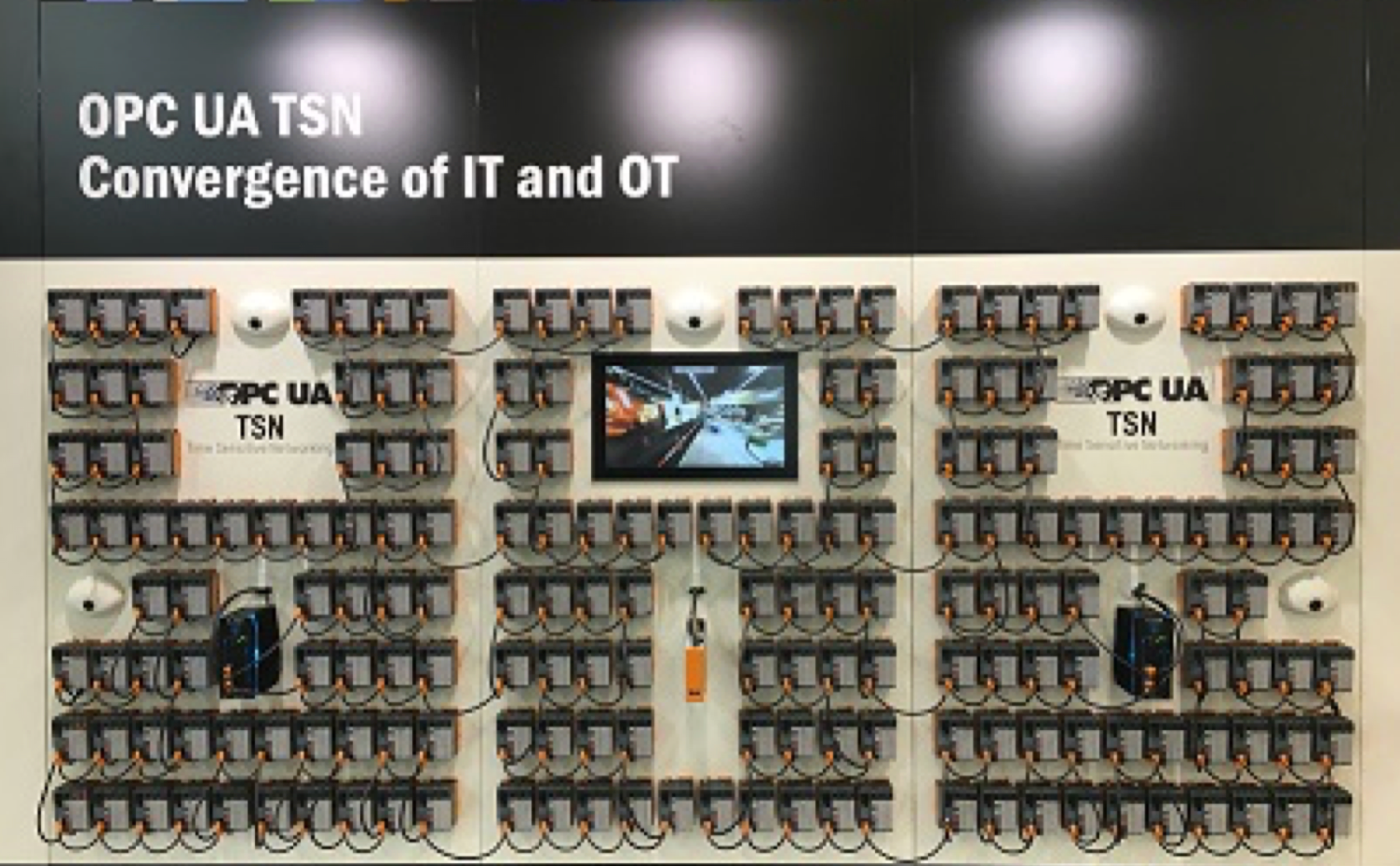}
    \caption{The R\&B testbed built for the SPS IPC Drives 2017 in Nuremberg.}
    \label{Fig:TSN:OPCUA}
  \end{figure}
}

\subsubsection{Schriegel et al.}
\cite{schriegel2018investigation} proposes a heterogeneous TSN network designed to study a distributed SDN control plane architecture in conjunction with other protocols like PROFINET RT, OPC UA, and Modbus. It aims to tackle the management and control challenges in heterogeneous networks with differing time-sensitivity requirements. A smart factory testbed featuring NXP bridges and PROFINET, OPC UA, and video stream communications was established. 
The NXP bridges support time synchronization, per-stream policing, and both Credit Based Shaper (IEEE 802.1Qav) and Time Aware Shaper (802.1Qbv). The configuration of TAS is set up based on the bandwidth percentage of each stream. The results indicate that the engineering effort for implementing high available sensor to cloud communication is reduced based on the proposed SDN solution. Based on the measurements, they also revealed that integrating with legacy network can limit bandwidth and minimum cycle time.

\eat{
  \begin{figure}[tb]
    \centering
    \includegraphics[width=\linewidth]{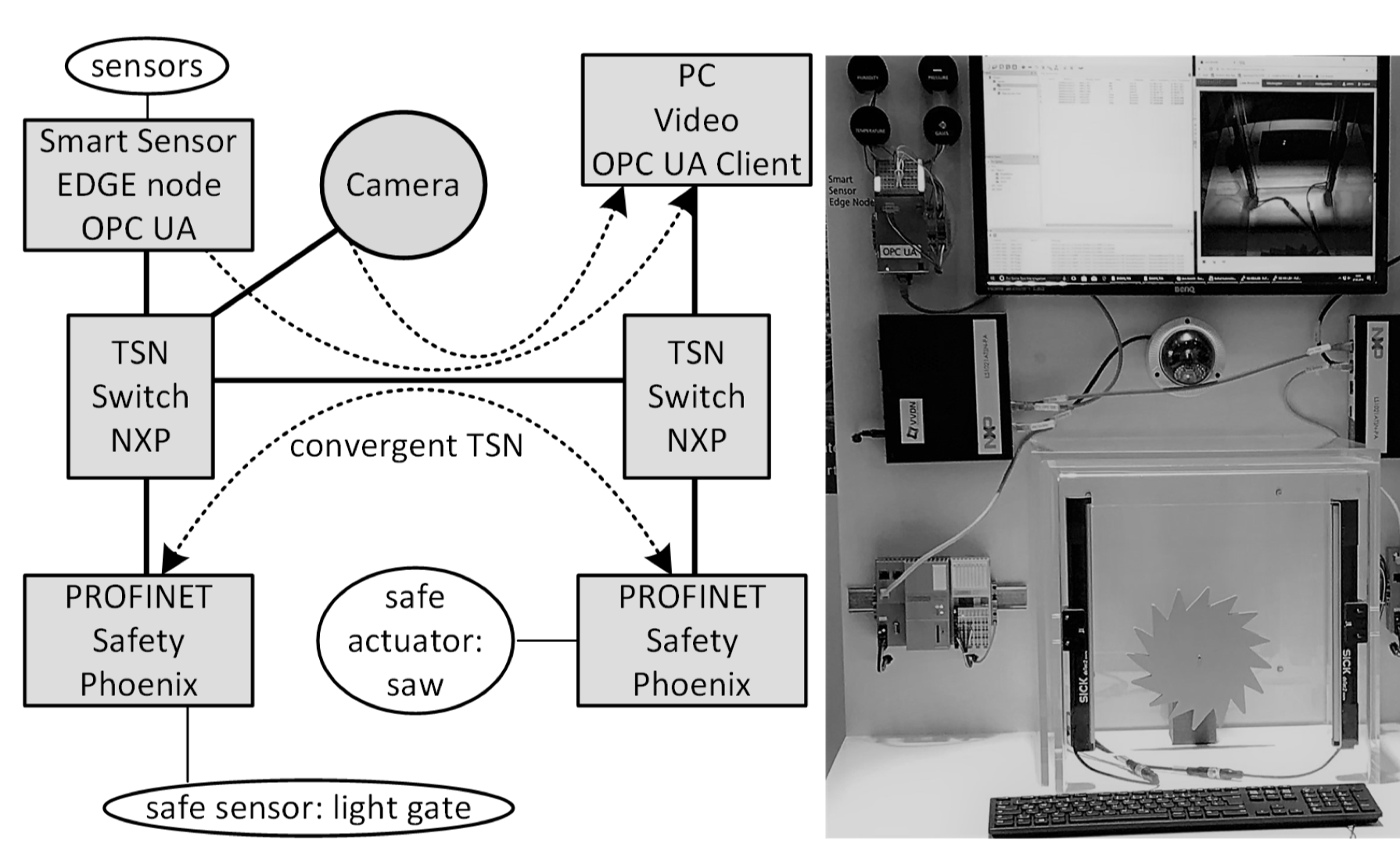}
    \caption{A smart factory testbed featuring NXP bridges.}
    \label{Fig:TSN:Smart}
  \end{figure}
}

\subsection{Wireless TSN Testbeds}

Wireless TSN extends IEEE 802.1 TSN capabilities to wireless media. The recent progress in 5G and IEEE 802.11 wireless technologies has sparked considerable interest in leveraging TSN capabilities wirelessly. This advancement can enhance safety, efficiency, and transparency on factory floors. While the use cases for wireless TSN inherently vary from those of wired networks, it's notable that features such as time synchronization and time-aware scheduling could facilitate a range of wireless industrial applications.

\subsubsection{w-iLab.t}
\begin{figure}[tb]
  \centering
  \includegraphics[width=.7\linewidth]{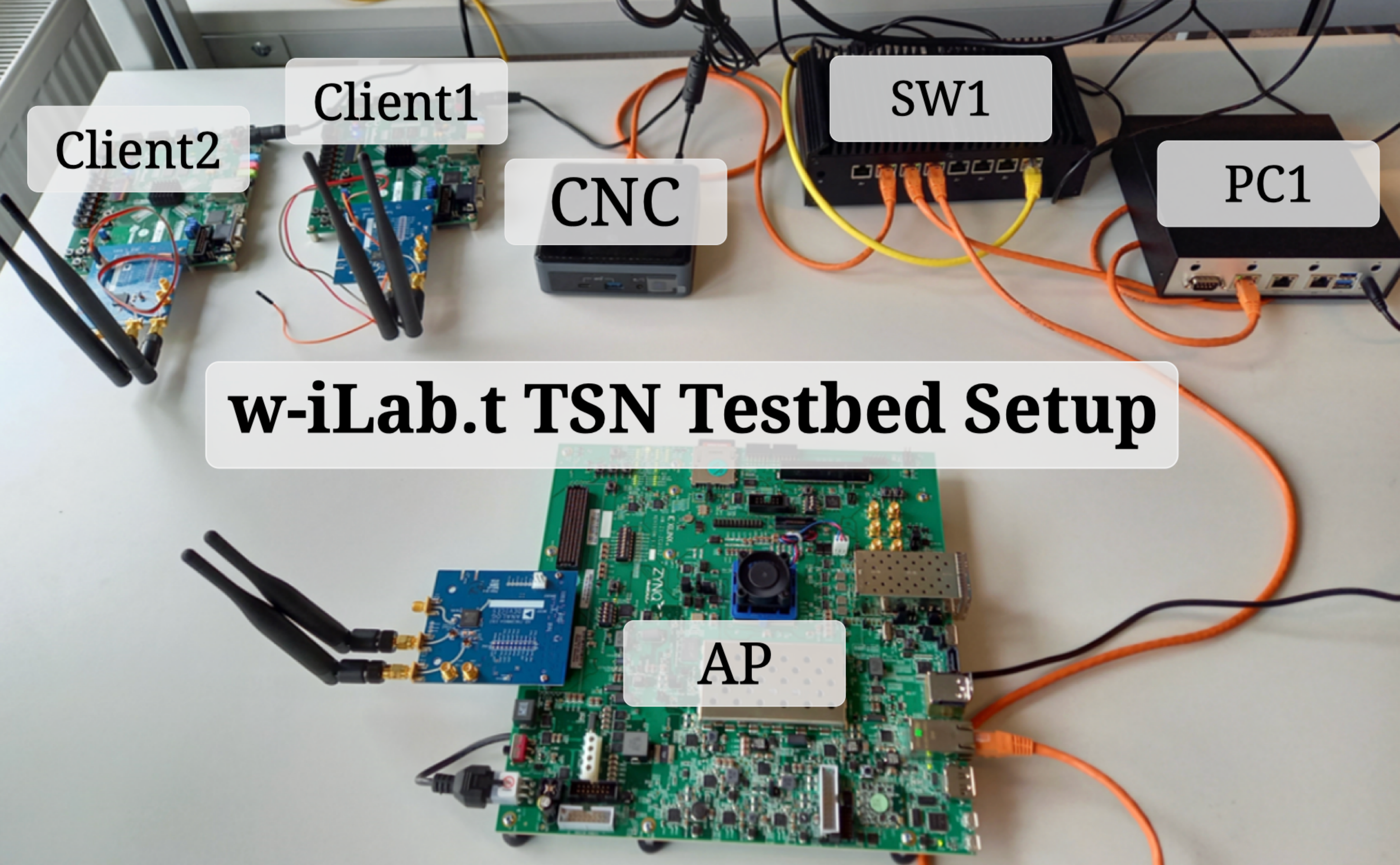}
  \caption{The w-iLab.t. wireless TSN testbed setup.}
  \label{Fig:TSN:wifi}
  \vspace{-0.15in}
\end{figure}

The w-iLab.t TSN testbed~\cite{miranda2023enabling}, integrating Ethernet and Wi-Fi technologies, demonstrates the effectiveness of a seamless, fine-grained traffic controller across both wired and wireless domains~\cite{haxhibeqiri2022safety}. It evaluates a SDN based control architecture for end-to-end TSN-enabled control over LAN and WLAN domains. The testbed incorporates one wired node, a bridge, a Wi-Fi Access Point (AP), and two Wi-Fi clients, as depicted in Fig. \ref{Fig:TSN:wifi}. The CNC and PC1 are Intel NUC nodes with Intel Core i7 CPUs and Intel I219-V Ethernet controllers. The bridge is an Evrtech KBOX3102 industrial bridge with an Intel Core i7 CPU and Intel I211-AT Ethernet controller. Notably, the bridge lacks TSN functions, but the TAS defined in IEEE 802.1Qbv is implemented on the AP using the OpenWiFi project. Synchronization on the AP is achieved by the LinuxPTP project. Experimental results show that synchronization needs a median of 1000 $ns$ and 2500 $ns$ (in 90\% cases) on the AP, confirming the schedule with a minimum cycle time of 5.12 $ms$ and a minimum granularity of 256 $us$ can be achieved on the testbed.

\subsubsection{Kehl et al.}
\cite{kehl2022prototype} set up a testbed prototype designed to integrate TSN and 5G for industrial use. 
It includes a pre-commercial system compliant with the 5G URLLC and a wired network supporting various TSN standards. Two industrial bridges, based on FPGA and supporting IEEE 802.1AS and 802.1Qbv, are used. Edge devices are implemented with Intel i210 NICs and a Real-Time Operating System (RTOS). TSN Qbv in the wired device helps mitigate jitter introduced by the wireless device in mobile robotics. Results show an average timing accuracy of 3 $\mu s$ in the integrated setup, with a maximum error of less than 8 $\mu s$. Furthermore, the application of IEEE 802.1Qbv reduced output jitter from 372 $\mu s$ to 922 $ns$.

\eat{
  \begin{figure}[tb]
    \centering \includegraphics[width=\linewidth]{Figs/tsn/5g.png}
    \caption{A testbed prototype designed to integrate TSN and 5G for industrial use.}
    \label{Fig:TSN:5g}
  \end{figure}
}

\section{IEEE 802.15.4-based IIoT Testbeds}\label{sec:15.4}
In this section, we provide detailed description on the IEEE 802.15.4-based IIoT testbeds, dividing them into general testbeds based on IEEE 802.15.4~\cite{petrova2006performance} and advanced testbeds based on IEEE 802.15.4e~\cite{de2016ieee} which is an extension pursuing higher real-time performance.

\begin{table*}[]
    \caption{Summary of IEEE 802.15.4-based Testbeds.}
    \centering
    \resizebox{1\linewidth}{!}{
        \begin{tabular}{|l|l|l|l|l|l|l|}
            \hline
            \rowcolor[HTML]{EFEFEF}
            \textbf{Name or Authors} & \textbf{Institution}                  & \textbf{Region} & \textbf{Protocol} & \textbf{Main Feature}                                    & \textbf{Main Hardwares}                 & \textbf{Num. Devices} \\ \hline
            MoteLab                  & Harvard University                    & U.S.            & 802.15.4          & Remote Web-based access                                  & CrossBow MicaZ                          & 30                    \\ \hline
            Trio                     & UC Berkeley                           & U.S.            & 802.15.4          & Outdoor solar powered motes                              & Telos                                   & 557                   \\ \hline
            Kansei                   & Ohio State Univ.                      & U.S.            & 802.15.4          & High fidelity sensor data generation                     & TMote Sky                               & 410                   \\ \hline
            WISEBED                  & Multiple EU univsities                & Europe          & 802.15.4          & Heterogeneous devices and networks                       & Mica2, TelosB, Sun Spot, iSense         & 550                   \\ \hline
            CONET                    & Univ. of Sevilla                      & Europe          & 802.15.4          & Modular architecture with mobile nodes                   & TelosB                                  & 21                    \\ \hline
            OpenWSN                  & UC Berkeley                           & U.S.            & 802.15.4e         & Open source full-stack implementation                    & TelosB, CC2538, GINA, LPC               & -                     \\ \hline
            Indriya                  & National Univ. of Singapore           & Asia            & 802.15.4          & Low maintance cost                                       & TelosB                                  & 127                   \\ \hline
            NetEye                   & Wayne State Univ.                     & U.S.            & 802.15.4          & Empolys TDMA for data reliability                        & TelosB                                  & 130                   \\ \hline
            SmartSantander           & Univ. of Cantabria                    & Europe          & 802.15.4          & City-scale testbed with real urban infrastructures       & Customized hardware based on ATmega1281 & -                     \\ \hline
            FIT IoT-LAB              & FIT                                   & Europe          & 802.15.4          & Large-scale heterogeneous deployments with remote access & WSN430, M3  A8                          & 2728                  \\ \hline
            PISA                     & Univ. of Pisa                         & Europe          & 802.15.4          & RPL routing protocol evaluation                          & CC2420                                  & 22                    \\ \hline
            ROVER                    & Cisco                                 & Europe          & 6TiSCH            & Cheap testbed solution                                   & CC2538                                  & 8                     \\ \hline
            WUSTL-WSN                & Washington Univ.                      & U.S.            & 802.15.4(e)       & Sensor-equipped multi-hop network                        & TelosB                                  & 70                    \\ \hline
            CityLab                  & Univ. of Antwerp                      & Europe          & 802.15.4          & Support multiple radios in one node                      & X86 PCEngines APU embedded device       & 32                    \\ \hline
            Indriya2                 & National Univ. of Singapore           & Asia            & 802.15.4          & Upgraded software and architecture from Indriya          & TelosB, CC2650                          & 102                   \\ \hline
            DistributedHART          & Wayne State Univ.                     & U.S.            & WirelessHART      & Distributed scheduling for WirelessHART                  & TelosB                                  & 130                   \\ \hline
            Cracking TSCH            & State Univ. of New York at Binghamton & U.S.            & 802.15.4e         & Crack the channel hopping sequence of TSCH               & TelosB                                  & 50                    \\ \hline
            REACT                    & Washington Univ.                      & U.S.            & WirelessHART      & Reactive scheduling and routing reconfiguration          & TelosB                                  & 60                    \\ \hline
            FlockLab 2               & ETH Zurich                            & Europe          & 802.15.4          & Non-intrusive hardware-based real-time tracing           & Customized multi-slot observer          & 15                    \\ \hline
            APaS                     & Univ. of Connecticut                  & U.S.            & 6TiSCH            & Full-stack implementation of 6TiSCH                      & CC2650                                  & 121                   \\ \hline
            g6TiSCH                  & Orange Labs                           & Europe          & 6TiSCH            & Multi-PHY wireless networking                            & OpenMote B                              & 36                    \\ \hline
            1KT                      & Univ. of Warsaw                       & Europe          & 802.15.4          & Large-scale indoor testbed                               & CC2650                                  & 1000                  \\ \hline
        \end{tabular}
    }
\end{table*}
\subsection{IEEE 802.15.4-based Testbeds}

\subsubsection{MoteLab}

MoteLab is a web-based sensor network testbed developed and deployed at Harvard University~\cite{werner2005motelab}. The testbed is composed of 30 CrossBow MicaZ  motes supporting IEEE 802.15.4 radio standard, providing a realistic environment for wireless sensor network research.

Developed to address the logistical challenges, MoteLab provides streamlined access to a large and stationary network of real sensor network devices. 
The testbed consists of a set of permanently-deployed sensor nodes connected to a central server. This server handles reprogramming and data logging while providing a web interface for creating and scheduling tasks running on the testbed. The web interface and preemptive scheduler allow a large community to share access to the lab, eliminating the difficulties inherent in cooperative scheduling. MoteLab is freely distributed and built on top of readily-available software tools, making it easy for other organizations to set up their own testbeds.

\subsubsection{Trio}
Trio is a large-scale outdoor wireless sensor network deployed over an area of approximately 50,000 square meters~\cite{dutta2006trio}. It consists of 557 Trio nodes, seven gateways, two repeaters, and a root server. The testbed's large scale and remote location necessitate the use of remotely-accessible tools for network management. To this end, the testbed utilizes the Nucleus network management system, which runs alongside testbed applications on the mote and includes a query system allowing users to determine which nodes are running at any particular time. 
The testbed is powered by solar energy, which presents both benefits and challenges. While the renewable energy source simplifies system operation, management, and maintenance, it also introduces instability due to the dynamics of solar power and the logistics of node initialization.

\eat{
\begin{figure}
  \centering \includegraphics[width=\linewidth]{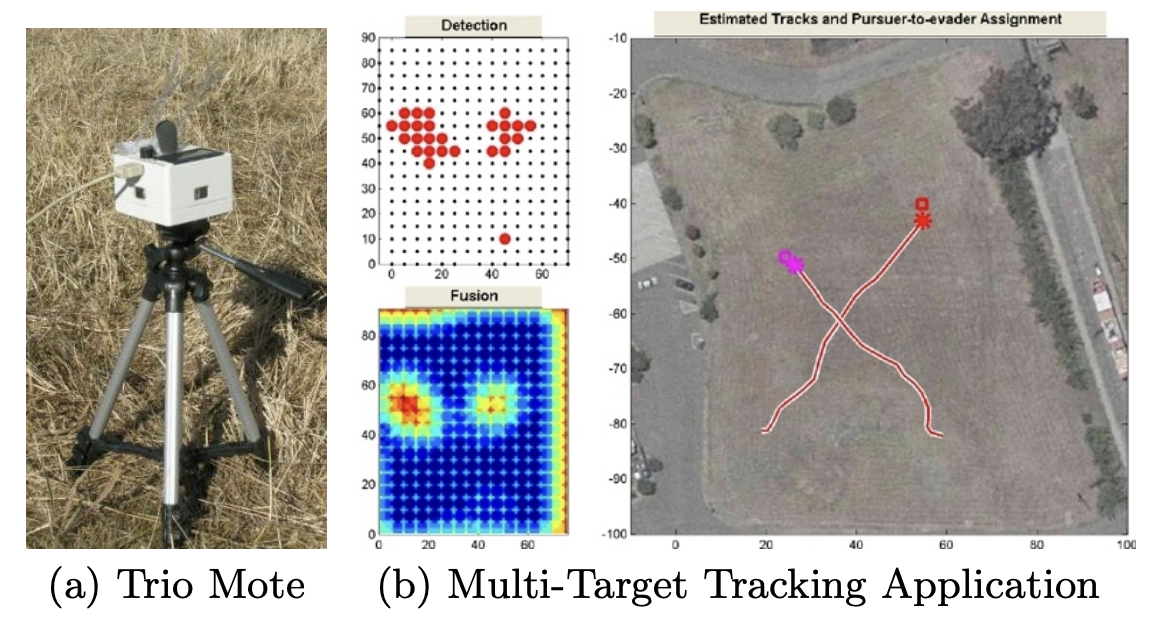}
  \caption{The Trio platform and an experimental multi-target tracking application that uses it.}
  \label{Fig:154:trio}
\end{figure}
}

The Trio testbed has been used for various research purposes, including the evaluation of multi-target tracking algorithms, the development and evaluation of programming tools, the evaluation of network management tools, and the exploration of various battery charging algorithms. 
The testbed has proven to be a valuable resource for WSAN research, despite the challenges associated with using a renewable energy source and managing a large-scale and outdoor network.

\subsubsection{Kansei}

\begin{figure}
  \centering \includegraphics[width=\linewidth]{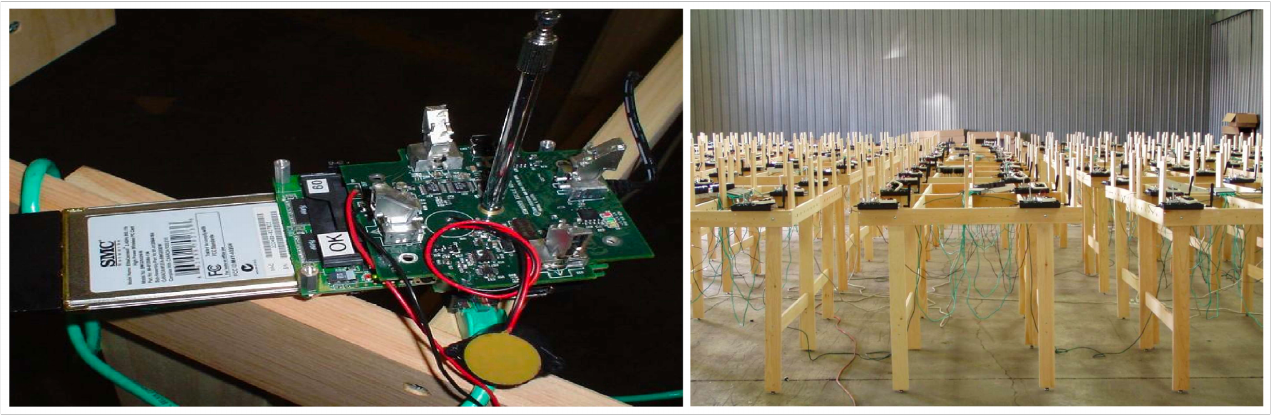}
  \caption{Kansei XSM node and the stationary array deployment.}
  \label{Fig:154:kansei}
  \vspace{-0.15in}
\end{figure}

Kansei is a large-scale testbed infrastructure designed for conducting experiments with both IEEE 802.11 and 802.15.4 mote networks \cite{ertin2006kansei}. The testbed is composed of 210 Extreme-Scale Motes (XSM), 150 TMote Sky nodes and 50 Trio motes. Fig.~\ref{Fig:154:kansei} shows the XSM node and its deployment on a 15×14 rectangular grid benchwork with 3ft spacing. The testbed features an experiment scheduler designed for flexible and dependable experimentation. The hardware and software platforms for Kansei have been set up to facilitate high-fidelity wireless network experimentation. This includes studying both indoor and outdoor wireless link properties and co-designing the network system to enable high-fidelity experimentation with reconfigurable network setup, such as node distribution density and wireless link reliability.

\subsubsection{WISEBED}

WISEBED is a large-scale WSAN testbed, consisting of 550 nodes distributed across nine geographically disparate sites in Europe \cite{chatzigiannakis2010wisebed}. The testbed is designed to support large-scale and diverse experiments with future internet technologies. It provides a Web interface for managing and demonstrating the experiment results. 
The WISEBED project consists of two different sites: 
the Lübeck Testbed and the Research Academic Computer Technology Institute (RACTI) in Patras. 

The Lübeck Testbed, operated by the University of Lübeck (UZL), consists of two testbeds. The first testbed uses approximately 50 Pacemate nodes, developed for services for athletes during a marathon. These nodes are equipped with Philips LPC2136 processors and a Xemics RF module running at 868 MHz. The second testbed consists of up to 50 iSense nodes by coalesenses GmbH. Each node in the stationary array consists of two hardware platforms: Extreme Scale Motes (XSMs) and Stargates.

The RACTI testbed in Patras spans over two locations at the University of Patras’ campus. The testbed is mainly used to monitor conditions inside these two buildings, including parameters such as temperature, light, humidity, acceleration, levels of magnetic fields, and barometric pressure. The hardware architecture used for the purposes of the testbed has three hierarchical levels: the sensor network level, the gateways used to interface the sensor network to the rest of the world, and the servers used to store information and administer the testbed. The testbed spans across 4 floors, covering almost one third of RACTI’s main building. The testbed consists of devices provided by Crossbow and Sun on the sensor network level, including 20 Crossbow Mica2 devices, 20 Crossbow TelosB devices, 45 Sun SPOT devices, and 60 iSense sensor nodes with a variety of sensor boards.

\subsubsection{CONET}

\cite{conet2011} provides a testbed developed for the EU-funded Cooperating Objects Network of Excellence (CONET) and located at the School of Engineering of Seville, Spain. The CONET testbed is a versatile platform designed for experiments integrating mobile robots and wireless sensor networks. It is equipped with a variety of sensors, including cameras, laser range finders, ultrasound sensors, GPS receivers, accelerometers, temperature sensors, and microphones, mounted on two different types of platforms: mobile robots and WSNs. The testbed includes five skid-steer holonomic Pioneer 3-AT all-terrain robots and a WSAN consisting of static and mobile nodes.

\eat{
\begin{figure}
  \centering \includegraphics[width=\linewidth]{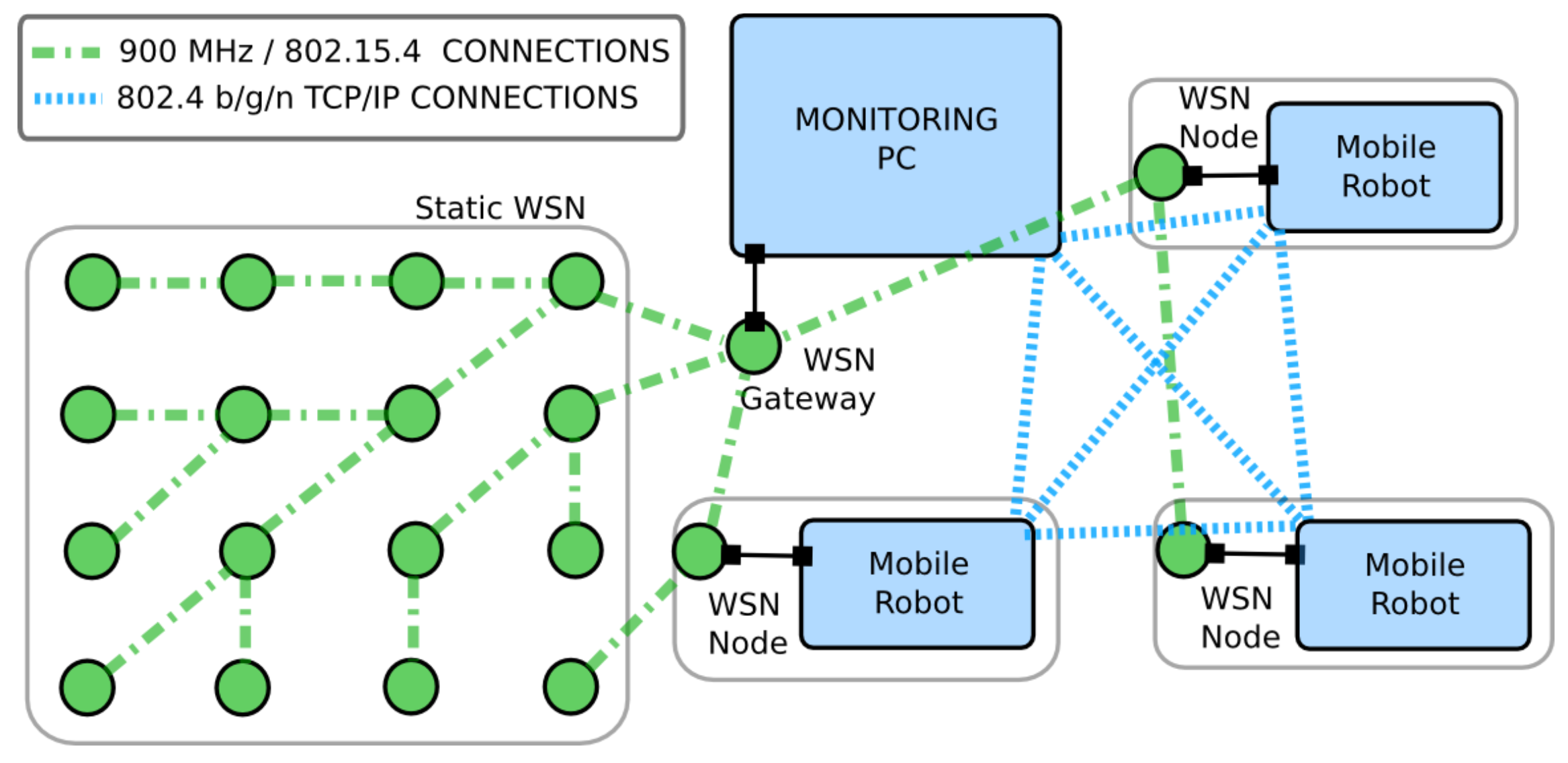}
  \caption{Connections among the elements in the CONET testbed.}
  \label{Fig:154:conet}
\end{figure}
}

Key features of the testbed include its open and modular architecture, which allows for easy extension to new hardware and software components and promotes code reuse. It supports different levels of decentralization, from fully distributed to centralized approaches, and provides full remote control through the Internet.
In terms of network protocols, TelosB, Iris, or MicaZ nodes use the IEEE 802.15.4 protocol while Mica2 nodes use an ad-hoc protocol operating in the 900 MHz radio band. 
The testbed is deployed in a room of more than 500 m$^2$, with static WSN nodes deployed at 21 predefined locations. The design allows for high flexibility in routing and network combinations, catering to a wide range of experimental needs.

\subsubsection{Indriya}

\begin{figure}
  \centering \includegraphics[width=\linewidth]{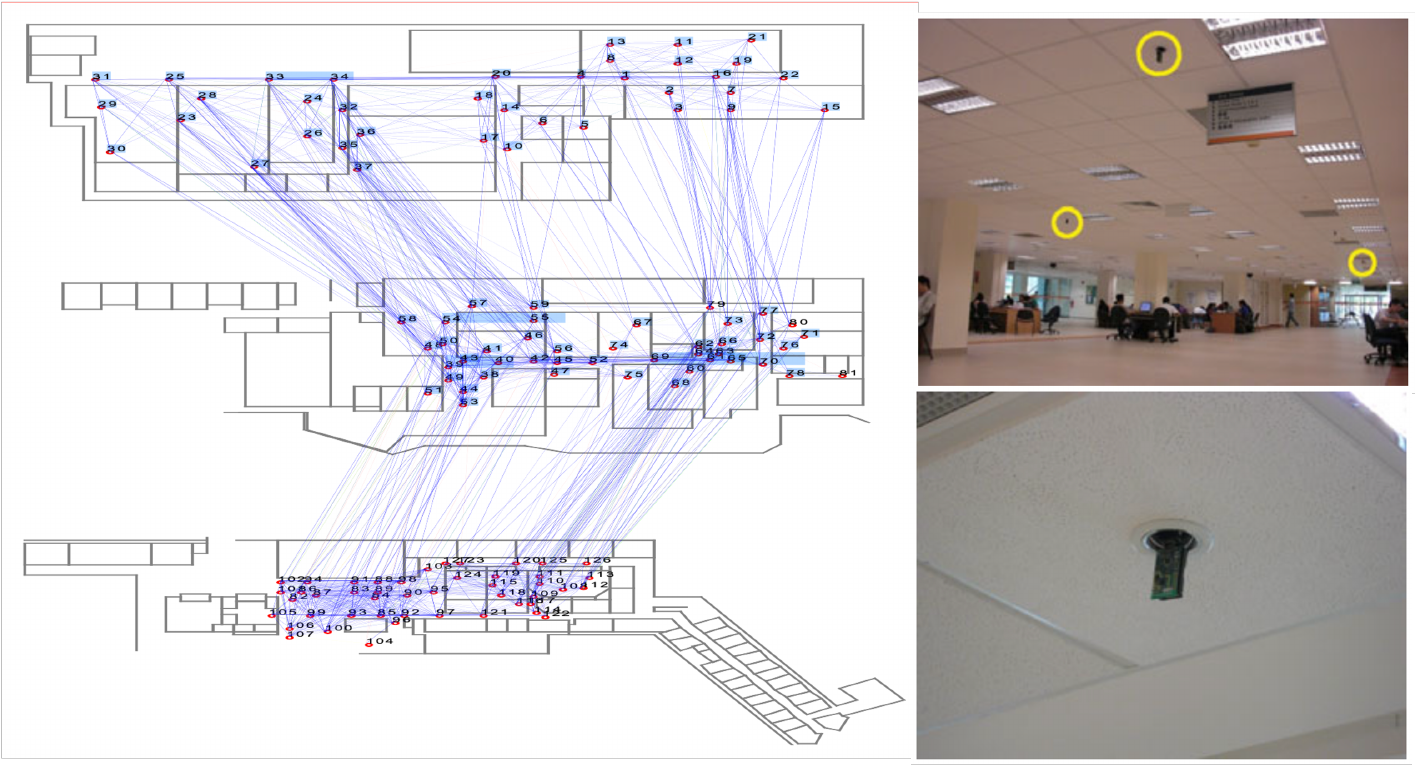}
  \caption{Deployment of the Indriya testbed and its USB-connected nodes.}
  \label{Fig:154:indriya}
  \vspace{-0.1in}
\end{figure}

Indriya, a large-scale and low-cost WSAN testbed, is deployed at the National University of Singapore~\cite{indriya2012}. The testbed is composed of 127 TelosB nodes, distributed across three floors of a building, as shown in Fig.\ref{Fig:154:indriya}. Each node is equipped with a CC2420 radio chip, which supports 16 channels of the IEEE 802.15.4 standard.
The deployment of Indriya is unique in its use of an active-USB infrastructure. This infrastructure acts as a remote programming back-channel, allowing for efficient updates and modifications to the network. Additionally, the USB infrastructure supplies electric power to the sensor devices, eliminating the need for battery replacements and reducing maintenance costs.

The nodes are connected to USB hubs, which are then connected to USB extenders. These extenders are linked to a server through Ethernet cables. This setup allows for the remote programming of the nodes and the collection of debugging information. The server runs the testbed management software, which schedules experiments, programs nodes, and collects experimental data.
The testbed focuses on the extensive study of all 16 channels of IEEE 802.15.4 supported by CC2420 devices. The objective is to understand performance differences and correlations that may exist among different channels. The results of these studies have been instrumental in designing WSAN routing and MAC protocols that exploit channel-diversity.

\subsubsection{NetEye}

NetEye is a high-fidelity, robust WSAN testbed designed for user-friendly scientific experimentation \cite{ju2012neteye}. It consists of 130 TelosB motes equipped with IEEE 802.15.4 radios and 15 Linux laptops with IEEE 802.11a/b/g radios. This setup allows for high-fidelity experimentation with both single-hop and multi-hop WSAN networking. The lab-scale deployment of NetEye is designed to mimic the network properties of building-scale and outdoor WSAN deployment.

NetEye provides users with a simple and powerful web portal, which streamlines the whole lifecycle of scientific experimentation, including experiment scheduling, status monitoring, and data exfiltration. It employs a stable and fault-tolerant state machine to improve the robustness of the testbed in the presence of failures. A health monitoring service, NetEye Doctor, monitors hardware and software status, providing real-time health monitoring information about the testbed. This information is integrated with the lifecycle of experiment scheduling, experiment status monitoring, and experiment data analysis for both robust experimentation and informed experiment analysis.

\subsubsection{SmartSantander}

The SmartSantander project proposes a unique and city-scale experimental research facility aimed at supporting typical applications and services for a smart city \cite{sanchez2014smartsantander}. This facility is designed to be large, open, and flexible, enabling it to federate both horizontally and vertically with other experimental facilities. It is intended to stimulate the development of new applications by various types of users, including those conducting experimental advanced research on IoT technologies and those performing realistic assessments of user acceptability tests. The project plans to deploy 20,000 sensors in the cities of Belgrade, Guildford, Lübeck, and Santander, leveraging a wide variety of technologies.

The Santander testbed, a part of this larger project, currently consists of around 2000 IEEE 802.15.4 devices deployed in a 3-tiered architecture. The devices  are equipped with a variety of sensors, such as temperature, humidity, light, and noise sensors, as well as parking sensors and magnetic sensors for traffic flow detection. 
The devices are deployed in various locations throughout the city, including parks, buildings, and streetlights, as well as on city buses. This diverse deployment allows for a wide range of experiments and use cases, from environmental monitoring and smart parking to traffic management and smart transportation.


\subsubsection{FIT IoT-LAB}

The FIT IoT-LAB is a large-scale, open-access experimental testbed designed to facilitate the testing and development of IoT solutions \cite{fit2015,fit-website}. It provides a comprehensive platform for researchers, students, and companies from around the world to conduct experiments in a controlled environment. Since its inception in 2012, the IoT-LAB has been utilized by over 5,000 users from more than 40 countries, conducting over 200,000 experiments.

The testbed offers a wide variety of hardware boards, with 18 different types available to cater to diverse experimental needs. Some of these include the IoT-LAB M3 and IoT-LAB A8-M3, both custom boards developed specifically for the IoT-LAB, the Microchip SAMR21, Zolertia Firefly, ST B-L072Z-LRWAN1, and the Nordic nRF52840DK. These boards support a range of wireless protocols such as IEEE 802.15.4, Sub-GHz, Bluetooth Low Energy (BLE), and LoRa, providing flexibility for various IoT applications.


\subsubsection{Pisa}

Pisa~\cite{pisa2016} is a multi-purpose platform designed for the validation and performance evaluation of IoT solutions. Deployed in the Department of Information Engineering at the University of Pisa, it consists of 22 nodes providing environmental measurements. Each node is made up of a TelosB sensor mote and a Raspberry Pi, connected via USB. The testbed allows for remote programming and troubleshooting via a dedicated backbone network. A preliminary experimental study conducted on this platform evaluated the trade-offs between passive and active link monitoring strategies in wireless sensor networks using RPL as the routing protocol.

\subsubsection{WUSTL-WSN}

The Cyber-Physical Systems Laboratory at Washington University in St. Louis has deployed a WSAN testbed based on IEEE 802.15.4~\cite{WUSTL-WSN}. This testbed is designed to facilitate advanced research in wireless sensor network technology and currently consists of 70 wireless sensor nodes (TelosB motes), placed throughout several office areas in Jolley Hall of WUSTL. The testbed deployment is based on the TWIST architecture~\cite{twist}, originally developed by the Telecommunications Group at the Technical University of Berlin. This hierarchical architecture consists of three different levels: sensor nodes, microservers (Raspberry Pis), and a desktop class host/server machine. The sensor nodes are connected to the microservers via USB cables, and the microservers are connected to the server machine over an Ethernet backbone.

The sensor nodes are equipped with sensors for light, radiation, temperature, and humidity. They are placed throughout the physical environment to take sensor readings and/or perform actuation. The server machine hosts a PostgreSQL database containing information about different sensor nodes and the microservers they are connected to.

\subsubsection{CityLab}

The CityLab testbed is a large-scale, multi-technology wireless testbed that enables experimentation in a real-world city environment~\cite{struye2018citylab}. The testbed is designed to address the challenges of interference in wireless networks, particularly in the context of smart cities where a wide variety of wireless technologies coexist.

The CityLab testbed is used to study the interference between heterogeneous systems operating in the same frequency band. This is particularly relevant for the IoT networks and WiFi networks, which often occupy the same 2.4 GHz frequency band. The interference control and spectrum coordination are crucial in such environments. The hardware equipment is hosted at 32 locations with another 22 sites planned. Each location has its own gateway attached to houses in the street or installed on a pole on a roof. There is currently only a single type of node (gateway) available in the City of Things testbed. It consists of of an apu2c4 embedded pc with two WiFi interfaces and a number of additional radio modules that are connected through USB, including IEEE 802.11, IEEE 802.15.4,
and sub-GHz protocols.

\subsubsection{Indriya 2}

Indriya 2~\cite{indriya2019} is an advanced WSAN testbed that improves upon its predecessor, Indriya \cite{indriya2012}, in several ways. The hardware deployment includes 74 TelosB motes and 28 CC2650 SensorTags, distributed over three floors and an outdoor area, all connected to the server via a Raspberry Pi. This is a significant upgrade from the MicaZ platform and direct server connection used in the original Indriya.

The testbed supports the 6LoWPAN and RPL network protocols, enabling the use of IP in low-power and lossy networks. Indriya 2 also boasts a more robust infrastructure with a better power supply system, a more stable network connection, and a user-friendly web-based interface. It offers improved support for long-term experiments, outdoor deployments, and advanced experiments involving multi-hop networks and mobile nodes.

\subsubsection{FlockLab2}

FlockLab 2 is a testbed designed for multi-modal testing and validation of wireless IoT devices, located at ETH Zurich \cite{trub2020flocklab}. It is an upgrade from the original FlockLab \cite{lim2013flocklab}, featuring 60 observer nodes distributed across three floors of a building. These observer nodes are capable of monitoring and interacting with a target node under test, providing a diverse testing environment.
The testbed supports a variety of testing modalities, including GPIO tracing, serial port logging, power measurements, and radio packet sniffing. It also provides a time synchronization service for precise timing of events across multiple observer nodes.

FlockLab 2 features a non-intrusive tracing system that leverages on-chip debugging capabilities, allowing for the monitoring of internal state variables of a target node's firmware without requiring explicit code instrumentation. This system is particularly useful for debugging and validating distributed wireless communication protocols.
The testbed is available for remote access, allowing researchers worldwide to conduct experiments on a large-scale, real-world wireless IoT network. The testbed's services and resources are accessible through a web interface, and it supports automated test execution for repeatability and scalability of experiments.

\subsubsection{1KT}

1KT is a low-cost and large-scale wireless IoT testbed designed for the experimental evaluation of IoT-oriented low-power wireless networking solutions~\cite{banaszek20211kt}. It comprises 1000 experimental devices deployed directly in human spaces of 168 rooms on all 5 floors of a sizable building. The testbed is designed to match the envisioned deployment conditions of IoT solutions, particularly in terms of scale.
The experimental devices in the 1KT testbed are custom devices featuring modern ARM Cortex-M3s, radios for both IEEE 802.15.4 and Bluetooth Low Energy (BLE), and a range of other functionalities, like continuous power consumption measurement. 


The 1KT testbed has been used to run a series of preliminary experiments to study its low-power wireless characteristics. The results suggest that 1KT establishes an entire spectrum of links, in terms of packet reception rate (PRR), a wide range of regions, in terms of density, and numerous multi-hop paths, making it particularly suitable for the evaluation of indoor networking protocols. The stable and repeatable results of long-term experiments indicate that 1KT is indeed capable of providing scientific-grade data.
\subsection{IEEE 802.15.4e-based Testbeds}

\begin{figure*}
\centering
\includegraphics[width=0.95\textwidth]{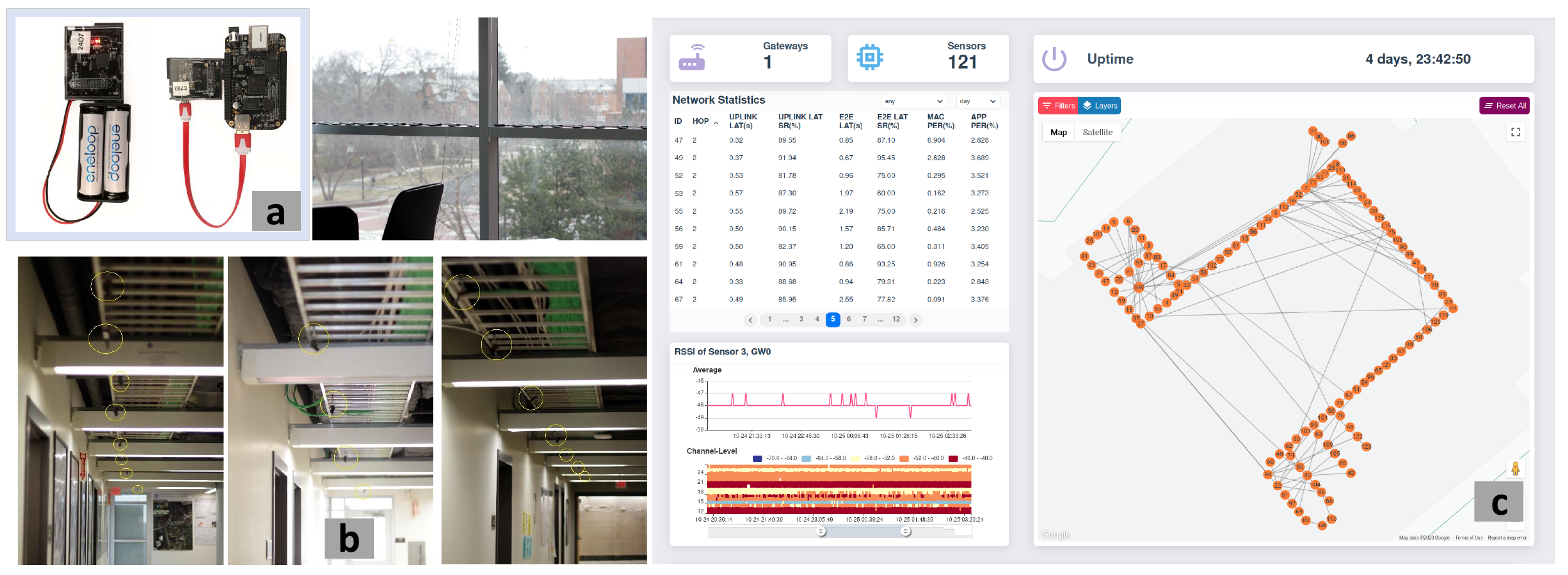}
\vspace{-0.1in}
\caption{6TiSCH testbed at UConn \cite{wang2021apas}. (a) Hardware platforms for the device (SensorTag) and gateway (BBB+SensorTag); (b) the testing environment where 122 devices and one gateway are mounted; (c) the cloud-based network management system.} 
\label{fig:apas-testbed}
\vspace{-0.1in}
\end{figure*}

\subsubsection{OpenWSN}

OpenWSN~\cite{openwsn2012} is a standards-based low-power wireless development environment that provides an open-source implementation of the IEEE802.15.4e standard, including the Time Synchronized Channel Hopping (TSCH) mode. The OpenWSN project aims to provide a platform for the development and testing of new protocols and applications in the domain of low-power wireless networks.

The OpenWSN project also provides a set of tools for testing and debugging the protocol stack. These tools include a network simulator, a packet sniffer, and a set of visualization tools. The network simulator allows for the simulation of large-scale networks, with support for different propagation models and node mobility models. The packet sniffer allows for the capture and analysis of packets transmitted over the air, while the visualization tools provide a graphical representation of the network topology and the packet flows in the network.

In terms of hardware setup, the OpenWSN project provides a reference implementation of the protocol stack on the OpenMote platform. The OpenMote is a small, low-cost, and low-power hardware platform designed for wireless sensor networks. It is based on the Texas Instruments CC2538 system-on-chip, which includes a 32-bit ARM Cortex-M3 microcontroller and IEEE 802.15.4 radio. 

The OpenWSN project has been used in a number of research projects and has been the subject of several academic publications. It has also been used in several industrial applications, including smart grid applications, building automation applications, and industrial automation applications.

\subsubsection{Rover}

The OpenVisualizer Rover testbed is a simple, easy-to-deploy, and cost-effective IoT testbed~\cite{brodard2016rover}. It is part of the OpenWSN project, which provides a free and open-source implementation of a standards-compliant protocol stack for the IoT, as well as all the necessary network management and debugging tools. 

The network management software, OpenVisualizer, is portable across popular operating systems and connects the OpenWSN low-power wireless mesh network to the Internet. In the current setup, motes (CC2538) are connected to the USB ports of the computer (Raspberry Pi 3) running OpenVisualizer. OpenVisualizer monitors the internal state of these motes and presents it through a web interface. 


\subsubsection{DistributedHART}

DistributedHART is a distributed real-time scheduling system designed for WirelessHART networks \cite{modekurthy2019distributedhart,modekurthy2021distributed}. It addresses the limitations of centralized scheduler of WirelessHART, which creates schedules centrally and in advance with a high degree of redundancy, leading to significant waste of energy, bandwidth, time, and memory. DistributedHART eliminates the need for creating and disseminating a centralized global schedule, reducing resource waste and enhancing scalability.

The authors of DistributedHART built a 130-node testbed to evaluate their proposed scheduling method using TelosB motes as device nodes and TinyOS as the real-time OS. The testbed is deployed in a controlled environment, and the devices used in the testbed are not specified in the paper. 

DistributedHART also introduces a higher network utilization at the expense of a slight increase in the end-to-end delay of all control loops. Through experiments on the testbed and large-scale simulation, the authors observed that DistributedHART consumes at least 85\% less energy compared to the existing centralized approaches.

\subsubsection{Cracking TSCH}

\cite{cheng2019cracking} focuses on the security vulnerabilities of Time-Synchronized Channel Hopping (TSCH) in IEEE 802.15.4e-based industrial networks. The authors argue that while TSCH simplifies network operations, it does so at the cost of security. 
In their case study, the authors use a publicly accessible implementation of TSCH and run experiments on a testbed consisting of 50 TelosB motes deployed in the second floor of the Engineering Building at their campus. They show that an attacker, with moderate computational capability, can crack the channel hopping sequence and predict future channel usage for attacks.


\subsubsection{REACT}

\cite{gunatilaka2020react} presents REACT, a reliable, efficient, and adaptive control plane for industrial WSANs. Instead of optimizing the performance of data plane, REACT specifically optimizes the network control plane. It significantly reduces the latency and energy cost of network reconfiguration, thereby improving the agility of WSANs in dynamic environments. To evaluate the effectiveness of REACT, authors built a WirelessHART testbed consisting of 50 TelosB motes located at WUSTL.

\subsubsection{APaS}

APaS (adaptive partition-based scheduling) is a framework designed for link scheduling of 6TiSCH networks~\cite{wang2021apas}. Instead of allocating network resources to individual devices, APaS partitions and assigns network resources to different groups of devices based on their distance to the network controller. This guarantees that the transmission latency of any end-to-end flow is within one slotframe length. APaS also employs online partition adjustment to improve its adaptability to dynamic network topology changes. 

Authors of APaS built a 122-node full-stack 6TiSCH testbed to evaluate their proposed resource partitioning framework~\cite{wang2021demo}. The 6TiSCH testbed is deployed at the 3rd floor of ITE Building at University of Connecticut, and it uses CC2650 SensorTag as the end devices, CC2652 as the access point and Beagle Bone Black (BBB) embedded Linux system as the gateway, creating a robust and efficient network for testing and evaluation. 6TiSCH stacks are implemented on top of TI RTOS. Fig.~\ref{fig:apas-testbed} shows the testbed hardware architecture and a Web-based user interface for network management.

In addition to APaS, the authors also proposed several resource management frameworks for handling network dynamics in real-time wireless networks (RTWNs)~\cite{zhang2018fd,zhang2019fully,gong2019reliable,shen2022distributed,wang2022harp,zhang2022reliable}. 
To evaluate the performance of the set of real-time scheduling frameworks, several RTWN testbeds~\cite{gong2017demo,gong2018demo} are also implemented based on OpenWSN~\cite{watteyne2012openwsn}. 

\subsubsection{g6TiSCH}

g6TiSCH is a generalized 6TiSCH architecture for multi-PHY wireless networking \cite{rady2021g6tisch}. g6TiSCH adds agility to the protocol stack by allowing nodes to use diverse physical layers within the same network and adapt their links depending on their conditions, while maintaining wire-like reliability.
To introduce this agility, the authors augment the 6TiSCH protocol stack in different ways. At the MAC layer, they add the physical layer to use for each time slot. At the 6LoWPAN adaptation layer, they demonstrate a generalized neighbor discovery mechanism where motes discover the network over different PHYs. At the RPL layer, they adapt the routing objective function with weighted link costs.

The resulting architecture is evaluated experimentally on a testbed of 36 OpenMote B boards deployed in an office building. Each OpenMote B can communicate using FSK 868 MHz, O-QPSK 2.4 GHz, or OFDM 868 MHz, a combination of long-range and short-range physical layers. The testbed is used to measure network formation time, end-to-end reliability, end-to-end latency, and battery lifetime.

\section{IEEE 802.11-based IIoT Testbeds}\label{sec:11}
In this section, we present a comprehensive description on a set of IEEE 802.11-based wireless testbeds. The description include four categories: industrial platform, simulators, academic SDR platform, FPGA SDR platform. Industrial platform targets on large-scale testbeds capable of working in industrial environment. Simulator is tool to replicate the behavior of wireless networks based on the IEEE 802.11 standard in a controlled and virtual environment. Academic SDR platform is testbed typically in lab scale and allows wireless communication systems to be reconfigured by using software running on a programmable hardware platform. FGPA SDR Platform has the feature of FPGA implementation on the testbed.

\subsection{Industrial Platforms}
The applications of industrial platforms using IEEE 802.11 standards have two major categories, i) industrial WiFi and ii) IEEE 802.11-based industrial wireless systems.

\subsubsection{Industrial WiFi}
Industrial WiFi is part of the industrial wireless local area network (IWLAN) and it fully follows the WiFi standards on both physical layer and MAC layer. Comparing to consumer grade WiFi, industrial WiFi enables high-level security, higher throughput, more spatial streams, larger coverage and being harsh environment ready. Besides the hardware features, industrial WiFi also supports ad-hoc mode besides the general infrastructure mode for more flexible network topology. Many companies provide industrial WiFi solutions, e.g., Phoenix Contact\cite{a2023_industrial}, Siemens\cite{IWLAN}, Cisco\cite{outdoor}, MOXA\cite{a2022_wlan}, etc. 

\subsubsection{IEEE 802.11-based industrial wireless systems}

\hfill

{\bf WIA-FA (ISA).}
Unlike legacy industrial WiFi system obeying both PHY and MAC protocols, WIA-FA, as a recently released international electrotechnical commission (IEC) standard, defines a three-layer stacks protocol: the PHY layer is based on IEEE 802.11; the DLL adopts a pure TDMA-based design; the application layer provides communication service and combination of multiple user application objects used to implement distributed industrial application~\cite{tramarin_2019_realtime}. Two testbeds, Large-Scale WIA-FA testbed and WIA-FA-Based Wireless programmable Logic Controller Testbed, compatible with the WIA-FA standard have been implemented in~\cite{8649759}.

\begin{figure}[bt]
    \centering
    \begin{minipage}[b]{0.23\textwidth}
    \includegraphics[width=\textwidth]{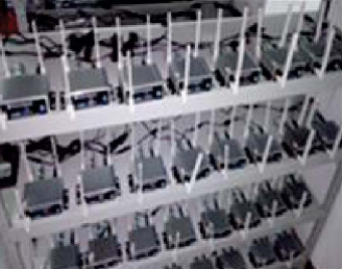}
        \caption{Part of the WIA-FA network.}
        \label{Fig:partofnodes}
    \end{minipage}
    \hfill
    \begin{minipage}[b]{0.23\textwidth}
    \includegraphics[width=\textwidth]{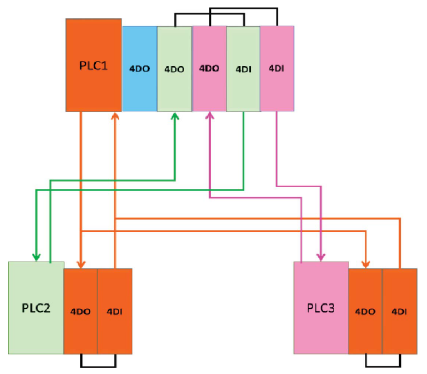}
        \caption{Distributed PLC IO control system.}
        \label{Fig:plcs}
    \end{minipage}
    \vspace{-0.1in}
\end{figure}

{Large-Scale WIA-FA testbed} developed 1000 nodes (shown in Fig.~\ref{Fig:partofnodes}) in order to conduct long-term experiments of WIA-FA networks. The proposed testbed is mainly used to evaluate the transmission reliability and delay of WIA-FA networks under different experiment setups. The WIA-FA network testbed is composed of one primary gateway, two redundant gateways, eight access devices, and 1000 field devices. 

{WIA-FA-Based Wireless programmable Logic Controller Testbed} includes both a WIA-FA network and a distributed IO control system of PLCs with the model number PSS4000/SNp, a product of the Pilz Company, Germany. The architecture of PLC IO control system is shown in Fig.~\ref{Fig:plcs}. The system includes threes PLCs with configurable scanning periods, four output ports denoted as 4DO, and four input ports denoted as 4DI. 


\eat{
\begin{figure}
  \centering \includegraphics[scale=0.55]{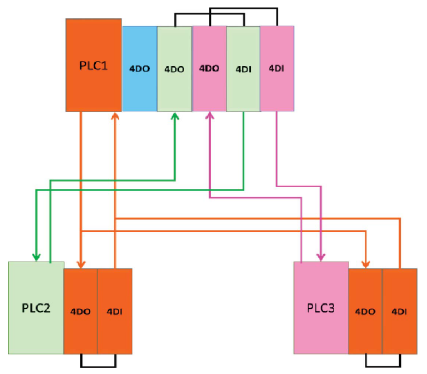}
  \caption{Distributed PLC IO control system.}
  \label{Fig:plcs}
\end{figure}
}

\eat{
\begin{figure}
  \centering \includegraphics[scale=0.8]{Figs/80211/wia-fa real testbed.png}
  \caption{WIA-FA-based distributed PLC IO control system.
(a) Schematic architecture of the testbed. (b) WIA-FA-based wireless
PLC testbed.}
  \label{Fig:realwiafa}
\end{figure}
}

{\bf RT-WiFi.} RT-WiFi~\cite{6728869} proposed a real-time high-speed communication solution based on WiFi protocol for testbed running on COTS application. Regular WiFi network is not suitable to support predictable real-time traffic since both the carrier sense multiple access with collision avoidance (CSMA/CA)
and random backoff mechanism introduce large packet delay. The TDMA-based data link layer is designed to address the problem by adapting centralized channel and time management to access the channels according to a strict time schedule. Since only one node can access a certain channel in a given time slot, RT-WiFi provides collision free and deterministic communication. 

\begin{figure}
  \centering \includegraphics[width=\linewidth]{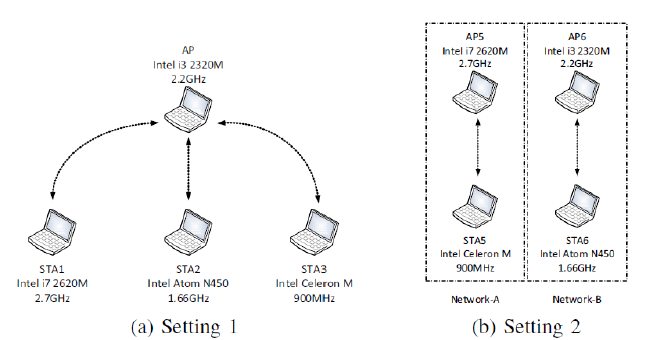}
  \caption{An overview of the testbed setting used in RT-WiFi.}
  \label{Fig:rtwifi}
  \vspace{-0.15in}
\end{figure}

Two testbed are conducted to perform performance comparison between RT-WiFi and regular WiFi. 
In the setting shown in Fig.~\ref{Fig:rtwifi}(a), the testbed consists of one AP and three stations which form a star network topology. All devices are equipped with the same Atheros AR9285 IEEE 802.11 compatible wireless interface under two environment settings. 
In the interference-free environment, where the testbed is deployed in a parking lot, 
the experimental results show that the latency variation in Wi-Fi network is up to 90 times to that of RT-WiFi network, and the maximum delay is more than 30 times compared to RT-WiFi.
In the office environment, the testbed is deployed in a building 
covered with more than 10 Wi-Fi APs, which spread to all the orthogonal Wi-Fi channels in ISM 2.4GHz band. The experimental results show that the mean delay in the office environment remains similar to the interference-free environment. 

Second testbed shown in Fig.~\ref{Fig:rtwifi}(b) is to evaluate the performance of the flexible channel access controller in the office environment. Network-A is a regular Wi-Fi network having traffic sent from STA5 to AP5. For network-B, data is published from AP6 to STA6 in four settings: 
regular Wi-Fi, RT-WiFi baseline, RT-WiFi with co-existence, RT-WiFi with co-existence and one in-slot-retry. The experimental results show that with the presence of Wi-Fi networks in the operation environment, the RT-WiFi data link layer keeps real-time characteristics while minimizing its influence on the surrounding regular Wi-Fi networks as well.

{\bf Det-WiFi.} \cite{cheng2017det} proposes a real-time TDMA MAC implementation for high-speed multihop industrial application. It is able to support high-speed applications and provide deterministic network features by combining the advantages of high-speed IEEE 802.11 physical layer and a software TDMA-based MAC layer. 

The Det-WiFi testbed setup includes three PCs, all of which are quipped COTS hardware Atheros AR9287 IEEE 802.11 compatible NIC. Each PC runs Ubuntu 14.04 as the operating system. In order to realize multihop deterministic characteristics, both default module frame format and CSMA mechanism in mac8011 framework are modified. The testbed is set under the real industrial environment and used to compare the deterministic performance between 802.11s network and Det-WiFi. 
Experimental results based on the testbed show that Det-WiFi achieves better deterministic performance compared with the 802.11s network under different packet payload size settings. 

\eat{
\begin{figure}
  \centering \includegraphics[width=0.8\linewidth]{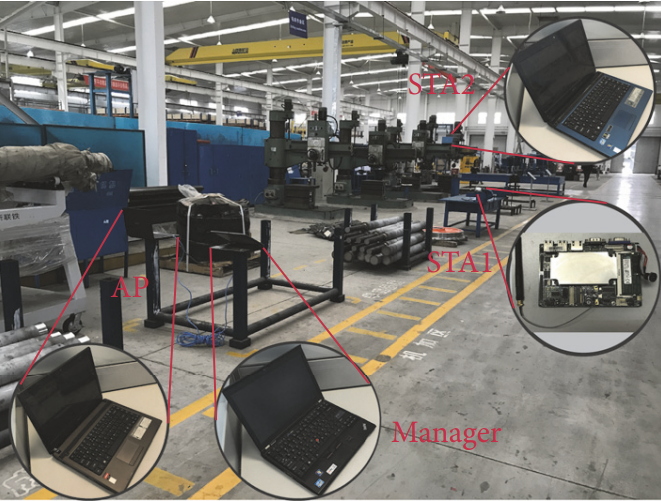}
  \caption{Testbed in real industrial environment.}
  \label{Fig:detwifi}
\end{figure}
}

\subsection{Simulators}
{\bf TETCOS.} TETCOS~\cite{tetcos_tetcos} developed by TETCOS LLP in India, is a comprehensive software simulation tool that enables users to design, model, and simulate Wi-Fi networks. It offers a range of devices such as AP and Wireless Nodes (STA) to create network configurations. The simulator supports WLAN protocols including IEEE 802.11 a/b/g/n/ac/p in PHY Layer and rate adaptation, CSMA/CA, infrastructure, MPDU and QoS based on EDCA in MAC Layer. A drag and drop GUI enables users to quickly create the network and set properties. The results of a simulation run are presented in a dashboard for convenient analysis. Graphics plots comprise application throughput, link throughput, buffer occupancy and TCP congestion windows. 

{\bf WLAN Toolbox.}  WLAN Toolbox~\cite{wlan} developed by MATLAB provides standards-compliant functions for the design, simulation, analysis, and testing of wireless LAN communications systems. It includes configurable physical layer waveform for IEEE 802.11be/az/ax/ac/ad/ah/j/p/n/g/a/b. It provides transmitter, channel modeling, and receiver operations, including channel coding, modulation, spatial stream mapping, and MIMO receivers. WLAN Toolbox extends support for system-level simulation by enabling the modeling of WiFi links with multiple nodes. Each node can have a configuration setting on application layer, MAC, and PHY layer parameters.The simulation results show performance metrics such as throughput, latency, and packet loss.

\subsection{Academic SDR Platforms}
Software-Defined Radio (SDR) refers to a radio communication system where traditional hardware components are replaced or supplemented by software algorithms running on a computer or embedded system. A typical SDR testbed setup includes a hardware interface (e.g., USRP, HackRF) taking care of radio reception and a software application (e.g., GNU Radio, MATLAB, and SDR Console) that has control over hardware performing signal processing related tasks, including symbol synchronization, channel equalization, carrier frequency offset compensation, and demodulation and decoding. 

{\bf SWiFi.} 
In both studies conducted at SWiFi \cite{vo2015swifi} and \cite{bloessl2013decoding}, the receiver is implemented using GNU Radio, enabling the capture and decoding of IEEE 802.11 packets. These two testbeds differ in their focus: the former targets at IEEE 802.11 a/b/g and includes a transmitter implementation, while the latter targets at IEEE 802.11 a/g/p and focuses solely on receiver implementation.
\cite{odquist2020software} has a receiver implemented in MATLAB and equip with USRP, and it has achieved a functionality of partially decoding IEEE 802.11n packets. 
A Beacon Frame Receiver for IEEE 802.11b, implemented in \cite{getz_2021_beacon}, utilizes FMCOMMS2 as an RF front-end to capture WiFi signals transmitted over the air. The receiver comprises an iio\textunderscore sys\textunderscore obj block and a Simulink model running in MATLAB, which performs demodulation and decoding of the incoming data stream and output the beacon information whenever packet is received.


{\bf Subramanian et al.} 
To accomplish a bidirectional transceiver that takes into account accurate timing requirements on standards-specified tasks,~\cite{7454680} designs an IEEE 802.11b standard-compliant Link Layer for MATLAB-based SDR testbed. Each node in the testbed has the setting of one USPR N210 connected to a PC host using a gigabit Ethernet cable. MATLAB is the software used to program the hardware and Ubuntu OS is the operating system running on the PC. There are two experiments have been done to evaluate the performance of the testbed. 

In the first experiment, the testbed contains two nodes and one bi-directional link. The experiment focuses on the performance evaluation when the transmit power lever of the DTx is decreased below standard levels. The experimental results confirm that the performance remains consistent for a wide range of transmit gains. 

In the second experiment, the testbed contains two bi-directional link incidents on one shared DRx and focuses on the modification of MAC parameters. The output shows that the latency of the two links deviates by only a small amount from the ideal line for varying payload sizes. The experimental result establishes the role and efficacy of the MAC layer in enabling and enforcing fairness among the two DTxs when accessing the common channel.

\eat{
\begin{figure}
  \centering \includegraphics[width=0.8\linewidth]{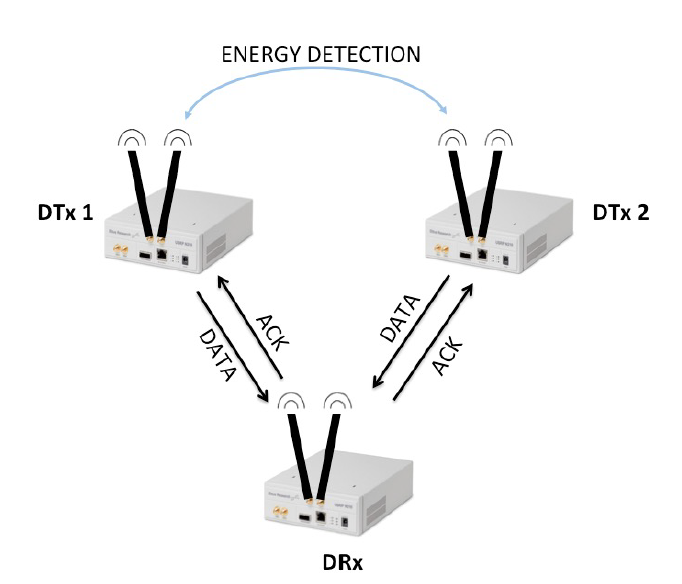}
  \caption{Three Node System with 2 DTxs and 1 DRx.}
  \label{Fig:threenodes}
\end{figure}
}

\subsection{FGPA-based SDR Platforms}

{\bf OpenWiFi.} A free and open source full stack IEEE 802.111a/g/n SDR platform is developed in OpenWiFi~\cite{9128614}. The platform is based on Xilinx Zynq SoC that includes a FPGA and ARM processor. Critical SIFS timing is achieved by implementing PHY and low-level MAC in FPGA. The corresponding driver is implemented in the embedded Linux running on the ARM processor. Due to the modular design, researchers could study and modify OpenWiFi easily. Through configuration and installation on both the FPGA and Linux, an OpenWiFi platform can be turned into a WiFi AP that gives access for stations to connect to the internet.


 {\bf SRT-WiFi.} SRT-WiFi is a SDR-based RT-WiFI solution~\cite{6728869} to tackle COTS hardware out-of-date issues. The design of SRT-WiFi is based on an advanced SDR platform (ZC706 development board with Zynq-7000 and AD9364). This advanced SDR system can run in real-time since the radio functions are achieved by the logic blocks in FPGA running at the speed as driven by an oscillator. With such a programmable real-time radio system, SRT-WiFi can achieve the key functions required to support high-speed real-time communications, and also provide an open-source platform to support ever-evolving IEEE 802.11 standards.

A multi-cluster SRT-WiFi testbed is conducted with the setting of two APs connected to a router to form a backbone network and two clusters, where each cluster contains two stations. The testbed shows that links between AP and station are capable of adapting proper rates to achieve good PDR when interference is generated and added to the network. The proposed heuristic task schedule running on the testbed offers an effective solution to perform scheduling for the multi-cluster SRT-WiFi network based on rate adaptation~\cite{9804669}.

{\bf Moreno et al.} \cite{9180885} designs and implements an open-source, low-cost and configurable mm-ware experimental platform for researchers to measure channel characteristics of mm-wave channels, compute Channel State Information (CSI) and allow the extension to multiple antenna systems. The testbed utilizes a software model to compute and generate IEEE 802.11ad compliant frames at the transmitter end. At the receiver end, FPGA implementations on the Packet Detector, CFO estimation/Correction, Boundary Detector, and CIR Computation blocks serve as the main signal process units that will output an estimated CIR corresponding to different indoor channel conditions. Final CIR information is plotted and saved by GNU radio powered by the RFNoC framework.

 \section{5G-based IIoT Testbeds}\label{sec:5g}
In this section, we provide a detailed description on a set of 5G-enabled testbeds deployed in various industrial scenarios. We divide the testbeds into two categories: large-scale and small-scale 5G-based testbeds. Large-scale testbeds comprise of more than one 5G network sites or span multiple physical locations. Small-scale IIoT testbeds are more compact and limited in size (lab-scale). 
Summaries of the reviewed large-scale and small-scale 5G testbeds are given in Table.~\ref{tab:5G_testbeds_large} and Table.~\ref{tab:5G_testbeds_small}, respectively. 




\begin{table*}[]
\caption{Summary of Large-Scale 5G-based IIoT Testbeds. }
\label{tab:5G_testbeds_large}
\begin{center}
\resizebox{\linewidth}{!}{
\begin{tabular}{|cc|c|c|c|c|c|}
\hline
\rowcolor[HTML]{EFEFEF} 
\multicolumn{2}{|c|}{\cellcolor[HTML]{EFEFEF}\textbf{Name or authors}}                                                                                           & \textbf{Institution}                                                      & \textbf{Region}           & \textbf{Use cases}                                                                                                                                                         & \textbf{Provider}                                                                                            & \textbf{Frequency}                                               \\ \hline
\multicolumn{1}{|c|}{}                                                                               & Kista site                                                & Ericsson                                                                  & Sweden                    & Robotics                                                                                                                                                                   & RAN: Ericsson                                                                                                & 28 GHz                                                           \\ \cline{2-7} 
\multicolumn{1}{|c|}{}                                                                               & Aachen site                                               & Fraunhofer IPT                                                            & Germany                   & Monitoring                                                                                                                                                                 & \begin{tabular}[c]{@{}c@{}}RAN: Ericsson, WNC\\ UE: Quectel\end{tabular}                                     & \begin{tabular}[c]{@{}c@{}}3.7 - 3.8 GHz,\\ 28 GHz\end{tabular}  \\ \cline{2-7} 
\multicolumn{1}{|c|}{\multirow{-3}{*}{5G-SMART}}                                                     & \begin{tabular}[c]{@{}c@{}}Reutlingen\\ site\end{tabular} & Bosch                                                                     & Germany                   & Processing                                                                                                                                                                 & RAN: Ericsson                                                                                                & \begin{tabular}[c]{@{}c@{}}3.7 - 3.8 GHz,\\ 28 GHz\end{tabular}  \\ \hline
\multicolumn{1}{|c|}{}                                                                               & UK site                                                   & BT Group                                                                  & UK                        &                                                                                                                                                                            & \begin{tabular}[c]{@{}c@{}}NSA Core: Samsung\\ RAN: Samsung\end{tabular}                                     & 3.6, 28 GHz                                                      \\ \cline{2-4} \cline{6-7} 
\multicolumn{1}{|c|}{}                                                                               & Norway site                                               & Telenor Norway                                                            & Norway                    &                                                                                                                                                                            & \begin{tabular}[c]{@{}c@{}}Core: Ericsson\\ RAN: Huawei, Ericsson\end{tabular}                               & 3.5, 26 GHz                                                      \\ \cline{2-4} \cline{6-7} 
\multicolumn{1}{|c|}{}                                                                               & Spain site                                                & 5TONIC laboratory                                                         & Spain                     &                                                                                                                                                                            & \begin{tabular}[c]{@{}c@{}}Core: Ericsson, Open5GCore\\ RAN: Ericsson\end{tabular}                           & 30 - 300 GHz                                                     \\ \cline{2-4} \cline{6-7} 
\multicolumn{1}{|c|}{}                                                                               & Greece site                                               & Univ. of Patras                                                           & Greece                    &                                                                                                                                                                            & \begin{tabular}[c]{@{}c@{}}Core: Open5GCore\\ RAN: srsRAN\end{tabular}                                       & 3.5 - 3.8 GHz                                                    \\ \cline{2-4} \cline{6-7} 
\multicolumn{1}{|c|}{}                                                                               & Portugal site                                             & Altice Labs                                                               & Portugal                  &                                                                                                                                                                            & \begin{tabular}[c]{@{}c@{}}Core: Open5GCore\\ RAN: OAI\end{tabular}                                          & \begin{tabular}[c]{@{}c@{}}250 MHz - \\ 3.8 GHz\end{tabular}     \\ \cline{2-4} \cline{6-7} 
\multicolumn{1}{|c|}{}                                                                               & Munich site                                               & Huawei MRC                                                                & Germany                   &                                                                                                                                                                            & Core\&RAN: Huawei                                                                                            & 3.41 GHz                                                         \\ \cline{2-4} \cline{6-7} 
\multicolumn{1}{|c|}{}                                                                               & Berlin site                                               & Fraunhofer FOKUS                                                          & Germany                   &                                                                                                                                                                            & \begin{tabular}[c]{@{}c@{}}Core: Open5GCore\\ Simulated RAN\&UE\end{tabular}                                 & N/A                                                              \\ \cline{2-4} \cline{6-7} 
\multicolumn{1}{|c|}{\multirow{-8}{*}{5G-VINNI}}                                                     & \begin{tabular}[c]{@{}c@{}}Luxembourg\\ site\end{tabular} & SATis5                                                                    & Luxembourg                & \multirow{-8}{*}{\begin{tabular}[c]{@{}c@{}}Public safety,\\ eHealth,\\ transportation,\\ AR/VR/MR,\\ emergency response,\\ energy distribution,\\ robotics.\end{tabular}} & \begin{tabular}[c]{@{}c@{}}Core: Open5GCore\\ RAN: Not applicable\end{tabular}                               & N/A                                                              \\ \hline
\multicolumn{1}{|c|}{}                                                                               & Oulu site                                                 & Oulu Univ.\&VTT                                                           &                           & \begin{tabular}[c]{@{}c@{}}Media, sports, eHealth\\ connected cars, VR,\\ factory of the future\end{tabular}                                                               & Core\&RAN: Nokia                                                                                             & 2.6, 3.5 GHz                                                     \\ \cline{2-3} \cline{5-7} 
\multicolumn{1}{|c|}{}                                                                               & Espoo site                                                & Aalto Univ.\&VTT                                                          &                           & \begin{tabular}[c]{@{}c@{}}AR/VR, gaming,\\ industrial internet\end{tabular}                                                                                               & Core\&RAN: Nokia                                                                                             & 2.6 GHz                                                          \\ \cline{2-3} \cline{5-7} 
\multicolumn{1}{|c|}{}                                                                               & Helsinki site                                             & Univ. of Helsinki                                                         &                           & Smart city                                                                                                                                                                 & \begin{tabular}[c]{@{}c@{}}Core: Openair-CN/Aalto/Nokia\\ RAN: Nokia\end{tabular}                            & 2.6 GHz                                                          \\ \cline{2-3} \cline{5-7} 
\multicolumn{1}{|c|}{}                                                                               & Tampere site                                              & Tampere Univ.                                                             &                           & Autonomous vehicles                                                                                                                                                        & Core\&RAN: Nokia                                                                                             & \begin{tabular}[c]{@{}c@{}}700 MHz, \\ 2.1, 2.6 GHz\end{tabular} \\ \cline{2-3} \cline{5-7} 
\multicolumn{1}{|c|}{}                                                                               & Turku site                                                & \begin{tabular}[c]{@{}c@{}}Turku Univ. of\\ Applied Sciences\end{tabular} &                           & \begin{tabular}[c]{@{}c@{}}Media and entertainment,\\ smart grid,\\ factory of the future\end{tabular}                                                                     & RAN: Nokia                                                                                                   & \begin{tabular}[c]{@{}c@{}}700 MHz,\\ 2.3, 3.5 GHz\end{tabular}  \\ \cline{2-3} \cline{5-7} 
\multicolumn{1}{|c|}{\multirow{-6}{*}{5GTNF}}                                                        & INX                                                       & VTT                                                                       & \multirow{-6}{*}{Finland} & \begin{tabular}[c]{@{}c@{}}Logistics, AR, robotics\\ production automation,\\ mobile control panels\end{tabular}                                                           & Core\&RAN: Nokia                                                                                             & 2.1, 2.6, 3.5 GHz                                                \\ \hline
\multicolumn{1}{|c|}{}                                                                               & Surrey                                                    &                                                                           & UK                        & AR                                                                                                                                                                         & \begin{tabular}[c]{@{}c@{}}Core: Vodafone\\ RAN: Nokia\\ UE: OAI/Nokia/Samsung\end{tabular}                  & 3.5, 28, 60 GHz                                                  \\ \cline{2-2} \cline{4-7} 
\multicolumn{1}{|c|}{\multirow{-2}{*}{5G-DRIVE}}                                                     & JRC Ispra                                                 & \multirow{-2}{*}{EURESCOM et al.}                                         & Italy                     & V2X                                                                                                                                                                        & RAN: Nokia/LimeNET/OAI/Amarisoft                                                                             & 3.5, 5.9 GHz                                                     \\ \hline
\multicolumn{2}{|c|}{Liverpool 5G}                                                                                                                               & Univ. of Liverpool                                                        & UK                        & eHealth, VR/AR                                                                                                                                                             & 5G solution: Blu Wireless                                                                                    & 59-63 GHz                                                        \\ \hline
\multicolumn{1}{|c|}{}                                                                               & Univ. of Bristol                                          & Univ. of Bristol                                                          &                           & No specific                                                                                                                                                                & \begin{tabular}[c]{@{}c@{}}Core: Open5GS/Nokia\\ RAN: Accelleran/Nokia\end{tabular}                          & \begin{tabular}[c]{@{}c@{}}2.6, 3.5,\\ 26, 60 GHz\end{tabular}   \\ \cline{2-3} \cline{5-7} 
\multicolumn{1}{|c|}{}                                                                               & 5GEM                                                      & Lancaster Univ.                                                           &                           & Manufacturing, logistics                                                                                                                                                   & 5G solution: Vodafone                                                                                        & 3.5 GHz                                                          \\ \cline{2-3} \cline{5-7} 
\multicolumn{1}{|c|}{\multirow{-3}{*}{\begin{tabular}[c]{@{}c@{}}5GUK\\ Test Networks\end{tabular}}} & 5G-ENCODE                                                 & Zeetta Networks                                                           & \multirow{-3}{*}{UK}      & Manufacturing industry                                                                                                                                                     & \begin{tabular}[c]{@{}c@{}}5G solution: Nokia/Airspan/Toshiba\\ UE: Quectel\end{tabular}                     & 3.6 GHz                                                          \\ \hline
\multicolumn{2}{|c|}{5G Innovation Hub North}                                                                                                                    & \begin{tabular}[c]{@{}c@{}}Luleå Univ.\\ of Technology\end{tabular}       & Sweden                    & \begin{tabular}[c]{@{}c@{}}Drones, robotics, eHealth\\ agricultural machinery\end{tabular}                                                                                 & \begin{tabular}[c]{@{}c@{}}5G solution: Ericsson\\ UE: Sony/Samsung/Xiaomi\end{tabular}                      & 3.5 GHz                                                          \\ \hline
\multicolumn{2}{|c|}{5G Security Test Bed}                                                                                                                       & CTIA et al.                                                               & U.S.                      & \begin{tabular}[c]{@{}c@{}}Logistics, smart city,\\ factory processing, AR\end{tabular}                                                                                    & Core\&RAN: Ericsson                                                                                          & 2.6, 39 GHz                                                      \\ \hline
\multicolumn{1}{|c|}{}                                                                               & POWDER                                                    & Univ. of Utah et al.                                                      &                           & No specific                                                                                                                                                                & \begin{tabular}[c]{@{}c@{}}Core: OAI/Open5GS/Free5GC\\ RAN: OAI/srsRAN\\ UE: Quectel\end{tabular}            & 3.5 GHz                                                          \\ \cline{2-3} \cline{5-7} 
\multicolumn{1}{|c|}{}                                                                               & AERPAW                                                    & NC State et al.                                                           &                           & UAVs                                                                                                                                                                       & RAN: Ericsson/OAI/srsRAN                                                                                     & 1.7, 2.1, 3.4 GHz                                                \\ \cline{2-3} \cline{5-7} 
\multicolumn{1}{|c|}{\multirow{-3}{*}{PAWR}}                                                         & COSMOS                                                    & Rutgers Univ. et al.                                                      & \multirow{-3}{*}{U.S.}    & Smart city                                                                                                                                                                 & OAI                                                                                                          & sub-6, 28, 60 GHz                                                \\ \hline
\multicolumn{1}{|c|}{}                                                                               & Athens                                                    & NCSR “Demokritos”                                                         & Greece                    & UAVs                                                                                                                                                                       & \begin{tabular}[c]{@{}c@{}}Core: Athonet/OAI EPC/Amarisoft EPC\\ RAN: Nokia/OAI/Amarisoft/RunEL\end{tabular} & 2.6, 3.6 GHz                                                     \\ \cline{2-7} 
\multicolumn{1}{|c|}{}                                                                               & Berlin                                                    & Fraunhofer FOKUS                                                          & Germany                   & Media and video                                                                                                                                                            & \begin{tabular}[c]{@{}c@{}}Core: Open5GC\\ RAN: Nokia/Huawei\end{tabular}                                    & 700 MHz, 2.6, 3.7 GHz                                            \\ \cline{2-7} 
\multicolumn{1}{|c|}{}                                                                               & Limassol                                                  & \begin{tabular}[c]{@{}c@{}}Space Hellas\\ Cyprus Ltd\end{tabular}         & Cyprus                    & Maritime communication                                                                                                                                                     & \begin{tabular}[c]{@{}c@{}}Core: Athonet/Amarisoft/Open5GS/NextEPC\\ RAN: Amarisoft\end{tabular}             & 2 GHz                                                            \\ \cline{2-7} 
\multicolumn{1}{|c|}{}                                                                               & Malaga                                                    & Univ. of Malaga                                                           & Spain                     & \begin{tabular}[c]{@{}c@{}}Mission critical\\ communication\end{tabular}                                                                                                   & \begin{tabular}[c]{@{}c@{}}Core: Athonet/Polaris EPC\\ RAN: Nokia/OAI/RunEL\end{tabular}                     & 6 GHz                                                            \\ \cline{2-7} 
\multicolumn{1}{|c|}{\multirow{-5}{*}{5GENESIS}}                                                     & Surrey                                                    & Univ. of Surrey                                                           & UK                        & Massive IoT                                                                                                                                                                & Not available                                                                                                & 700 MHz, 2.6, 3.7 GHz                                            \\ \hline
\multicolumn{1}{|c|}{}                                                                               & Erlangen                                                  &                                                                           &                           & Positioning, V2X                                                                                                                                                           &                                                                                                              & 3.7–3.8 GHz                                                      \\ \cline{2-2} \cline{5-5} \cline{7-7} 
\multicolumn{1}{|c|}{}                                                                               & Nürnberg                                                  &                                                                           &                           & Positioning                                                                                                                                                                &                                                                                                              & 3.7-3.8, 26, 28 GHz                                              \\ \cline{2-2} \cline{5-5} \cline{7-7} 
\multicolumn{1}{|c|}{\multirow{-3}{*}{5G Bavaria}}                                                   & Rosenheim                                                 & \multirow{-3}{*}{Fraunhofer IIS}                                          & \multirow{-3}{*}{Germany} & V2X                                                                                                                                                                        & \multirow{-3}{*}{Not available}                                                                              & NA                                                               \\ \hline
\multicolumn{2}{|c|}{\begin{tabular}[c]{@{}c@{}}DEMAG Research\\ Factory - MHD\end{tabular}}                                                                     & Demag                                                                     & Europe                    & Intralogistics                                                                                                                                                             & Nokia solution                                                                                               &                         3.7 - 3.8 GHz                                         \\ \hline
\end{tabular}
}
\end{center}
\end{table*}

\subsection{Large-Scale 5G-based IIoT Testbeds}

\subsubsection{5G-SMART}
5G-SMART~\cite{5gsmart}, started in June 2019, is a project funded by the European Commission and coordinated by Ericsson and ABB at Ericsson headquarters in Stockholm, Sweden. It aims at unlocking the value of 5G for smart manufacturing through demonstrating, validating and evaluating its potential in real manufacturing environments. 
To achieve this, 5G-SMART emphasises the three following groups of use cases which introduce the new challenges with respect to the seamless integration of 5G into a manufacturing system. i) Use cases targeting time-critical process optimization inside a factory. ii) Use cases targeting non-time critical in-factory communication for large number of devices. iii) Use cases targeting remote operation and massive information exchange. 

To investigate and validate the suitability of 5G for these use cases, three 5G trial have been setup within three factories. i) An Ericsson factory in Kista, Sweden, to implement the 5G-enhanced industrial robotics use cases and test spectrum coexistence aspects between the indoor and outdoor networks. ii) A factory shop-floor at the Fraunhofer IPT in Aachen, Germany, targets 5G-enabled real-time monitoring of work pieces and machining process. iii) A commercial semiconductor manufacturing plant of Bosch in Reutlingen, Germany, targets validation of use cases such cloud-based AGV control and Industrial LAN connectivity over 5G. 

\begin{figure}
  \centering \includegraphics[width=\linewidth]{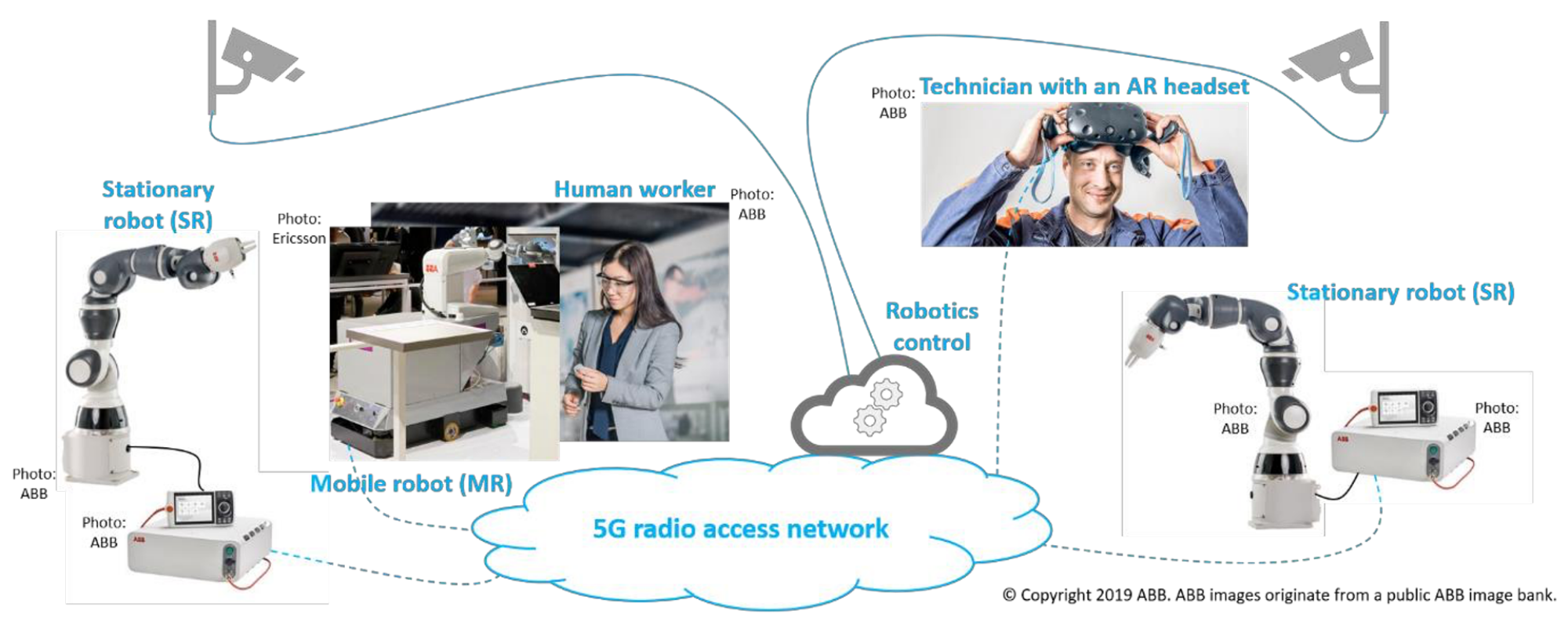}
  \caption{Setup illustration for use cases in Kista site of 5G-SMART.}
  \label{Fig:5g:smart}
  \vspace{-0.15in}
\end{figure}

\noindent {\bf Kista site.}
At the 5G-SMART trial site in Kista~\cite{5gsmart_d2.1}, three use cases (see Fig.~\ref{Fig:5g:smart}) are investigated for 5G-enabled industrial robotics, revolving around a factory-automation-like scenario (e.g., interactions between industrial robots, or between robots and human workers). 
Given the critical operational demands of the use cases (e.g., motion control), the main research question is that whether 5G systems can offer high-reliability and low-latency communication performance. 

Based on the functional architecture of the use cases, a 5G non-standalone (NSA) system operating at 28 GHz (5G band n257) using 200 MHz of bandwidth is deployed within an Ericsson factory. 
A 4G Radio Dot is installed on the ceiling to cover the complete testing area, where the robotics-related equipment is deployed. The 5G mmWave radio is wall-mounted and similarly covers the testing area. The base stations are connected to a local 5G core network which is co-located with the Edge cloud platform and installed nearby the testing area and the radios. Time-division duplexing (TDD) is used so that the spectrum is time-shared between UL and DL using the so called DDDSU pattern (4 DL slots and 1 UL slot). 
4G is primarily used for control traffic, such as for initial network access and paging, while 5G is exploited to carry data traffic. 
All network components are operated with commercial software. 

All the traffic running over the 5G network are served in a best-effort manner due to expectedly good latency performance and plentiful spectrum availability where the capacity and network capability in the factory can be seen as resource over-provisioning. 
The use case validations show that the 5G network coverage is excellent, and the performance, in terms of throughput and latency, is very stable over the shopfloor. The network performance is expected to support all the use cases being validated at the Kista trial site. 

\noindent {\bf Aachen site.}
At the Aachen trial site~\cite{5gsmart_d3.4}, two use cases are investigated for 5G-enabled monitoring of workpieces and machines: the wireless acoustic emission sensor system and the multi-sensor platform. The first use case demands a machine-near solution with a low latency and fast reaction times, the second use case requires the acquisition and transmission of several different signals resulting in varying connectivity requirements in terms of latency. 
Both a 5G NSA system and a 5G standalone (SA) system have been deployed forming a non-public network (NPN) on a shopfloor with an area of 3000 m$^2$ at the Fraunhofer Institute for Production Technology. 
The 5G RAN uses 100 MHz of bandwidth in the locally licensed mid-band TDD spectrum using DDDSU pattern at the 3.7–3.8 GHz spectrum (5G band n78). 
Measurement results are carried out from a single 5G terminal device running in the system where the radio network was loaded additional traffic created by up to three other background UEs. 
Different RAN configuration parameters related to scheduling and link robustness are experimented but no detailed descriptions on the applied resource management are delivered. 
The use case validations show that process and shopfloor monitoring applications can be realized via 5G and will provide a valuable tool to rapidly ramp up new processes and achieve a high product quality in a short time. 
monitoring equipment. 

\noindent{\bf Reutlingen site.}
At the Reutlingen trial site~\cite{5gsmart_d4.1}, two use cases are realized in the Bosch semiconductor factory: a cloud-based mobile robotics use case and a TSN / Industrial Local Area Network (LAN) over 5G (wired-to-wireless link replacement) use case. Both use cases pose the need for low latency and high reliable radio connectivity provided by the 5G system. 
A 5G SA system is deployed in the cleanroom factory floor of around 8000 sqm. The 5G NPN system operates in the locally licensed spectrum at 3.7–3.8 GHz (5G band n78) using a bandwidth of 100 MHz. The RAN network is realized with the Ericsson Radio Dot System and two radio cells are deployed with in total 16 radio dots. 
Some preliminary use case validation results show that controller-to-controller (C2C) applications can be realized via 5G, although the C2C communication performance over the wired Ethernet network outperforms that over the 5G network due to shorter round-trip time. 

\subsubsection{5G-VINNI}
5G Verticals Innovation Infrastructure (5G-VINNI~\cite{5gvinni}) project deploys multiple testbed facilities, to enable future projects and outside parties to experiment with 5G use cases. The 5G-VINNI facility is composed of several facility sites spanning many European countries. 5G VINNI’s combination of multiple interworking 5G RAN and 5G core infrastructures and E2E service orchestration provides a platform for testing and trialing industry use cases. 
Below, we describe the UK facility site solution as an example and readers can refer to~\cite{5gvinni_d2.1} for the detailed descriptions of other 5G-VINNI facility sites. 

\noindent{\bf UK site.}
The UK facility offers a full E2E 5G system with sub-6GHz and 28GHz radio at BT’s research centre at Adastral Park, Suffolk, UK~\cite{ghassemian2020experience}. 
The facility can validate a wide set of use cases, sufficient to demonstrate capability required to support a broad spectrum of 5G services, test against KPIs majoring on bandwidth, throughput, reliability, coverage and latency. 
Some example use cases supported at the UK site include public safety, transportation and connected vehicles, media production and distribution, and health and social care. 

All the 5G Core and RAN components at the UK site are supplied by Samsung. For 5G-NR implementation in sub-6GHz band, 100MHz of spectrum (between 3.6GHz and 3.7GHz) is utilized and NSA configuration is applied using LTE in Band 1 and the LTE anchor connection. For the mmWave band, 5GTF-based components is used, operating in the upper part of the 26GHz band. Both 4G base station (eNB) and 5G base station (gNB) connect to a fully virtualized 5G Core, implemented on servers in the central data centre. Nokia FlowOne solution is taking care of the E2E service orchestration function providing network slicing to perform network resource management.

The 5G-VINNI facility sites have undergone a rigorous validation procedure that included several categories of use case tests. Such test cases, executed using state-of-the-art open source (e.g., Openslice) and commercial testing tools, have measured the current capabilities of the 5G-VINNI infrastructure that can be used by verticals as of today. Some initial KPI results from 5G-VINNI facility sites can be found in~\cite{5gvinni_white_paper} and the values are in line with the expected status of development of the 5G-VINNI infrastructure, e.g., UL maximum throughput 104.27 Mbit/s.

\subsubsection{5GTNF}
The 5GTNF (5G Test Network Finland)~\cite{5gtnf} is a network of testing sites with 5G-capable networks to test new 5G applications and services in a controlled environment prior to commercialization. 
The goal of 5GTNF is to provide an ecosystem of testing and research facilities and to coordinate regulatory and technical work in Finland to ensure that there is a cohesive approach to 5G development. 
5GTNF is a joint effort from industry, academia and Finnish government and the infrastructure comprises of multiple interconnected sites around Finland. 
Many vertical projects focusing on different use cases of 5G and beyond networks are developed on the the 5GTNF site(s). 
Below, we provide a brief description of the 5G VIIMA project built on the Oulu site as an example, as detailed technical specifications for these 5GTNF projects are not currently available at the time of writing. 


\noindent{\bf 5G VIIMA.}
The 5G VIIMA (5G Vertical Integrated Industry for Massive Automation) project~\cite{5gviima} researches and experiments Industry 4.0 relevant 5G technologies and services for indoor and outdoor use. 
The project includes practical experiments inside a factory, in a controlled outdoor/indoor industry campus and with energy grids in four different locations in three Finnish cities (Tampere, Espoo, and Oulu). 
The four 5G VIIMA trial sites include Port Oulu industry campus, Kalmar industry campus, ABB smart grids and Nokia digital factory. 


\subsubsection{5G-DRIVE}
5G-DRIVE (5G HarmoniseD Research and TrIals for serVice Evolution between EU and China~\cite{5gdrive}), a project led by EURESCOM involving 17 European partners from 11 countries, aims to trial and validate the interoperability between EU \& China 5G networks operating at 3.5 GHz bands for enhanced Mobile Broadband (eMBB) and 3.5 \& 5.9 GHz bands for V2X scenarios. The project carries out its R\&D activities at three locations: Surrey (UK) site, Espoo (Finland) site and JRC Ispra (Italy) site where the Espoo site provides 5G testing facilities under the 5GTNF framework~\cite{kostopoulos20195g}. 

\begin{figure}
  \centering \includegraphics[width=0.9\linewidth]{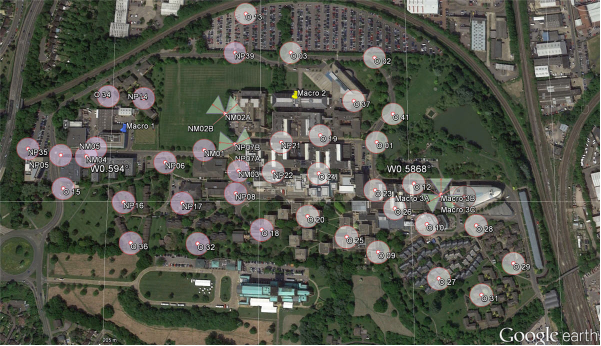}
  \caption{Surrey Trial site in 5G-DRIVE project.}
  \label{Fig:5g:surrey}
  \vspace{-0.15in}
\end{figure}

\noindent{\bf Surrey site.}
The Surrey site, operated by the 5G Innovation Centre (5GIC) at the University of Surrey, focuses on the development and evaluation of the eMBB scenario. The Surrey testbed is connected to the Vodafone Core Network, and covers a 4 km$^2$ area for the testing of 5G technologies (as shown in Fig.~\ref{Fig:5g:surrey}). 
This end-to-end testbed incorporates a different range of frequency bands (3.5 GHz, 28 GHz, and 60 GHz) and allows the testing and trialing of new air-interface solutions. Supported by a mix of wireless and fibre optic backhaul connectivity, trials can be matched to meet industry requirements. 

\noindent{\bf JRC Ispra site.}
The JRC Ispra trial site is a research campus featuring 36 km of roads under real-life driving conditions, nine Vehicle Emissions Laboratories (VELA 1-9), with a focus on the experimental evaluation of V2X scenarios both at the laboratory and field test levels.
The JRC owns various LTE eNodeBs of both commercial and experimental grade (Nokia FlexiZone indoor/outdoor units and USRP-based 4G eNodeBs), as well as commercial ITS-G5 Road-Side Units (RSUs) and On-Board Units (OBUs) for LTE-V2X/ITS-G5 coexistence testing. The aim of these tests is to study some of the RF/PHY/MAC phenomena stemming from co-channel coexistence of road ITS technologies through experimental means. Main evaluation metrics used in these evaluation include packet error rate and transmission latency. The results complied with the service requirements for V2X services set out in 3GPP TS 22.185~\cite{5gdrive_d4.4}.

\subsubsection{Liverpool 5G}
The Liverpool 5G Health and Social Testbed~\cite{liverpool5g_d} brought together the challenges currently facing health and social care services and the opportunities created by 5G mmWave technology to develop a private mesh network to support services in the Kensington area of Liverpool. 
The aim of this deployment is to support a range of social and healthcare applications for the community, from simple health sensor and monitoring services through to VoIP, full HD or 4K streaming for remote consultations and low latency VR tools. 

The network in the Liverpool trial is provided by Blu Wireless and uses mmWave frequencies in the 59 to 63 GHz frequency range. 34 DN101LC devices, each of which is a single radio 60GHz wireless node designed by Blu Wireless, are deployed at lamp-post sites. Each node is collocated with a Wi-Fi AP to provide WLAN connectivity, with some nodes also supporting ZigBee low-power networking. Each node is also collocated with a 5G small cell running in the n77 band and standalone mode~\cite{mackay2022modelling}.  
Based on the demonstration of a social gaming app created by CGA Simulation, which is the only use case with stringent latency requirement in Liverpool 5G, Blu Wireless mmWave links deliver round-trip times of around 0.4ms per link satisfying the requirement for VR and AR applications. 

\subsubsection{5GUK}
The 5GUK Test Networks project~\cite{5guk}, funded by the UK Government’s Department for Digital, Culture, Media and Sport (DCMS), consists of a collaboration between the 5G Innovation Centre at the University of Surrey, the University of Bristol~\cite{fernandez2021validating} and King's College London. 
The 5GUK test networks are used to trial 5G applications and technologies, with over 25 projects using it beyond the 5G Programme. 
2 projects from the industrial competition, 5GEM and 5G-ENCODE, were tested the benefits of using 5G to boost productivity in the manufacturing sector.

\noindent{\bf 5GEM.}
This project~\cite{5gem} focuses on the use of 5G in manufacturing to connect machines allowing real-time feedback, control, analysis and remote expert support, by conducting testbeds at the Ford Motor Company’s Dunton site and the TWI’s Cambridge facility. The use cases explored by 5GEM include real time process analysis and control demonstrated at Ford and intelligent maintenance demonstrated at TWI. 
A Vodafone mobile private network is installed at each site, while the specific network configuration is not publicly available. 

\noindent{\bf 5G-ENCODE.}
The 5G-ENCODE project~\cite{5gencode}, led by Zeetta Networks, is launched to design and deliver a private 5G network within the National Composites Centre. This will be used to explore new business models and 5G technologies, including network slicing and splicing, within an industrial environment. The 5G SA ORAN network in this project was deployed with a 100MHz bandwidth and several different use cases are delivered including asset tracking in-factory/out-of-factory, automated preforming control, and liquid resin infusion. Different 5G solutions are deployed in individual use cases with various performance measurements. In general, the project proved the value of 5G to industrial and manufacturing processes, and also identified many lessons learned. More details can be found in~\cite{5gencode_d}.

\subsubsection{5G Innovation Hub North}
Luleå University of Technology has in collaboration with Telia, Ericsson and TietoEVRY established five testbeds for 5G in Norrbotten and Västerbotten~\cite{5ginnovation}. 
All 5G testbeds use frequencies on the 3.5 GHz band and are open to all industry players, service providers, universities, and research institutes to test ideas, services and products. 
The testbeds have stable performance with downlink speeds of over 1 Gbps and uplink speeds of about 100 Mbps. The delay in the 5G environments is currently between 10 and 25 ms, depending on which of the testbeds is used~\cite{5glulea}.
An ERICSSON'S 5G RC car test example is also given in~\cite{5ginnovation_car} and the real-time performance of a 5G network with demanding video is evaluated.  

\subsubsection{5G Security Test Bed}
Washington – CTIA formally launched its 5G Security Test Bed (STB)~\cite{5gsecurity} in 2022, to develop and improve security features for industrial wireless networks and consumers. 
The 5G STB focuses on security use cases driven by operator priorities and government recommendations, and it prioritizes those use cases recommended by the FCC’s CSRIC technical advisory body, e.g., logistics management, automated factory processing, unmanned aerial systems and autonomous vehicles. 

The 5G STB is built on both SA and NSA 5G network architecture using state-of-the-art equipment and facilities. In NSA architecture, an Ericsson RAN housed at the University of Maryland directs user equipment (such as 4G and 5G smartphones) traffic to the Ericsson LTE Evolved Packet Core (EPC). 
In the SA 5G architecture, UE connects through the Ericsson 4G and 5G New Radios at the University of Maryland to the Ericsson Dual Mode 5G core at MITRE. 
The SA architecture enables a hybrid mode that simultaneously supports VoLTE (4G) voice calls and pure 5G data, as well as an evolving suite of 5G security functions. 
Some initial testing results on 3GPP Technical Specifications for network slicing validate various properties of networks partitioned through network slicing, including authentication, segmentation, security, and data integrity~\cite{5gstb_report}.

\subsubsection{PAWR}
NSF's PAWR program~\cite{PAWR} is currently supporting the deployment and initial operations of three advanced wireless research platforms conceived by the U.S. academic and industrial wireless research community. 
The four platforms funded to date by NSF's PAWR program are: COSMOS~\cite{COSMOS}, POWDER-RENEW~\cite{POWDER}, AERPAW~\cite{AERPAW}, and ARA~\cite{ARA}. ARA is a rural broadband testbed currently under construction in Ames, Iowa. 

\noindent{\bf COSMOS.} 
COSMOS (Cloud enhanced Open Software defined MObile wireless testbed for city-Scale deployment~\cite{raychaudhuri2020challenge}) is aimed at design, development, and deployment of a city-scale advanced wireless testbed in order to support real-world experimentation on next-generation wireless technologies and applications. The COSMOS architecture has a particular focus on ultra-high bandwidth and low latency wireless communication tightly coupled with edge cloud computing. As such it enables the design of beyond-5G and 6G networks. Researchers are able to run experiments remotely on the COSMOS testbed by logging into a web-based portal which provides various facilities for experiment execution, measurements, and data collection.
The COSMOS testbed is deployed in Upper Manhattan, covering one square mile in West Harlem, and its full deployment release to public is announced in January, 2024. Results of the COSMOS experimental trial via the Boldyn Networks fiber plant can be found in several recent papers, e.g.,~\cite{huang2023field,chen2023programmable,kostic2022smart,choudhury2022experimental}. 

\begin{figure}
  \centering \includegraphics[width=0.9\linewidth]{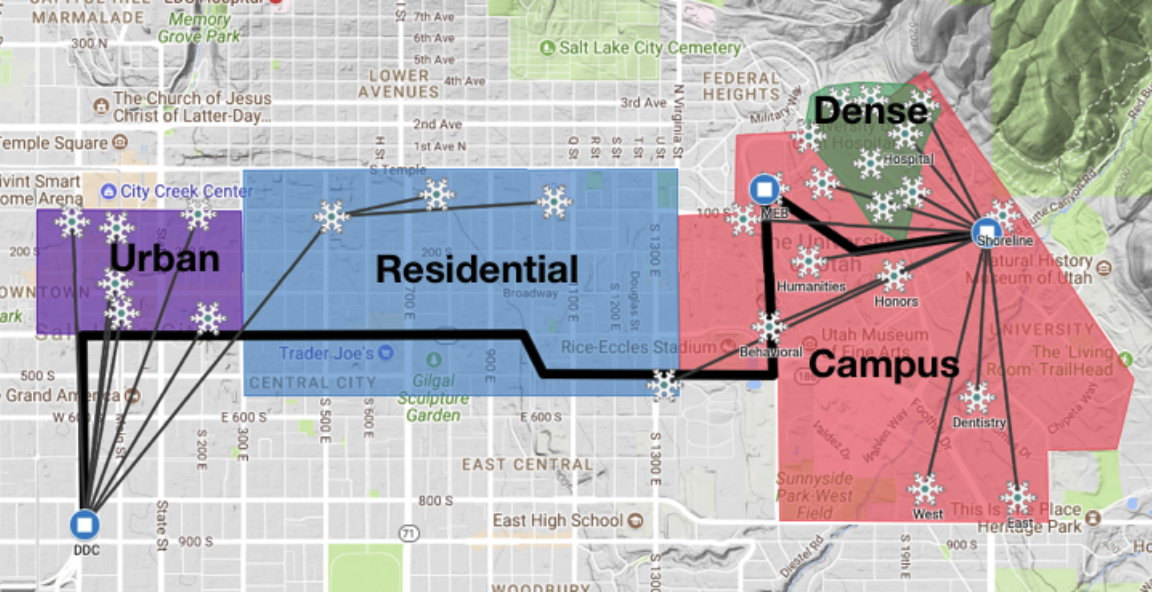}
  \caption{NSF-supported Powder city-scale footprint, including a campus, residential, urban and dense deployments. Snowflakes represent base stations.}
  \label{Fig:5g:powder}
  \vspace{-0.15in}
\end{figure}

\noindent{\bf POWDER.}
Powder (the Platform for Open Wireless Data-driven Experimental Research~\cite{breen2020powder}) is flexible infrastructure enabling a wide range of software-defined experiments on the future of wireless networks. It supports 
range of use cases by using the \textit{profile} mechanism to “package” hardware and software building blocks to creating starting points for a range of research with a variety of ready-to-go experimental recipes, including simulated RAN, controlled RF environment, over-the-air environment.
The entire Powder system is software-defined since Powder software building blocks include a variety of SDR stacks (such as GNU Radio, OpenAirInterface (OAI)~\cite{nikaein2014openairinterface}, and srsRAN~\cite{srsran}), core mobile networking stacks (such as free5GC~\cite{free5gc}, and OpenAirInterface Core). 
Some example use cases supported by Powder using its profile mechanism include RF monitoring, wireless communication (e.g., waveforms and coding techniques, RF propagation modeling, novel wireless architectures), and mobile communication (e.g., O-RAN ecosystem).

\noindent{\bf AERPAW.}
AERPAW (Aerial Experimentation and Research Platform for Advanced Wireless~\cite{marojevic2020advanced}) was named as the third PAWR testbed in 2019 and is designed with a focus on enabling experimentation at the intersection of advanced wireless technology, and Unmanned Aerial Vehicles (UAVs), or drone, technology. 
To satisfy the core requirements of AERPAW (e.g., remote usability, embedding in the real-world, and equipment flexibility), AERPAW uses SDR, and a custom vehicle and vehicle control platform fabricated from off-the-shelf hardware, open-source code, and purpose-built AERPAW software. 
AERPAW is composed of a number AERPAW Hardware Nodes (AHNs), each of which comprises multiple SDRs (National Instrument USRPs), and a companion computer to control them. Some AHNs are “Fixed”, mounted on 60-foot towers, others are “Portable”, battery-powered, possible to position at arbitrary locations, and also of being mounted on UAVs. 
Experiments on AERPAW use the AHNs to create complete working systems of srsRAN, OAI, GNU Radio, python scripts to collect raw in-phase and quadrature (I/Q) samples, pre-planned UAV trajectories, joint autonomous SDR and UAV control algorithms. 

\subsubsection{~5GENESIS}
5GENESIS (5th Generation End-to-end Network, Experimentation, System Integration, and Showcasing~\cite{xylouris2021experimentation}), started at October 2018, is an EU-funded project with the goal of 5G KPIs validation for various 5G use cases, in both commercial and industrial scenarios. 
The 5GENESIS facility is based on five end-to-end experimental platforms, distributed across Europe, having complementary features, yet all implementing the full 5G stack. 

The Athens 5G platform~\cite{5genesis_d4.3} comprises three dispersed sites in the Athens metropolitan area which features 5G and 4G radio access technologies (RATs) deployed in both indoor and outdoor environments combining software network technologies (i.e. NFV and SDN) and edge computing deployments. 
The industrial use case considered in this platform is UAV-aided surveillance deployed in the COSMOTE building (OTEAcademy site) in the north of Athens. The UAV applications range from low latency to high bandwidth scenarios, while requiring reliable communication channels for the security and safety of the related operations. 
The infrastructure layer of 5GENESIS at COSMOTE site include multiple components, i.e., 5G/4G RAN, core network equipment, and the edge clouds deployed in the edge data centers. 
For RAN, commercial 5G equipment based on Nokia AirScale platform is installed and operates with 5G NSA Core (including one AirScale BBU, 1 5G Base Transceiver Station and 2 n78 Airscale Micro RRH). 
The 5GC solution is based on Amarisoft’s 5GC implementation that is Rel. 15 and Rel. 16 compliant. A wide variety of COTS UEs are available and tested using the 3.5GHz band and NR TDD DDDSUUDDDD (4+2+4) SFS 3:8:3. 
Various KPIs, e.g., latency, capacity and throughput, are evaluated and detailed evaluation measurement results can be found in~\cite{5genesis_d6.2}. 

The mission of the Malaga platform~\cite{5genesis_d4.6} is to build a 5G multi-technology, multi-domain e2e platform that enables to validate the 5G KPIs for 3GPP Mission Critical Services (MCS) offered over e2e network slices. The Malaga platform comprises of five different sites distributed in three different locations for the evaluation of three use cases. In Malaga Police Department, a control room in the Malaga City Emergency Center is deployed to monitor the surveillance cameras and send control messages to UEs in the city. The RAN setup includes both indoor and outdoor radio deployment for LTE, 5G NSA and 5G SA. Radio access equipment deployed include Nokia 5G pico/mmW RRH, Nokia 4G Small Cell, Keysight 5G Wireless Test Platform, and prototype 5G RAN setups from OAI and RunEL et al. Detailed specifications for each of the RAN networks and CN deployment can be found in~\cite{5genesis_d4.6} and experimental results based on the 5G KPIs defined by the 5G PPP~\cite{5gppp} are delivered in~\cite{5genesis_d6.2}. 
For the descriptions of other platforms (Berlin~\cite{5genesis_d4.15}, Limassol~\cite{5genesis_d4.9} and Surrey~\cite{5genesis_d4.12}) deployed in 5GENESIS project, readers can refer to~\cite{5genesisdel}. 

\subsubsection{~5G Bavaria}

Fraunhofer IIS’s 5G Bavaria~\cite{5gbavaria} project aims to support the transition from research and standardization to application. 
It is primarily open to companies and enterprises, enabling them to evaluate new functionalities in a fully comprehensive 5G system context. The tools used to achieve this include simulation and emulation in the lab as well as application-specific 5G testbeds within real mobile communications environments. The 5G Bavaria Initiative encompasses a test center at Fraunhofer IIS in Erlangen and various testbeds in Bavaria. 

The 5G Bavaria test center at Fraunhofer IIS in Erlangen is a facility for testing 5G technologies using virtual simulation on the computer and realistic emulation in the lab. 
Link-level and system-level environments are available in the test center. Link-level simulations are used to optimize the performance of message transmission on a single mobile transmission path, while system-level simulations include all simultaneous transmissions in a communication network. For the lab-based emulation, extensive SDR platforms are available to emulate the properties of planned systems and components in hardware. More details on the available equipment in the emulation lab can be found in~\cite{5gbavariatest}. 

The 5G Industry 4.0 testbed~\cite{5gbavariaindustry} at Nürnberg, Germany is an open environment for testing specific customer use cases in industry and logistics. It provides truck accessible industrial facility (44 x 30 x 9 m) with reference systems for positioning and equipment for the emulation of industrial environments. 
It employs an SA 5G campus network with local frequency allocation at 3.7 GHz to 3.8 GHz (FR 1) and experimental radio license for the mm-wave bands n258 at 26 GHz and n257 at 28 GHz (FR 2). The RAN supports fully virtualized Open RAN architecture on commercial of-the-shelf (COTS) hardware with open interfaces and APIs. 

The 5G Bavaria automotive testbed~\cite{5gbavariaauto} at Rosenheim provides the infrastructure for testing 5G functionalities in a real traffic environment. 
Located on the southern outskirts of Rosenheim, the facility features a five-kilometer test area, covered by a closed 5G network supported by multiple base stations. 
The testbed is available for companies, research institutions and universities to test connected automotive applications (e.g., automated driving) with 5G. 

\subsubsection{~DEMAG Research Factory – MHD}
The DEMAG Research Factory – Material Handling Demonstrator (MHD) testbed~\cite{demag} offers the possibility to test 5G technology in a real industrial environment for intralogistics and material handling use cases. A 5G network and corresponding intralogistics equipment provide a platform in a manufacturing environment to quantitatively evaluate application requirements for 5G communication. 
The components for 5G infrastructure in the MH-demonstrator hall are delivered by NOKIA, including a 5G SA system based on Nokia Digital Automation Cloud Nx gNBs and On-Premise Core Initially based on 3GPP rel. 15. 
Further detailed description of the testbed is not revealed.

\begin{table*}[htb]
\caption{Summary of Small-Scale 5G-based IIoT Testbeds.}
\label{tab:5G_testbeds_small}
\begin{center}
\resizebox{\linewidth}{!}{
\begin{tabular}{|cc|c|c|c|c|c|c|c|}
\hline
\rowcolor[HTML]{EFEFEF} 
\multicolumn{2}{|c|}{\cellcolor[HTML]{EFEFEF}\textbf{Name or authors}}                                                                    & \textbf{Year} & \textbf{Inst.}                                                                       & \textbf{Region} & \textbf{Scenario}    & \textbf{Archt.} & \textbf{Provider}                                                        & \textbf{Orientation}               \\ \hline
\multicolumn{2}{|c|}{Mekikis et al.}                                                                                                      & 2020          & Iquadrat                                                                             & Europe          & Monitoring           & -               & -                                                                        & NFV experiments                    \\ \hline
\multicolumn{1}{|c|}{\begin{tabular}[c]{@{}c@{}}5G Lab\\ Germany\end{tabular}} & \begin{tabular}[c]{@{}c@{}}Rischke\\ et al.\end{tabular} & 2021          & TU Dresden                                                                           & Europe          & Robotics, AGV        & SA/NSA          & \begin{tabular}[c]{@{}c@{}}Core: Nokia/Open5GS\\ RAN: Nokia\end{tabular} & KPI measurement                    \\ \hline
\multicolumn{2}{|c|}{Aijaz et al.}                                                                                                        & 2021          & \begin{tabular}[c]{@{}c@{}}Bristol Research and\\ Innovation Laboratory\end{tabular} & Europe          & Generic              & NSA             & RAN: OAI                                                                 & Private network demonstration      \\ \hline
\multicolumn{2}{|c|}{Xu et al.}                                                                                                           & 2023          & CAS                                                                                  & Asia            & Generic              & -               & Unknown                                                                  & Modbus adaptation with 5G          \\ \hline
\multicolumn{2}{|c|}{Charpentier et al.}                                                                                                  & 2023          & University of Antwerp                                                                & Europe          & Transport, logistics & SA/NSA          & Telenet                                                                  & MEC for T\&L                       \\ \hline
\multicolumn{2}{|c|}{Boeding et al.}                                                                                                      & 2023          & Univ. of Nebraska-Lincoln                                                            & North America   & Smart grid           & SA              & RAN: OAI                                                                 & Experimental evaluation            \\ \hline
\multicolumn{2}{|c|}{John et al.}                                                                                                         & 2023          & \begin{tabular}[c]{@{}c@{}}Lubeck Univ. of\\ Applied Sciences\end{tabular}           & Europe          & Control automation   & SA              & RAN: OAI                                                                 & Industrial reference demonstrators \\ \hline
\end{tabular}
}
\end{center}
\end{table*}


\subsection{Small-Scale 5G-based IIoT Testbeds}
Research on 5G networks has witnessed explosive growth in recent years. In addition to the large-scale 5G testbed projects outlined in the previous section, numerous researchers have constructed small-scale or lab-sized 5G testbeds in their work~\cite{foukas2018experience,shorov20195g,koutlia2019design,esmaeily2020cloud,dreibholz2020flexible,wang2021development,nakkina2021performance,douarre2022design,wei20225gperf,zhang2023real,pineda2023design,mamushiane2023deploying,amini20235g,bahl2023accelerating,chepkoech2023evaluation}. Among these, a significant portion is not tailored for IIoT scenarios but rather serves general purposes. These include exploring how to design a 5G testbed, investigating the impact of different testbed elements on performance, or observing network performance in typical scenarios. 
Given the focus of our survey on IioT scenarios, in this section, we mainly describe the 5G testbeds specifically designed for industrial settings.

\subsubsection{Mekikis et al.}
In~\cite{mekikis2019nfv}, an NFV-enabled testbed based on the SEMIoTICS architecture~\cite{petroulakis2019smart} is built to enable secure and dependable smart sensing and actuation in IIoT applications. The testbed implements an end-to-end SDN/NFV architecture, complete with the local cloud, SDN networking, and field layers that demonstrate smart actuation, monitoring, and analytics functionalities. 
In the testbed, one four-core 64-bit server with 16 GB RAM acts as the controller and two six-core 64-bit servers with 32 GB RAM act as the compute nodes. Two Odroid C2 single-board computers act as the field layer virtualized IoT gateway. Note that, this paper employs IEEE 802.15.4 radio module to implement the radio access network interconnecting sensors with the gateway. Smart monitoring and actuation services are implemented in the form of VNFs, each in a dedicated tenant network that compete for resources. 

Bandwidth reservation is performed for each end-to-end slice in the testbed and on-line VNF migration is applied if any hypervisor is oversubscribed by a service slice. 
The IIoT testbed is evaluated in terms of the maximum throughput of two VNFs and the end-to-end latency of the critical actuation VNF. 
It concludes that sub-millisecond latency is achievable if services are directly hosted at the edge gateway and the network workload is less than $50\%$. 

\subsubsection{Rischke et al.}
In~\cite{rischke20215g}, a campus network testbed, for both 5G SA and 5G NSA structure, is constructed to measure the one-way packet delays between a 5G end device via a radio access network (RAN) to a packet core with sub-microsecond precision as well as for measuring the packet core delay with nanosecond precision. 
For the packet core, Open5GS~\cite{open5gs} (an open-source 5G core) is used in the case of SA and a proprietary packet core by Nokia is used for NSA. For the RAN, Nokia ABIA and ASIA hardware is used for the 4G part and Nokia ABIL and ASIK hardware is used for the 5G part. The radio transmits in the n78 band from 3.7Ghz to 3.8 Ghz with a bandwidth of 100 MHz and a transmitter power output of 17 dbm. The radio operates in the TDD mode with slot format $1/4$ for UL/DL. The RAN operates with a basic configuration without aggregation. Automation use case is assumed and Constant Bit Rate (CBR) traffic is generated using MoonGen~\cite{emmerich2015moongen} software. 

Comprehensive measurement results using various metrics (e.g., one-way delay, packet loss and downtime) are carried out and the performance comparison between SA network NSA network is also delivered. A wide range of packet delays is observed in the measurement for both SA and NSA structures, especially in the RAN (e.g., $95$\% of the SA download packet delays are in the $4–10$ ms range). Given the stringent timing requirement of some critical industrial applications (e.g., industrial robotics), the worst-case packet delay should be bounded with high reliability (e.g., $99.999$\%). Therefore, real-time 5G RAN scheduling mechanisms~\cite{dinh2022dynamic,zhang2023contention} need to be researched and developed so as to prioritize the guarantee on the packet latency.

\begin{figure}
  \centering \includegraphics[width=0.8\linewidth]{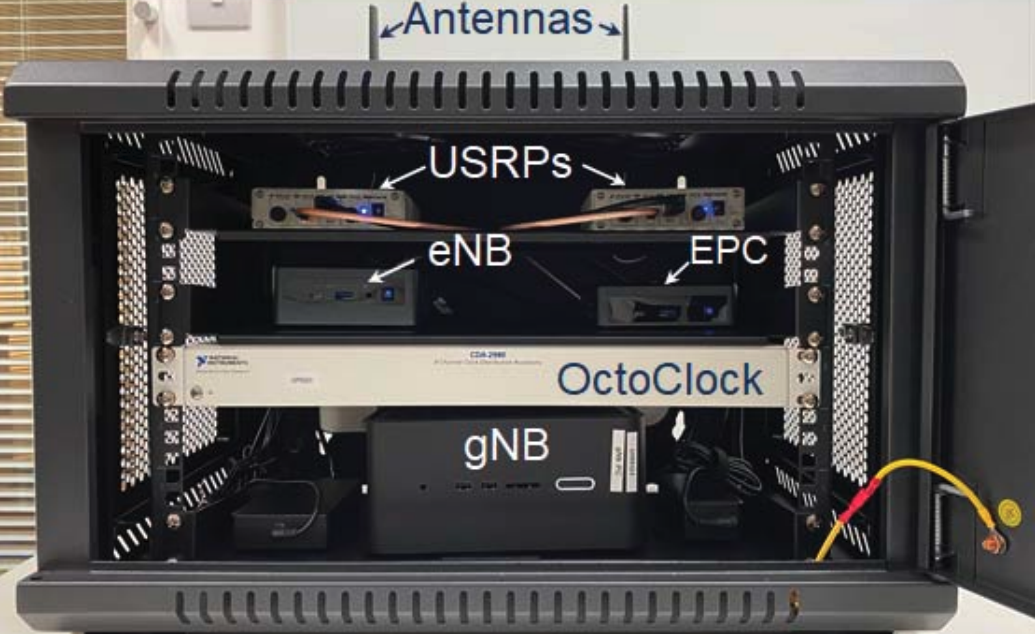}
  \caption{Prototype of the 5G NSA network-in-box testbed in~\cite{aijaz2021open}.}
  \label{Fig:5g:network-in-box}
  \vspace{-0.15in}
\end{figure}

\subsubsection{Aijaz et al.}
In~\cite{aijaz2021open}, authors demonstrates an open and programmable 5G network-in-a-box solution for private deployments (see Fig.~\ref{Fig:5g:network-in-box}). The testbed is based on OAI stack and general-purpose hardware, for operation in 5G NSA as well as 4G LTE modes.  The demonstration also shows the capability of operation in different sub-6 GHz frequency bands and provides empirical evaluations of 5G NSA performance, in terms of end-to-end latency and throughput. 
In the NSA mode, the 5G gNB co-exists with a 4G eNB and a 4G core network, also known as evolved packet core (i.e., EPC). The two base stations are connected via a 1 Gigabit Ethernet (GE) interface. The eNB and the gNB front-ends are based on Ettus USRP B210 connected via USB 3.0 interfaces to respective servers. The eNB/gNB front-ends use wideband antennas. Ettus OctoClock-G CDA-2990 is used  for time synchronization of the eNB and the gNB USRPs. The 5G network-in-a-box connects to 4G/5G COTS devices, including Quectel RM500Q-GL module and Samsung S10 5G. 

The 5G NSA implementation uses 40 MHz bandwidth and a sub-carrier spacing of 30 kHz. In the evaluation, the average round-trip latency of 5G NSA is 10.01 msec and a minimum value of 6.9 msec has been observed. Throughput evaluation based on TCP iPerf test shows a mean throughput for anchor bands 3 (1.8 GHz) and 7  (2.6 GHz) of 22.14 Mbps and 22.21 Mbps, respectively. A maximum throughput of 30.7 Mbps has also been observed. 

\subsubsection{Xu et al.}
\cite{xu20235g} studies protocol adaptation between Modbus (Modbus RTU and Modbus TCP) and 5G, and the authors develop an industrial wireless control system prototype including three parts: the programmable logic controller (PLC), the 5G-based protocol adapter, and the actuator. The 5G module in the protocol adapter is an industrial-level module using sub-6GHz band under both NSA and SA modes. The paper does not provide any further information about the specific 5G device used. The testbed-based evaluation results show that the speed and reliability of Modbus TCP adaptation are better than those of Modbus RTU adaptation. 

\subsubsection{5G EdgeApps.}
\cite{charpentier2023enhancing} deﬁnes a 5G testbed tailored to transport and logistic (T\&L) vertical services, which are designed and developed using the concept of 5G-based Edge Network Applications (EdgeApps) deﬁned within the European project VITAL-5G~\cite{trichias2021vital}. The testbed supports T\&L actors to experiment and validate their services within the real-life 5G-based T\&L environment (e.g., sea ports, river ports, and warehouses). The 5G EdgeApps testbed, deployed in the port of Antwerp, is based on the infrastructure of Telenet’s Innovation center supporting both 5G NSA and SA modes. The 5G SA cell is radiated by only N78 Time Division Duplex (TDD) RU, while the 5G NSA network is radiated by both N78 TDD RU and N1 Frequency Division Duplex (FDD) RU. 

The use case evaluated on the testbed is called Assisted Vessel Transport, which is built upon two vertical services, i.e., Remote Vessel Monitoring and Assisted Vessel Navigation. 5G connectivity and slicing are used to control semi-autonomous vessels in the challenging environment of a busy port area. The most important KPIs for this use case are i) latency less than 20 ms, ii) a throughput for the camera’s streams of around 300 Mbps. With the EdgeApps running at the edge on top of the testbed, some preliminary results validate performance improvement over 4G which is promising for T\&L services~\cite{charpentier2023enhancing}.

\subsubsection{Boeding et al.}
In~\cite{boeding2024toward}, an OAI-based in-lab 5G Core Network and gNB testbed is deployed for testing with off-the-shelf UE in smart grid networks with different OT protocols such as GOOSE, Modbus, and DNP3. The study outlines the latency impact of communication over 5G for time-critical and non-critical applications regarding their transition toward private 5G-based OT network implementations. 

The 5G testbed is located at the Advanced Telecommunication Engineering Laboratory (TEL) at the University of Nebraska-Lincoln. The gNB is implemented using GNURadio, with the signal processing aspects performed by a bare-metal (non-Dockerized) implementation on a server with operating system and kernel specifications identical to those of the CN. The radio frontend utilized is an Ettus USRP B210, operating on band n78 in Time Division Duplex (TDD) mode at a center frequency of 3.6192 GHz and a 40 MHz wide channel. The UE functionality is provided by a Raspberry Pi 4 with Ubuntu 22.04 LTS installed, along with a Quectel RM520N-GL 5G HAT Modem. 
The e2e latency measurements on the testbed for different test cases reveal some shortcomings in the case of GOOSE packets, where simple encapsulation may exceed the protocol’s time-critical nature. Readers can refer to~\cite{boeding2024toward} for more experiment results.

\begin{figure}
  \centering \includegraphics[width=\linewidth]{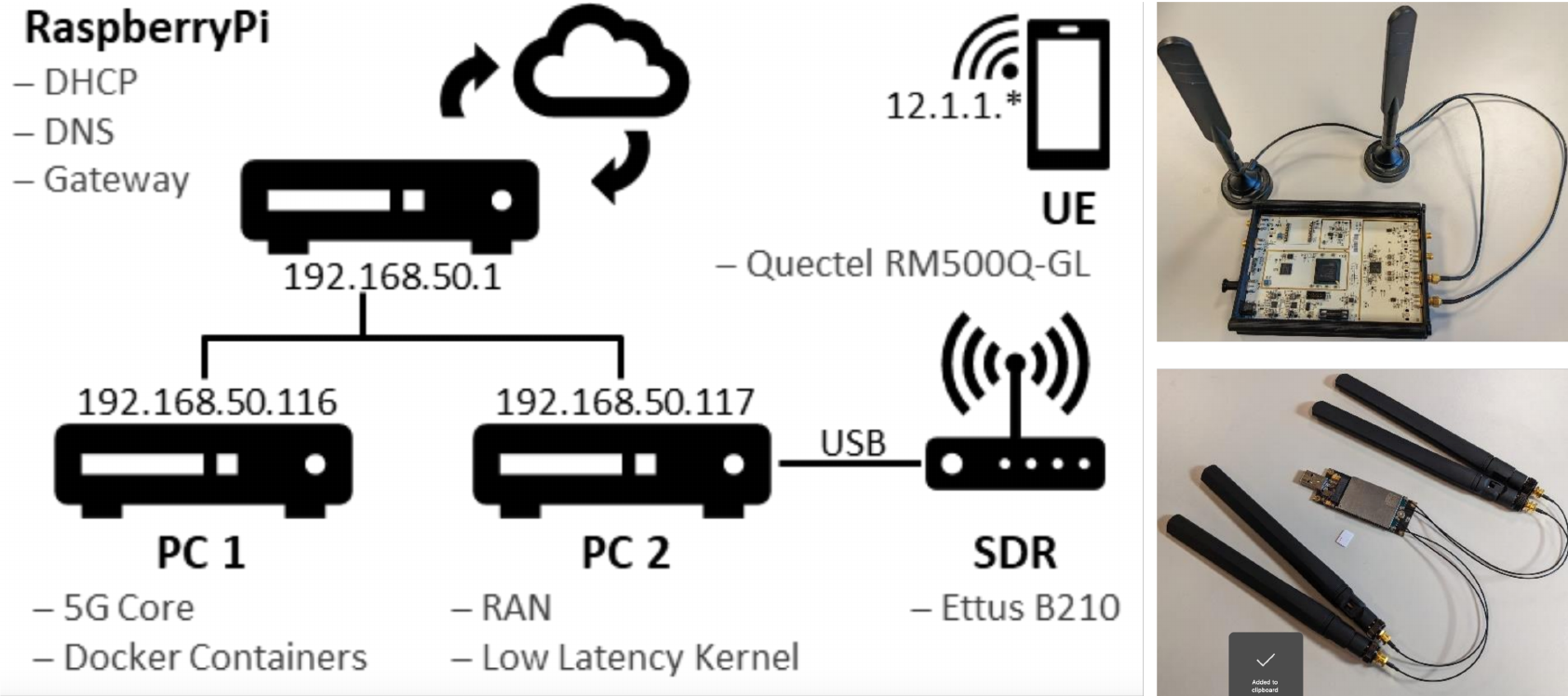}
  \caption{Structure of the 5G testbed at the Technische Hochschule Lübeck – University of Applied Sciences~\cite{john2022reference}.}
  \label{Fig:5g:reference}
  \vspace{-0.15in}
\end{figure}

\subsubsection{John et al.}
In~\cite{john2023two}, to demonstrate the advantages of high data throughput and low latency in 5G networks, a 5G testbed is deployed for experimental evaluation with two demonstrators (high throughput and low latency) in the industrial context. Fig.~\ref{Fig:5g:reference} shows the structure of the 5G SA testbed deployment and some real hardware used in the testbed, including USRP B210 and Quectel RM500Q-GL. 

To demonstrate the high throughput capabilities of 5G, uncompressed 4K images are transmitted continuously from the server (Core network) to the client (UE) through a TCP socket. The measured downlink throughput is below 50 Mbit/s with a median of 34.61 Mbit/s, serving as a benchmarking purpose. For the low latency demonstrator, an inverted pendulum, which is well-known controlling problem, is assumed where the system’s controller runs on a virtual server in the 5G core and sends control messages to the UE. A median latency (round-trip time) of 10.21 ms is measured with the demonstrator and the system is controllable.

 \section{Suggestions and Practices}\label{sec:lesson}
Based on the knowledge acquired from the comprehensive survey on IIoT testbeds, in this section, we provide suggestions and good practices for researchers and practitioners who are interested to design and develop an IIoT testbeds. An effective IIoT testbed development entails multiple steps, each varying in complexity and cost. As researchers and practitioners consider embarking on the development of an IIoT testbed, they must first address the following two questions. 

\vspace{0.05in}
\noindent {\bf Q1: What is the Purpose of the Testbed?} During the testbed design phase, researchers must clearly define the purpose of the testbed, that is, its primary intended use. An IIoT testbed can serve various purposes, including 1) exploration or discovery of new methodologies, 2) demonstration of system performance or validation of existing research findings, and 3) educational purposes. Based on these distinct objectives, the design of the testbed may differ. For instance, a testbed intended for exploration or discovery may require specific configurations based on existing assumptions and hypotheses, such as specific communication interference settings. A testbed intended for demonstration purposes may involve implementing one's own designed methods into the built testbed, which may introduce additional workload. An educational testbed may primarily rely on standard settings during development but may require careful selection of representative and easily demonstrable testbed types to fulfill its educational objectives.

\vspace{0.05in}
\noindent {\bf Q2: Build Your Own or Remotely Access Others?} Many research institutions, government bodies, or companies now offer projects that provide free or paid access to remotely accessible testbeds, greatly facilitating research efforts. However, researchers must weigh the advantages and disadvantages of building their own testbed versus accessing public ones remotely. From the cost perspective, developing certain testbeds may require expensive equipment (e.g., high-end 5G gNBs) or a large number of devices (e.g., large-scale WSN systems), making the use of remotely accessible testbeds significantly more cost-effective. Additionally, some projects offer detailed usage instructions and manuals, thereby enhancing usability. However, from the availability standpoint, remotely accessible testbeds often rely on virtualization techniques for multi-user sharing, which may not guarantee availability in terms of usage duration, device availability, or continuity. Users may also need to schedule appointments for each usage instance, resulting in inconvenience. Furthermore, some shared testbeds do not provide interfaces or allow users to adjust or modify the software running on the testbed or its parameters, which can be inconvenient for researchers requiring customized testbed configurations for research validation purposes.

Once the aforementioned questions have been addressed and the decision to build an IIoT testbed has been made, the next step is to develop a testbed that meets the specific requirements. Some key considerations include:
1) Determining the communication protocol (e.g., TSN, IEEE 802.15.4, IEEE 802.11 or 5G) based on the target industrial scenario or research purpose. This survey paper discusses most commonly used communication protocols in emerging IIoT systems, covering both wired and wireless transmission protocols. 
2) Selecting appropriate software and hardware. For educational or demonstrative purposes, commercial off-the-shelf (COTS) solutions containing hardware and software may be suitable, offering convenience albeit at a potentially higher cost and limited customization. Alternatively, programmable and customizable hardware devices such as SDR and SDN may require consideration of specific hardware models based on functionality requirements. Open-source projects for software may vary in terms of functionality and support for different communication protocols. 
3) Deciding on network configuration, contingent upon the used communication protocol, device quantity, network topology, and architecture. 
 \section{Conclusion}
\label{sec:concl}
In this survey article, we have delved into the diverse landscape of IIoT testbeds and their critical role in shaping the future of industrial automation and innovation. Through an extensive literature review, we explored various types of testbeds deployed based on different communication protocols suitable for individual industrial application scenarios. Finally, we present good practices and suggestions for researchers and practitioners who are interested to design and develop their own IIoT testbed to perform function and performance validation, as well as new scientific and engineering discovery.  

\section{Acknowledgement}

The work is supported in part by the NSF Grant CNS-1932480, CNS-2008463, CCF-2028875 and CNS-1925706. 

\bibliographystyle{IEEEtran}

\bibliography{refs}


\end{document}